\documentclass[twocolumn]{aastex63}

\usepackage{graphicx}
\usepackage{longtable}
\usepackage{mdwlist}

\usepackage{natbib}
\usepackage{multirow}
\usepackage{upgreek}
\usepackage{wasysym}
\usepackage{txfonts}
\usepackage{gensymb}
\usepackage{morefloats}
\usepackage{threeparttable}
\usepackage{rotating}
\usepackage{graphicx}
\usepackage{color}

\usepackage{lineno}
\shorttitle{3CR\,196.1: H$\alpha$ emission associated with an X-ray cavity}
\shortauthors{Jimenez-Gallardo et al.}

\begin{document}
\title{The cavity of 3CR\,196.1: H$\alpha$ emission spatially associated with an X-ray cavity} % @tit
\correspondingauthor{Ana Jimenez-Gallardo}
\email{ana.jimenezgallardo@unibo.it}

\author[0000-0003-4413-7722]{A. Jimenez-Gallardo}
\affiliation{Dipartimento di Fisica, Universit\`a degli Studi di Torino, via Pietro Giuria 1, I-10125 Torino, Italy}
\affiliation{European Southern Observatory, Alonso de C\'ordova 3107, Vitacura, Regi\'on Metropolitana, Chile}
\affiliation{INAF-Osservatorio Astrofisico di Torino, via Osservatorio 20, 10025 Pino Torinese, Italy}
\affiliation{Dipartimento di Fisica e Astronomia dell’Universit\`a degli Studi di Bologna, via P. Gobetti 93/2, I-40129 Bologna, Italy}

\author[0000-0002-3140-4070]{E. Sani}
\affiliation{European Southern Observatory, Alonso de C\'ordova 3107, Vitacura, Regi\'on Metropolitana, Chile}

\author[0000-0001-5742-5980]{F. Ricci}
\affiliation{Dipartimento di Fisica e Astronomia dell’Universit\`a degli Studi di Bologna, via P. Gobetti 93/2, I-40129 Bologna, Italy}
\affiliation{INAF- Osservatorio di Astrofisica e Scienza dello Spazio di Bologna, via Gobetti 93/3, I-40129 Bologna, Italy}

\author[0000-0002-5941-5214]{C. Mazzucchelli}
\affiliation{Núcleo de Astronomía, Facultad de Ingeniería y Ciencias, Universidad Diego Portales, Av. Ejército 441, Santiago, 8320000, Chile}
\affiliation{European Southern Observatory, Alonso de C\'ordova 3107, Vitacura, Regi\'on Metropolitana, Chile}

\author[0000-0002-0690-0638]{B. Balmaverde}
\affiliation{INAF-Osservatorio Astrofisico di Torino, via Osservatorio 20, 10025 Pino Torinese, Italy}

\author[0000-0002-1704-9850]{F. Massaro}
\affiliation{Dipartimento di Fisica, Universit\`a degli Studi di Torino, via Pietro Giuria 1, I-10125 Torino, Italy}
\affiliation{Istituto Nazionale di Fisica Nucleare, Sezione di Torino, I-10125 Torino, Italy}
\affiliation{INAF-Osservatorio Astrofisico di Torino, via Osservatorio 20, 10025 Pino Torinese, Italy}
\affiliation{Consorzio Interuniversitario per la Fisica Spaziale, via Pietro Giuria 1, I-10125 Torino, Italy}

\author[0000-0003-3684-4275]{A. Capetti}
\affiliation{INAF-Osservatorio Astrofisico di Torino, via Osservatorio 20, 10025 Pino Torinese, Italy}

\author[0000-0002-9478-1682]{W. R. Forman}
\affiliation{Center for Astrophysics $|$ Harvard \& Smithsonian, 60 Garden Street, Cambridge, MA 02138, USA}

\author[0000-0002-0765-0511]{R. P. Kraft}
\affiliation{Center for Astrophysics $|$ Harvard \& Smithsonian, 60 Garden Street, Cambridge, MA 02138, USA}

\author[0000-0002-5941-5214]{G. Venturi}
\affiliation{Instituto de Astrof\'isica, Facultad de F\'isica, Pontificia Universidad Cat\'olica de Chile, Casilla 306, Santiago 22, Chile}
\affiliation{INAF - Osservatorio Astrofisico di Arcetri, Largo E. Fermi 5, I-50125 Firenze, Italy}

\author[0000-0002-7326-5793]{M. Gendron-Marsolais}
\affiliation{European Southern Observatory, Alonso de C\'ordova 3107, Vitacura, Regi\'on Metropolitana, Chile}
\affiliation{Instituto de Astrof\'isica de Andaluc\'ia (IAA-CSIC), Glorieta de la Astronom\'ia, 18008 Granada, Spain}

\author[0000-0002-3585-2639]{M. A. Prieto}
\affiliation{Departamento de Astrof\'isica, Universidad de La Laguna, E-38206 La Laguna, Tenerife, Spain}
\affiliation{Instituto de Astrof\'isica de Canarias (IAC), E-38200 La Laguna, Tenerife, Spain}

\author[0000-0002-9889-4238]{A. Marconi}
\affiliation{INAF - Osservatorio Astrofisico di Arcetri, Largo E. Fermi 5, 50125 Firenze, Italy}
\affiliation{Dipartimento di Fisica e Astronomia, Universit\`a di Firenze, via G. Sansone 1, 50019 Sesto F.no, Firenze, Italy}

\author[0000-0003-0032-9538]{H. A. Pe\~na-Herazo}
\affiliation{East Asian Observatory, 660 North A’ohōkū Place, Hilo, HI 96720, USA}

\author{S. A. Baum}
\affiliation{University of Manitoba, Dept. of Physics and Astronomy, Winnipeg, MB R3T 2N2, Canada}

\author[0000-0001-6421-054X]{C. P. O'Dea}
\affiliation{University of Manitoba, Dept. of Physics and Astronomy, Winnipeg, MB R3T 2N2, Canada}

\author[0000-0002-3754-2415]{L. Lovisari}
\affiliation{INAF- Osservatorio di Astrofisica e Scienza dello Spazio di Bologna, via Gobetti 93/3, I-40129 Bologna, Italy}

\author[0000-0001-8121-6177]{R. Gilli}
\affiliation{INAF- Osservatorio di Astrofisica e Scienza dello Spazio di Bologna, via Gobetti 93/3, I-40129 Bologna, Italy}

\author[0000-0002-5201-010X]{E. Torresi}
\affiliation{INAF- Osservatorio di Astrofisica e Scienza dello Spazio di Bologna, via Gobetti 93/3, I-40129 Bologna, Italy}

\author[0000-0002-5646-2410]{A. Paggi}
\affiliation{Dipartimento di Fisica, Universit\`a degli Studi di Torino, via Pietro Giuria 1, I-10125 Torino, Italy}
\affiliation{Istituto Nazionale di Fisica Nucleare, Sezione di Torino, I-10125 Torino, Italy}
\affiliation{INAF-Osservatorio Astrofisico di Torino, via Osservatorio 20, 10025 Pino Torinese, Italy}

\author[0000-0001-8382-3229]{V. Missaglia}
\affiliation{Dipartimento di Fisica, Universit\`a degli Studi di Torino, via Pietro Giuria 1, I-10125 Torino, Italy}
\affiliation{Istituto Nazionale di Fisica Nucleare, Sezione di Torino, I-10125 Torino, Italy}
\affiliation{INAF-Osservatorio Astrofisico di Torino, via Osservatorio 20, 10025 Pino Torinese, Italy}

\author[0000-0002-5445-5401]{G. R. Tremblay}
\affiliation{Center for Astrophysics $|$ Harvard \& Smithsonian, 60 Garden Street, Cambridge, MA 02138, USA}

\author[0000-0003-1809-2364]{B. J. Wilkes}
\affiliation{Center for Astrophysics $|$ Harvard \& Smithsonian, 60 Garden Street, Cambridge, MA 02138, USA}

\begin{abstract} %@abs
We present a multifrequency analysis of the radio galaxy 3CR\,196.1 ($z = 0.198$), associated with the brightest galaxy of the cool core cluster CIZAJ0815.4-0303. This nearby radio galaxy shows a hybrid radio morphology and an X-ray cavity, all signatures of a turbulent past activity, potentially due to merger events and AGN outbursts. We present results of the comparison between {\it Chandra} and VLT/MUSE data for the inner region of the galaxy cluster, on a scale of tens of kpc. We discovered H$\alpha$ + [N II]$\lambda6584$ emission spatially associated with the X-ray cavity (at $\sim$10 kpc from the galaxy nucleus) instead of with its rim. This result differs from previous discoveries of ionized gas surrounding X-ray cavities in other radio galaxies harbored in galaxy clusters and could represent the first reported case of ionized gas filling an X-ray cavity, either due to different AGN outbursts or to the cooling of warm ($10^4<T\leq10^7$ K) AGN outflows. We also found that the H$\alpha$, [N II]$\lambda\lambda6548,6584$ and [S II]$\lambda\lambda6718,6733$ emission lines show an additional redward component, at $\sim$1000 km$\,$s$^{-1}$ from rest frame, with no detection in H$\beta$ or [O III]$\lambda\lambda4960,5008$. We believe the most likely explanation for this redward component is the presence of a background gas cloud since there appears to be a discrete difference in velocities between this component and the rest frame.

\end{abstract}


\section{Introduction}
\label{sec:intro}

The extragalactic fraction of the Third Cambridge Catalog of radio sources and its revised versions (3C, 3CR, 3CRR, \citealt{Edge1959,Bennett1962,Spinrad1985,Laing1983}) constitute one of the most valuable samples of powerful radio-loud active galactic nuclei (AGN) to explore feedback processes occurring between radio galaxies and their environments (see, e.g., \citealt{McNamara2007, McNamara2012}, \citealt{Fabian2012} and \citealt{Kraft2012}).

One of the main pieces of evidence of radio mode feedback occurring in AGN in galaxy clusters and groups is the presence of X-ray cavities, first reported by \citet{Boringer1996} and \citet{Churazov2000}, due to the interaction of jets with the intracluster medium (ICM; see reviews by \citealt{McNamara2007}, \citealt{Fabian2012} and \citealt{Gitti2012}, and works by \citealt{Birzan2004}, \citealt{Cavagnolo2010}, and \citealt{McNamara2012}).
Some of the most remarkable examples of cavities include those of Perseus (see \citealt{Fabian2003,Fabian2006} and \citealt{Graham2008}), M87 (see e.g., \citealt{Forman2005,Forman2017}), Hydra A (\citealt{Nulsen2005a}), Hercules A (\citealt{Nulsen2005b}), MS0735.6+7421 (\citealt{McNamara2005,McNamara2009}), NGC 5813 (\citealt{Randall2011,Randall2015}), A2052 (\citealt{Blanton2011}), A2597 (\citealt{Tremblay2012}), and NGC\,4636 (\citealt{Jones2002}).

Although it is common to find H$\alpha$ emission either surrounding X-ray cavities (see e.g., \citealt{Blanton2011}, \citealt{Tremblay2015}, \citealt{Balmaverde2018}), or spatially associated with radio emission (known as the ``alignment effect"; see e.g., \citealt{Fosbury1986}, \citealt{Hansen1987}, \citealt{Baum1988}, \citealt{McCarthy1988}, \citealt{Baum1990}, \citealt{Tremblay2009} and \citealt{Baldi2019}), the effect of radio jets on extended emission-line regions (EELR, regions of line-emitting gas on scales of tens of kpc, see e.g., \citealt{Baum1988}) is not yet fully understood.

A recent analysis of radio, optical, and X-ray emission in 3CR\,196.1 showed it is a promising source to investigate how radio jets affect EELRs, thanks to the discovery of an X-ray cavity at $\sim$10 kpc from the nucleus toward the southwest (see \citealt{Ricci2018}). Moreover, recent observations from the Multi-Unit Spectroscopic Explorer (MUSE; \citealt{Bacon2010}) at the Very Large Telescope (VLT) obtained as part of the MUse RAdio Loud Emission line Snapshot (MURALES) survey (\citealt{Balmaverde2021}) revealed that the H$\alpha$ + [N II]$\lambda6584$ emission extends beyond the radio emission and appears, in projection, spatially associated with the X-ray cavity.

To test where such optical emission stands with respect to the X-rays, we compared the optical, X-ray, and radio emission of 3CR\,196.1 up to a few tens of kpc, taking advantage of new VLT/MUSE observations. This emission had not been detected in previous {\it HST} data originally published in \citet{Tremblay2009}, due to {\it HST}'s lower sensitivity compared to MUSE. This spatial association of ionized gas with an X-ray cavity, instead of with its borders has been rarely reported in the literature (see e.g., A1668, \citealt{Pasini2021}). We focus on the comparison of X-ray and optical observations, previously investigated only separately by \citet{Ricci2018} and \citet{Balmaverde2021}, respectively.

%Cavity systems are often surrounded by belts (Smith et al. 2002), arms (Young et al. 2002; Forman et al. 2005, 2007), filaments, sheets (Fabian et al. 2006), and frag-ile tendrils of gas maintained against thermal evaporation, perhaps, by magnetic fields threaded along their lengths (Nipoti & Binney 2004; Forman et al. 2007). This structure is usually composed of cooler gas and is associated with H emis-sion that may be tracing circulation driven by rising radio lobes and cavities (Fabian et al. 2003b).

This manuscript is organized as follows. A brief description of 3CR\,196.1 is given in \S~\ref{sec:target}. Astrometric registration of all images is reported in \S~\ref{sec:reduc}, together with details about data reduction procedures for optical and X-ray data sets. Results and their discussion are presented in \S~\ref{sec:results} and \S~\ref{sec:discussion}, respectively, while \S~\ref{sec:conclus} is devoted to our conclusions. Further details on the astrometric registration and the ionized gas kinematics can be found in Appendices \ref{app:regis} and \ref{sec:maps}.

We adopted cgs units for numerical results and assumed a flat cosmology with $H_0=69.6$ km s$^{-1}$ Mpc$^{-1}$, $\Omega_{M}=0.286$ and $\Omega_{\Lambda}=0.714$ \citep{bennett14}, unless otherwise stated. At the source redshift (i.e., $z=0.198$) the physical scale is 3.299 kpc/arcsec. Standard astronomical orientation (north upwards and east to the left) is adopted throughout the paper.

\section{The strange case of 3CR\,196.1}
\label{sec:target}

3CR\,196.1 is the radio galaxy associated with the brightest cluster galaxy of the galaxy cluster CIZA\,J0815.4-0303 (\citealt{Kocevski2007}). From an optical perspective, this radio galaxy is classified as a low-excitation radio galaxy (\citealt{Buttiglione2010}), while at radio frequencies it presents a hymor radio structure (see \citealt{Gopal2000}), with a classical edge-darkened morphology, typical of FR\,Is, toward the southwest and a radio lobe, typical of FR\,IIs on the opposite side (\citealt{Fanaroff1974}).

Near-infrared (\citealt{Madrid2006}) and optical continuum observations (\citealt{Baum1988,deKoff1996}) revealed that the host galaxy of 3CR\,196.1 shows an elliptical morphology elongated from northeast to southwest, in the same direction as the radio jet. Optical images also revealed the presence of periodic shells possibly due to past merger events (\citealt {Zirbel1996}), in agreement with centroid shifts measured using optical isophotes of its host galaxy (\citealt{deKoff1996}). Additionally, using {\it HST} emission line images, \citet{Baldi2019} measured a lower limit for the total ionized gas mass for 3CR\,196.1 of 3.5$\cdot$10$^5$ M$_\odot$, using the nuclear electron density. The analysis of newly obtained MUSE data focused on the [O III]$\lambda5007$ properties by \citet{Speranza2021} revealed a broad blue [O III]$\lambda5007$ component with a maximum velocity of $\sim$-640 km$\,$s$^{-1}$, extending toward the northeast, in the same direction as the radio jet.

\citet{Ricci2018} confirmed the presence of an X-ray surface brightness depression, previously suggested by \citet{Massaro2012}, by carrying out a multiwavelength analysis of archival radio, optical, and X-ray observations. This surface brightness depression indicates the presence of a ``butterfly-shaped" cavity at $\sim$10 kpc from the nucleus (see left panel of Fig. \ref{fig:chandra}). The temperature of the galaxy cluster ($\sim$4.2 keV) drops to $\sim$2.8 keV in the inner 30 kpc, suggesting that 3CR\,196.1 is the brightest cluster galaxy of a cool core cluster.

At scales of tens of kpc, the H$\alpha$ + [N II]$\lambda6584$ emission observed with the {\it Hubble Space Telescope} is spatially aligned with the radio jet and lies in the base of the X-ray cavity. Lastly, \citeauthor{Ricci2018} concluded that 3CR\,196.1 could be an example where cold gas is being uplifted by radio lobes and that it could have undergone several AGN outbursts at multiple epochs. They also suggested its galaxy cluster may have experienced a merger episode, revealed by the spiral pattern seen in the X-ray observation, characteristic of gas sloshing (see, e.g., \citealt{Markevitch2007}).

%(i.e., a galaxy cluster with low central entropy, decreasing central temperatures, and short central cooling times, see \citealt{Hudson2010})

\section{Data analysis procedures}
\label{sec:reduc}

X-ray data employed for this work are $\sim$8 ks archival Chandra observations (ObsID 12729). The MUSE dataset was obtained from the MURALES survey (ID 0102.B-0048(A); two 10 minute exposures). Details on these observations as well as on the \textit{Chandra} and MUSE data reduction and analyses can be found in \citet{Ricci2018} and \citet{Balmaverde2021}, respectively.
%The 8.4 GHz radio map was obtained from the Very Large Array (VLA\footnote{https://science.nrao.edu/facilities/vla/archive/index}) archive.

\subsection{Astrometric registration}

Astrometric registration was carried out by comparing 8.4 GHz Very Large Array (VLA\footnote{https://science.nrao.edu/facilities/vla/archive/index}), Pan-STARRS\footnote{https://catalogs.mast.stsci.edu/panstarrs/} $r$-band, MUSE, and \textit{Chandra} images. An image showing the sources used for the registration can be found in Appendix \ref{app:regis}.

We decided against registering the \textit{Chandra} image since the nucleus of 3CR\,196.1 is not detected in the \textit{Chandra} image and the typical shift applied to \textit{Chandra} images is $<$0.7\arcsec\ (\citealt{Massaro2011}), which corresponds to a shift of less than 2 pixels in our X-ray images and, thus, it is sufficiently accurate for our comparison. Additionally, no registration was performed on the radio since the radio core is not detected in the available radio maps and their positional uncertainty is $<0.03\arcsec$ (adopting a conservative VLA uncertainty of 10\% of the beam size\footnote{https://science.nrao.edu/facilities/vla/docs/manuals/oss/performance/positional-accuracy}). Thus, the final astrometric registration of MUSE data relied on the Pan-STARRS $r$-band image since all sources in the MUSE data were also detected in Pan-STARRS. A detailed description of the strategy adopted to register the data can be found in Appendix \ref{app:regis}.

%We only registered the MUSE data, since its astrometry could be offset by several arcseconds, and decided against registering the \textit{Chandra} observation, since the typical offset in \textit{Chandra} images is $<$0.7\arcsec\ (i.e., less than two pixels in the X-ray image and less than 2 kpc; \citealt{Massaro2011}), which is accurate enough for our analysis. 

 %MUSE data were registered to the Pan-STARRS $r$-band image since all sources in the MUSE data were also detected in Pan-STARRS. We measured the centroids of 3CR\,196.1 and other four field sources in Pan-STARRS and in the white light collapsed MUSE image and applied the average of the shifts to the MUSE cube. The final shift was 3.1\arcsec\ (corresponding to $\sim$10 kpc), which we verified provided a good alignment for the rest of field sources, with an rms of 0.18\arcsec. The result of the astrometric registration is shown in the left panel of Fig. \ref{fig:chandra}, where we show the 0.7 - 2 keV X-ray emission with H$\alpha$ + [N II]$\lambda6584$ contours overlaid in blue, the position of the host marked with a green cross and the direction of the radio jet indicated in cyan.

\begin{figure*}
\begin{center}
\includegraphics[height=6.9cm,angle=0]{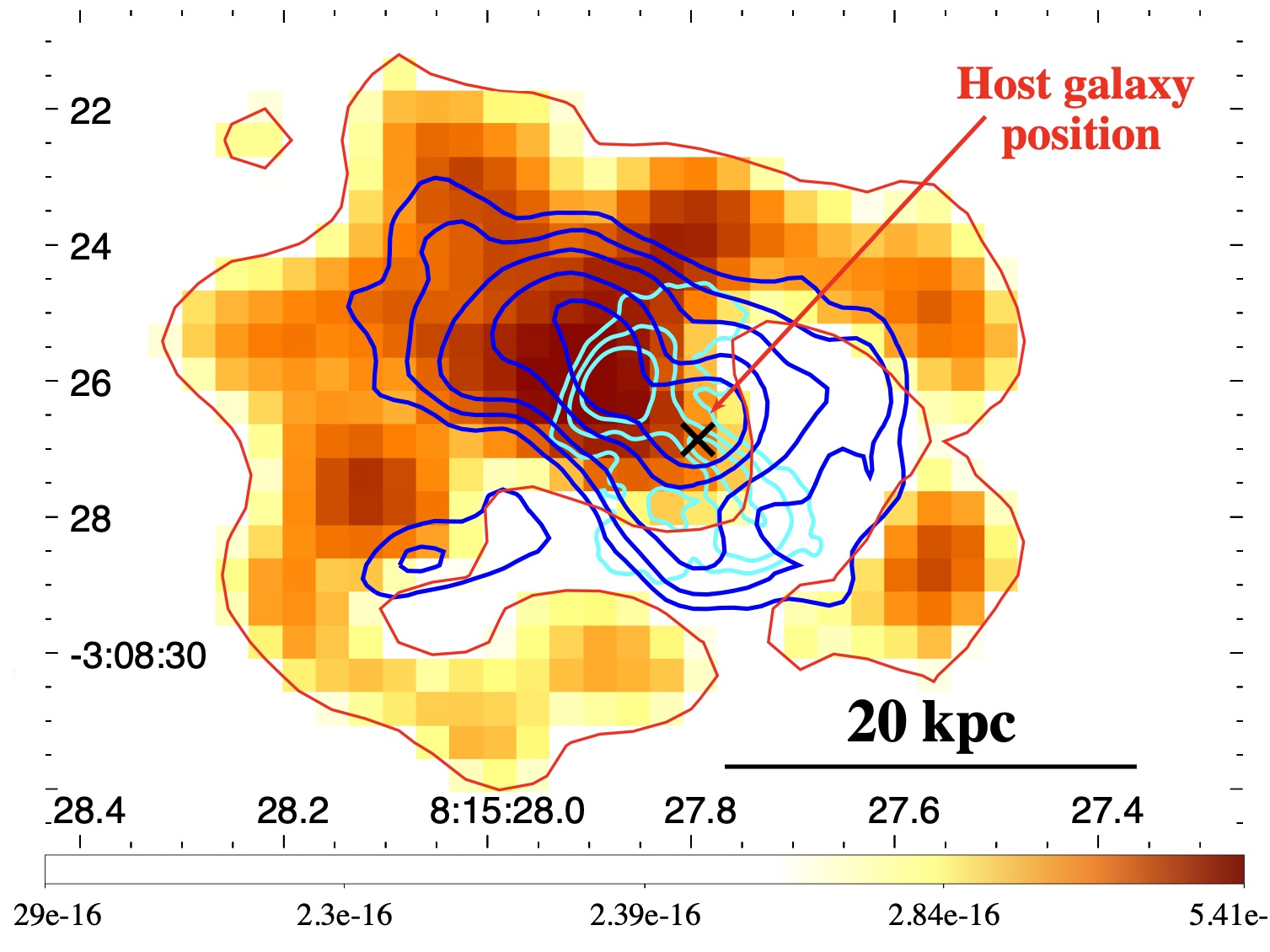}
\includegraphics[height=6.5cm,angle=0]{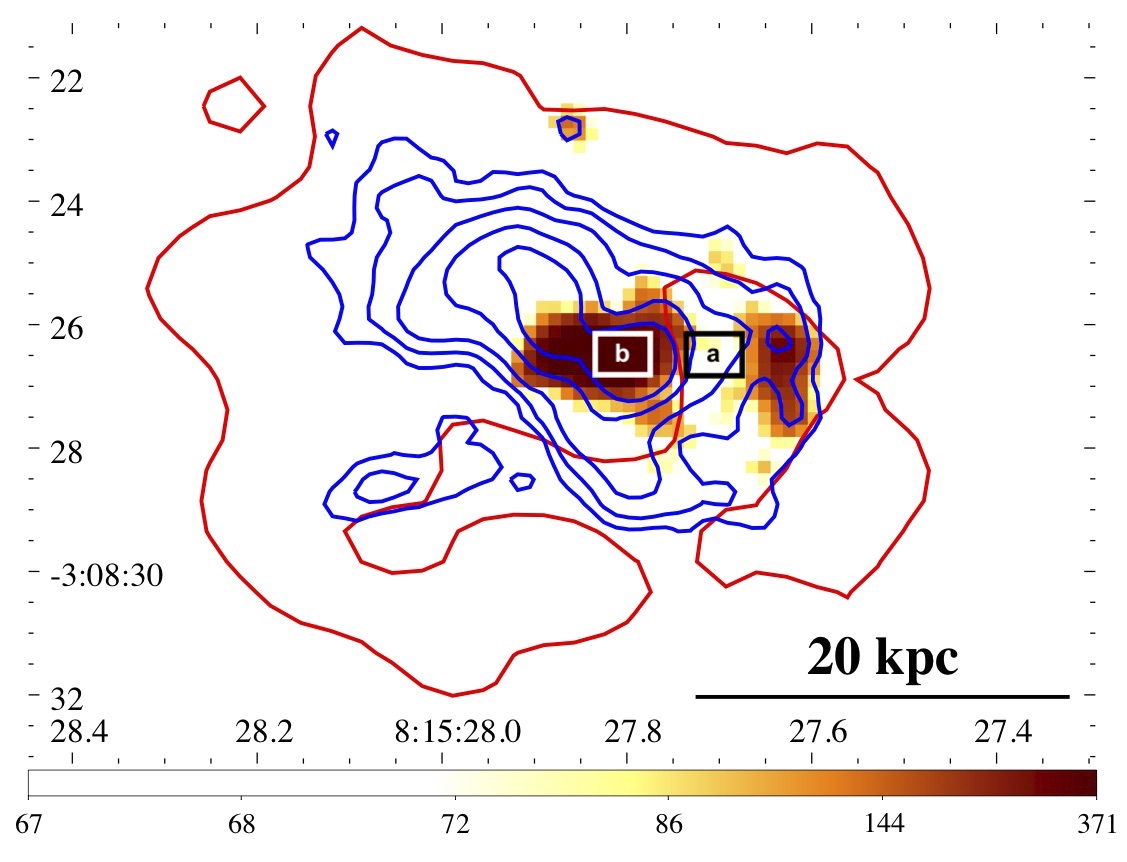}
\caption{Left: Exposure-corrected 0.7 -- 2 keV \textit{Chandra} image of 3CR\,196.1 with H$\alpha$ + [N II]$\lambda6584$ (blue) and 8.4 GHz VLA (cyan) contours. The position of the
centroid of the optical host, which corresponds to the radio core position is shown as a black cross. The $\textit{Chandra}$ image has a pixel scale of 0.492\arcsec\ (corresponding to $\sim$1.6 kpc) and was smoothed with a 1.5\arcsec\ radius Gaussian kernel. X-ray contours are drawn at 0.25$\,\cdot\,$10$^{-15}$ erg$\,$cm$^{-2}\,$s$^{-1}$. H$\alpha$ + [N II]$\lambda6584$ contours were drawn at 5, 10, 20, 50, and 100 times the rms. 8.4 GHz VLA contours were drawn at 5, 20, and 50 times the rms level of the background. H$\alpha$ and the [N II]$\lambda6584$ emission presents an offset with respect to the borders of the X-ray cavity. The northeastern radio lobe is co-spatial with the X-ray emission peak, while the southwestern lobe is co-spatial with the X-ray cavity.
Right: Redward [N II]$\lambda6584$ emission with the same
exposure-corrected 0.7 -- 2 keV \textit{Chandra} contours (red) and H$\alpha$ + [N II]$\lambda6584$ contours (blue) overlaid as in the left panel. The (a) and (b) regions marked in the figure show the extraction areas of the spectra in Fig. \ref{fig:spectra}}.
\label{fig:chandra}
\end{center}
\end{figure*}

\subsection{Chandra data analysis}

We performed the X-ray data reduction following the standard procedures of the {\it Chandra} Interactive Analysis of Observations (CIAO; \citealt{Fruscione2006}) v4.11 threads\footnote{http://cxc.harvard.edu/ciao/threads/}, adopting the \textit{Chandra} Calibration Database v4.8.4.1 (for an in-depth analysis of the X-ray observation see \citealt{Ricci2018}).

X-ray flux maps were created by taking into account the exposure time and the effective area. X-ray maps presented in this work were restricted to 0.7 - 2 keV to highlight the soft diffuse X-ray emission by using monochromatic exposure maps set to the nominal energies of 1.2 keV. The flux map was converted from units of counts s$^{-1}$ cm$^{-2}$ to cgs units by multiplying each event by the nominal energy of the band, assuming that every event in the same band has the same energy (see \citealt{Massaro2015}).

\subsection{MUSE data analysis}
\label{sec:musedata}
Regarding the MUSE data analysis, we obtained the fully reduced data cube from the ESO archive\footnote{http://archive.eso.org/scienceportal/home}. We measured the redshift by using the Penalized Pixel-Fitting code (\citealt{Cappellari2017}) to fit the stellar continuum and absorption features in a 2\arcsec\ radius circular region centered on the host position. Following this procedure, we measured a value for the redshift of $z=0.1982$, which we adopted throughout our analysis.

\subsubsection{Large scale ionized gas}

We subtracted the continuum spaxel by spaxel from the archival MUSE reduced data in two different ranges, blue range, from 5370 to 6170 $\AA$, and red range, from 7170 to 8680 $\AA$, by fitting the continuum in each range using a power law. We fitted the H$\beta$ and [O III]$\lambda\lambda4960,5008$ lines in the continuum subtracted blue range and the [O I]$\lambda6300$, H$\alpha$, [N II]$\lambda\lambda6548,6584$ and [S II]$\lambda\lambda6718,6733$ lines in the continuum subtracted red range. In general, we fitted each line with a single Gaussian component, fixing the wavelength separation between lines as well as the line ratio of [N II]$\lambda\lambda6548,6584$ and [O III]$\lambda\lambda4960,5008$ components, to their theoretical value of 1/3 (adopting CASE B, see \citealt{Osterbrock2006}).

\subsubsection{Central region}

Although a single Gaussian was adequate to fit the spectral components in most spaxels, we found that the spectra in the central area and the X-ray cavity region have additional components in H$\alpha$, [N II]$\lambda6584$ and [S II]$\lambda\lambda6718,6733$. The position of the redward [N II]$\lambda6584$ emission is shown in the right panel of Fig. \ref{fig:chandra}.

To illustrate these redward components, we chose two regions in the central area of 3CR\,196.1, regions {\it a} and {\it b}, shown in the right panel of Fig. \ref{fig:chandra}. Both these regions were chosen to be close to the central region so the signal-to-noise was the highest and with sizes large enough to match the seeing (i.e., 1\arcsec), but not so large that the kinematics of the ionized gas would compromise the identification of the different spectral components. Additionally, region {\it a} was chosen to contain only a single spectral feature for each emission line, while region {\it b} was chosen to illustrate a region with both ``rest-frame" and redward components.

\begin{figure*}
\begin{center}
\includegraphics[height=8.5cm,angle=0]{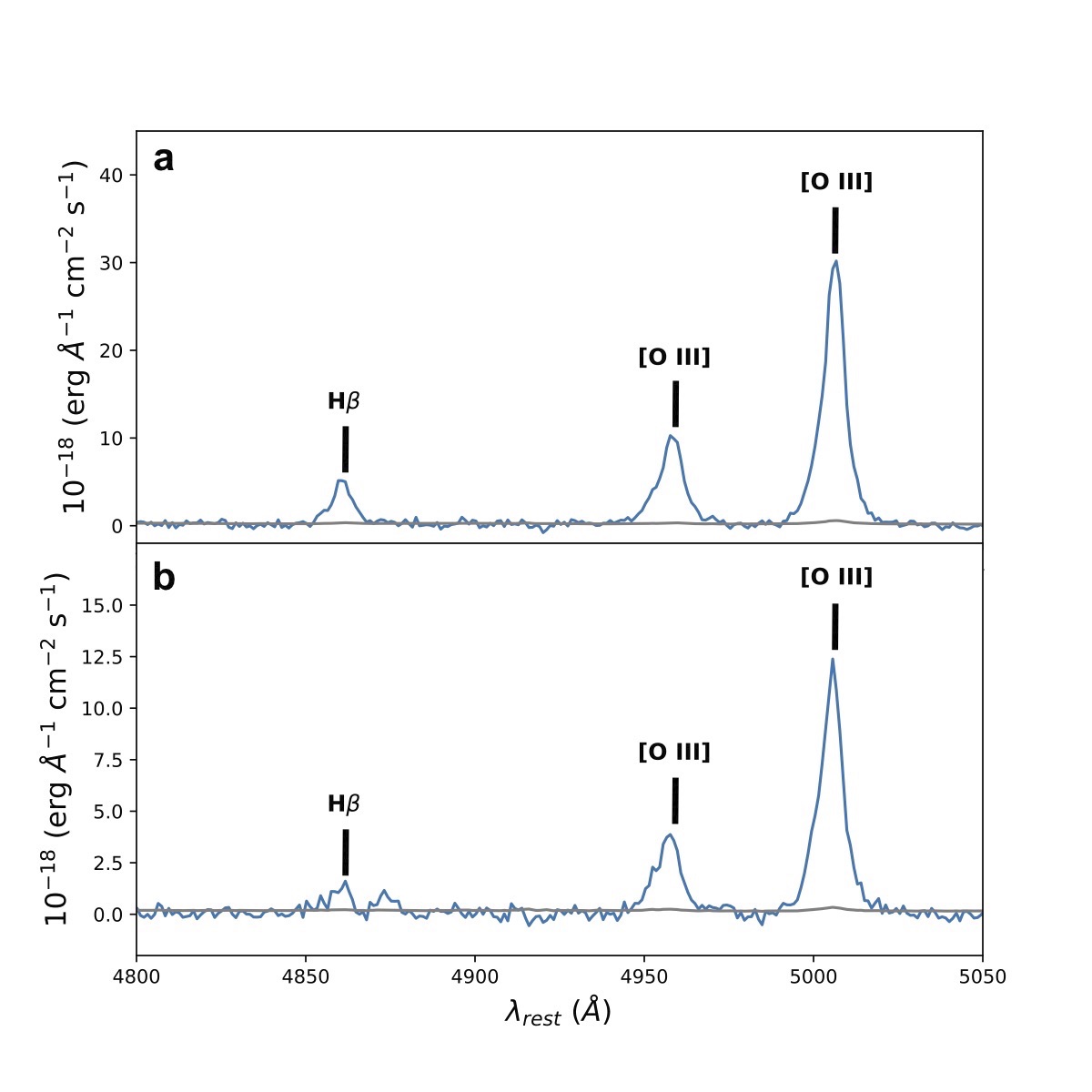}
\includegraphics[height=8.5cm,angle=0]{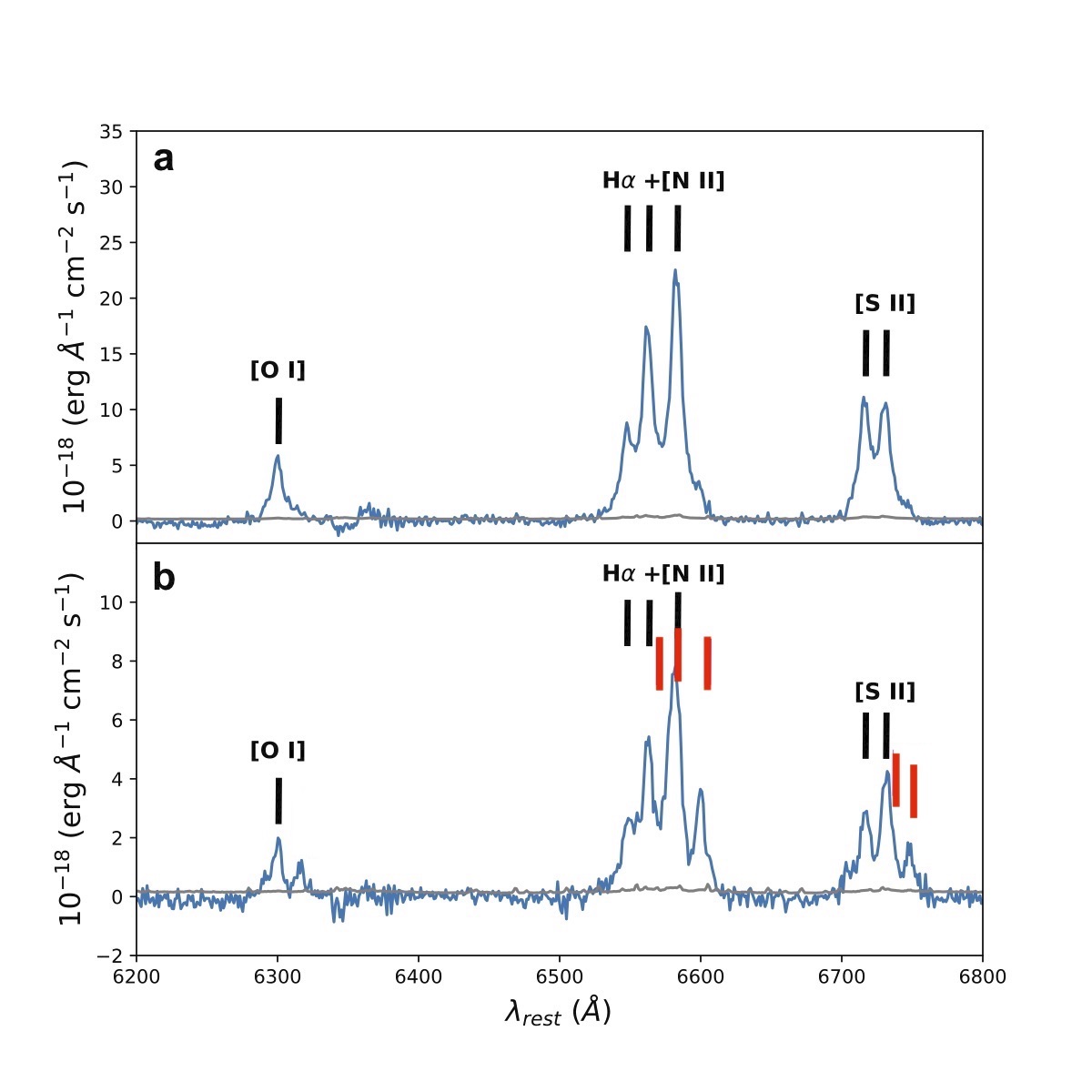}

\caption{Spectra extracted from regions {\it a} and {\it b} (top and bottom panels in Fig. \ref{fig:chandra}, right panel, respectively). The top spectrum could be fitted using a single Gaussian component for each line, while the bottom one shows an extra redward component in all lines except for H$\beta$ and [O III]$\lambda\lambda4960,5008$ lines. Black markers point to the rest-frame position of each line while red markers are located at 21 $\AA$ from rest-frame ($\sim$1000 km$\,$s$^{-1}$). Uncertainties in the spectra are shown in gray.}
\label{fig:spectra}
\end{center}
\end{figure*}

\begin{figure}
\begin{center}
\includegraphics[width=9.5cm,angle=0]{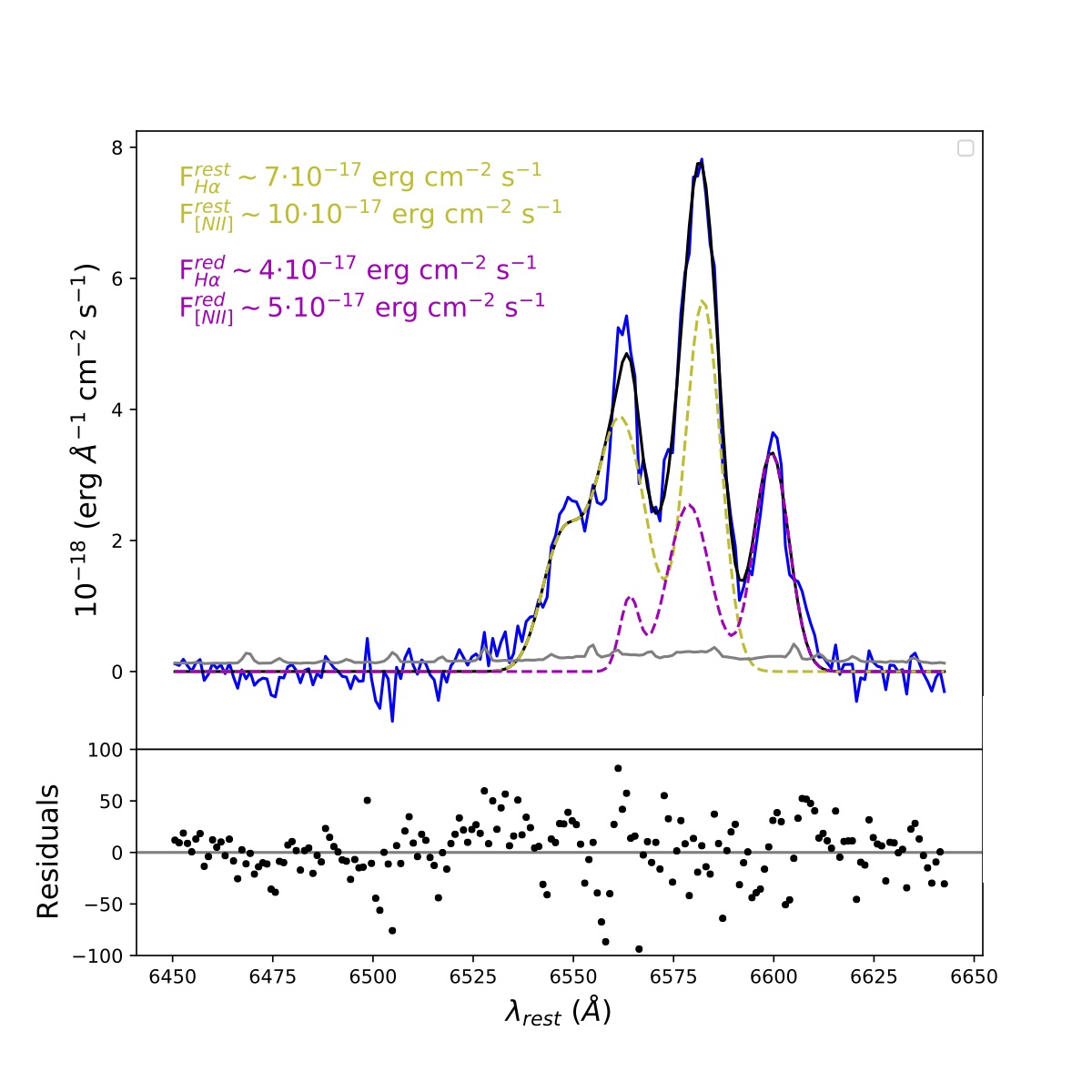}
\caption{Fit to the spectrum extracted from region {\it b} (i.e., a 1 arcsec$^2$ region, as shown in the right panel of Fig. \ref{fig:chandra}) using a H$\alpha$ + [N II]$\lambda6584$ at rest frame (yellow) and another one redshifted by $\sim$21 $\AA$. Uncertainties in the spectrum are shown in grey. Spectra in region {\it b} (see right panel of Fig. \ref{fig:chandra}) cannot be reproduced using a single triplet.}
\label{fig:cloudspec}
\end{center}
\end{figure}

A comparison of the spectra extracted from regions {\it a} and {\it b} is shown in Fig. \ref{fig:spectra}, where the rest-frame wavelengths of lines are marked in black and those of the redward components are marked in red. Although there appear to be possible additional spectral features corresponding to the H$\beta$ and [O I]$\lambda6300$ lines, these components are not detected (with a significance below 2$\sigma$). The spectral position of the reward components in region {\it b} in Fig. \ref{fig:spectra} corresponds to $\sim$21 $\AA$ (i.e., $\sim$1000 km$\,$s$^{-1}$) from rest-frame and was derived on the basis of the spectral fit performed on the spectrum extracted from region {\it b}. This fit, shown in Fig. \ref{fig:cloudspec}, was performed using two H$\alpha$ + [N II]$\lambda6584$ triplets, one at rest-frame and one redshifted by $\sim$21 $\AA$, and shows how the spectrum from region {\it b} cannot be reproduced without introducing an additional redward component.

These redshifted spectral features are spectrally blended with emission at the systemic redshift, i.e., the redward and ``rest-frame" components have an offset consistent with the separation between the H$\alpha$ and the [N II]$\lambda6584$ lines (see Fig. \ref{fig:spectra}). This offset, together with the fact that this redward component overlaps, in projection, with the AGN emission as well as with the low signal-to-noise of the [N II]$\lambda6548$ line, introduces a degeneracy in the intensities of the H$\alpha$, [N II]$\lambda\lambda6548,6584$ and [S II]$\lambda\lambda6718,6733$ lines that we cannot break without knowing a priori the ionization status of the source and the density of the gas. Thus, the limited spectral ($\sim$50 km$\,$s$^{-1}$) and spatial ($\sim$1\arcsec) resolutions of MUSE cubes prevent us from breaking the degeneracy, and from providing a thorough analysis of the kinematics and ionization state of the gas.

\begin{figure*}
\begin{center}
\includegraphics[height=6.5cm,angle=0]{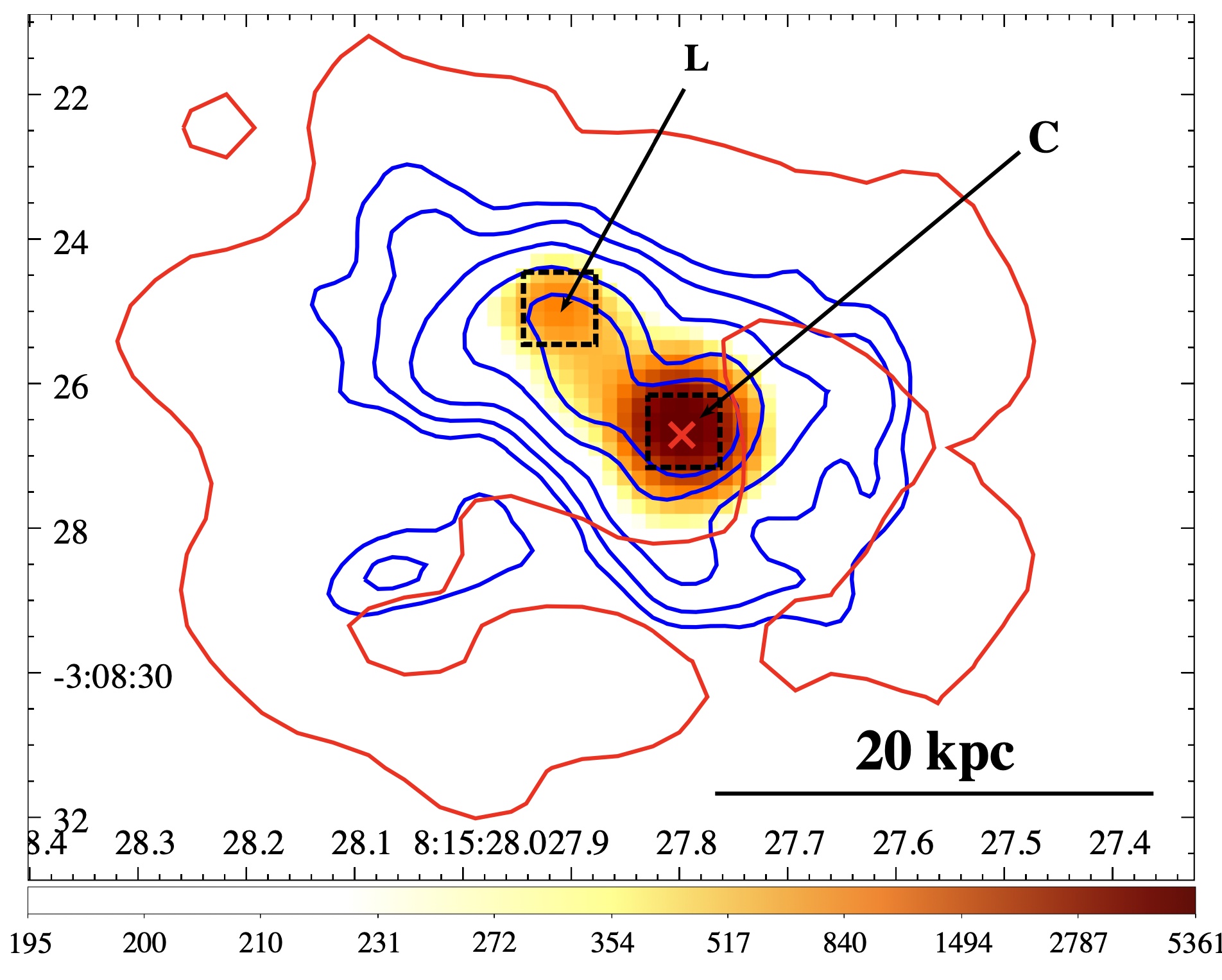}
\includegraphics[height=6.5cm,angle=0]{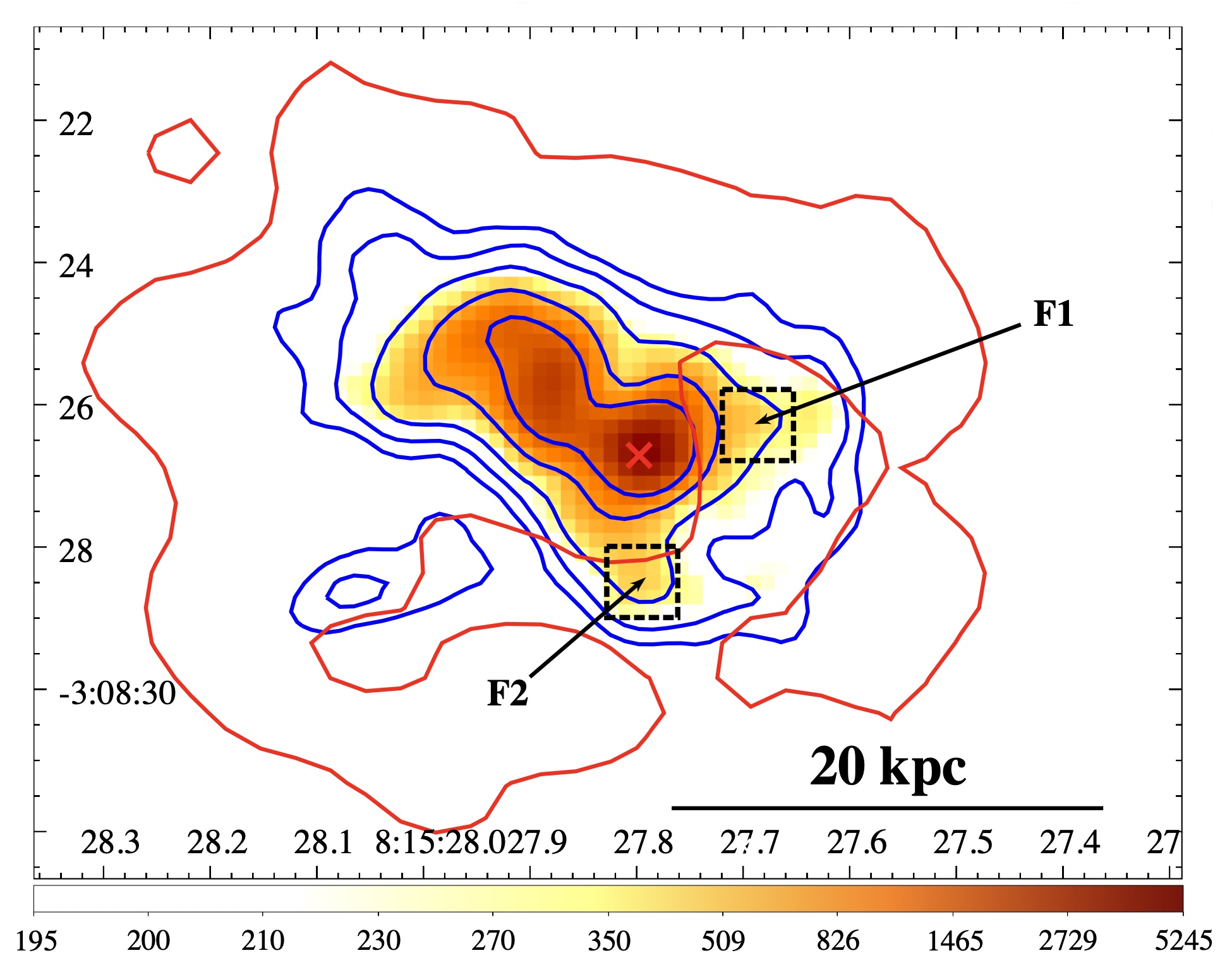}
\caption{[O III]$\lambda\lambda4960,5008$ (left) and [S II]$\lambda\lambda6718,6733$ (right) emission with exposure-corrected 0.7 -- 2 keV \textit{Chandra} contours (red) and H$\alpha$ + [N II]$\lambda6584$ contours (blue) overlaid. The position of the centroid of the optical host is indicated with a red cross. The 1 arcsec$^2$ extraction regions, L, C, F1, and F2, for the BPT diagram in Fig. \ref{fig:bpt} are shown as black dashed squares. $\textit{Chandra}$ contours were smoothed with a 1.5\arcsec\ Gaussian kernel radius and drawn in red at 0.25$\,\cdot\,$10$^{-15}$ erg$\,$cm$^{-2}\,$s$^{-1}$. H$\alpha$ + [N II]$\lambda6584$ contours were drawn at 5, 10, 20, 50, and 100 times the rms. Region L marks the region co-spatial with the northeastern radio lobe.}
\label{fig:other}
\end{center}
\end{figure*}

\section{Results}
\label{sec:results}

We detected H$\alpha$+[N II]$\lambda6584$ emission spatially associated with the X-ray cavity (see left panel of Fig. \ref{fig:chandra}) and, at the same time, did not detect H$\beta$ or [O III]$\lambda5007$ emission in the X-ray cavity region (see left panel of Fig. \ref{fig:other}). A comparison between X-ray and H$\alpha$+[N II]$\lambda6584$, [O III]$\lambda\lambda4960,5008$, and [S II]$\lambda\lambda6718,6733$ is shown in the left panel of Fig. \ref{fig:chandra} and the left and right panels of Fig. \ref{fig:other}, respectively. Additionally, flux maps, velocity, and velocity dispersion maps for all emission lines in the MUSE observed range are shown in Fig. \ref{fig:maps}, Appendix \ref{sec:maps}.

%While with the H$\alpha$+[N II]$\lambda6584$ emission extends up to $\sim$10 kpc, the detected [O III]$\lambda5007$ emission (see Fig. \ref{fig:other}, left panel) is restricted to the central and northeastern regions, extending up to $\sim$10 kpc, suggesting a low ionization state in the X-ray cavity region. 

The morphology of the H$\alpha$+[N II]$\lambda6584$ hints at differences in the density of the environment. In particular, the northeastern component (region L in the left panel of Fig. \ref{fig:other}) being co-spatial with the northeastern radio lobe and with the highest X-ray surface brightness peak indicates a denser environment toward the northeast than toward the southwest, where the X-ray cavity is located.

\begin{figure*}
\begin{center}
\includegraphics[height=6cm,angle=0]{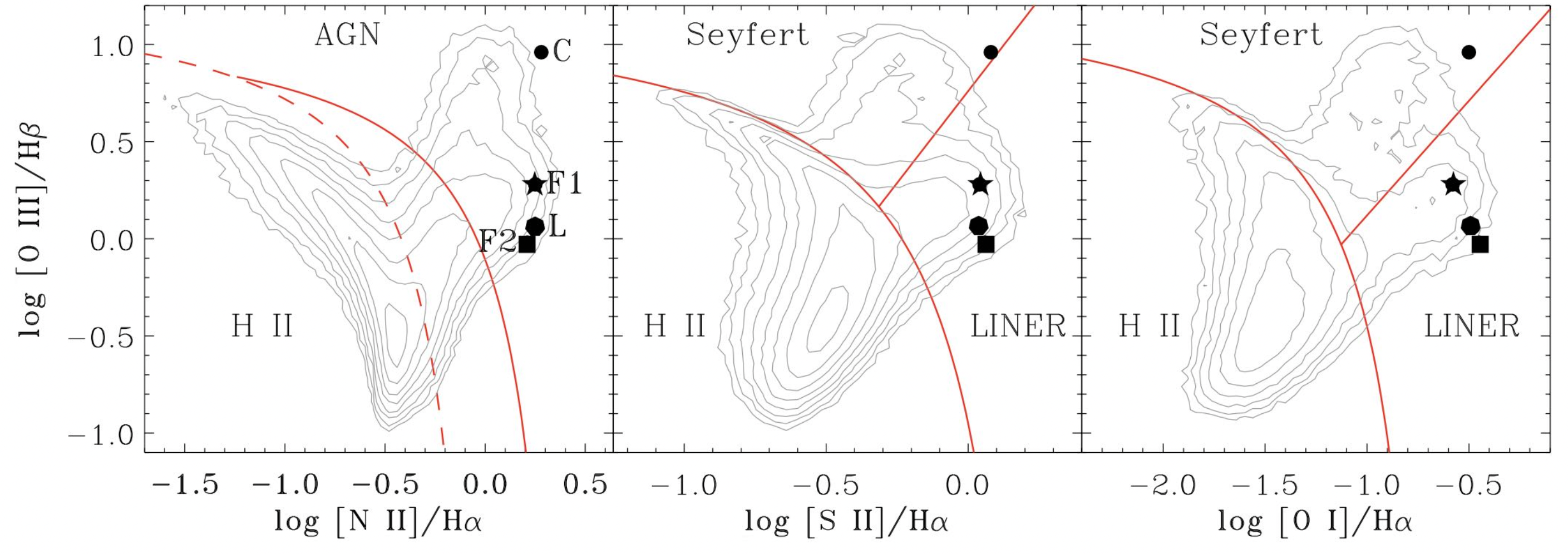}
\caption{Location of 3CR\,196.1 regions core, northeastern radio lobe and blue and red filaments (C, L, F1 and F2, respectively) in the BPT diagnostic diagrams. The size of regions was chosen to match the seeing of MUSE observations (i.e., $\sim$1\arcsec). Extraction regions are shown in Fig. \ref{fig:other}. The red solid curves represent the \citet{Kewley2001} theoretical upper bound for pure star formation, the red straight line shows the \citet{Kewley2006} separation between AGN and LINERs, while the dashed red curve in the [N II]$\lambda6584$ BPT is the \citet{Kauffmann2003} empirical classification separating star-forming galaxies and AGN. Contours represent the iso-densities of all SDSS/DR7 emission line galaxies (\citealt{Capetti2011}). Extended ionized gas structures in 3CR\,196.1 have an ionization state compatible with the LINER region of the BPT diagnostic diagrams, while the ionization state at optical host position is compatible with Seyfert-like ionization.}
\label{fig:bpt}
\end{center}
\end{figure*}

%In particular, the northeastern component (region L in left panel of Fig. \ref{fig:other}), co-spatial with the northeastern radio lobe and with the highest X-ray surface brightness peak, has an ionization state compatible with the LINER region of the BPT diagnostic diagrams (see \citealt{Baldwin1981} and \citealt{Kewley2006}), with $\log$([O III]/H$\beta$) $\sim$0.3; $\log$([N II]/H$\alpha$) $\sim$0.2; $\log$([S II]/H$\alpha$) $\sim$0.005 and $\log$([O I]/H$\alpha$) $\sim$-0.5.

\begin{figure*}
\begin{center}
\includegraphics[height=3cm,angle=0]{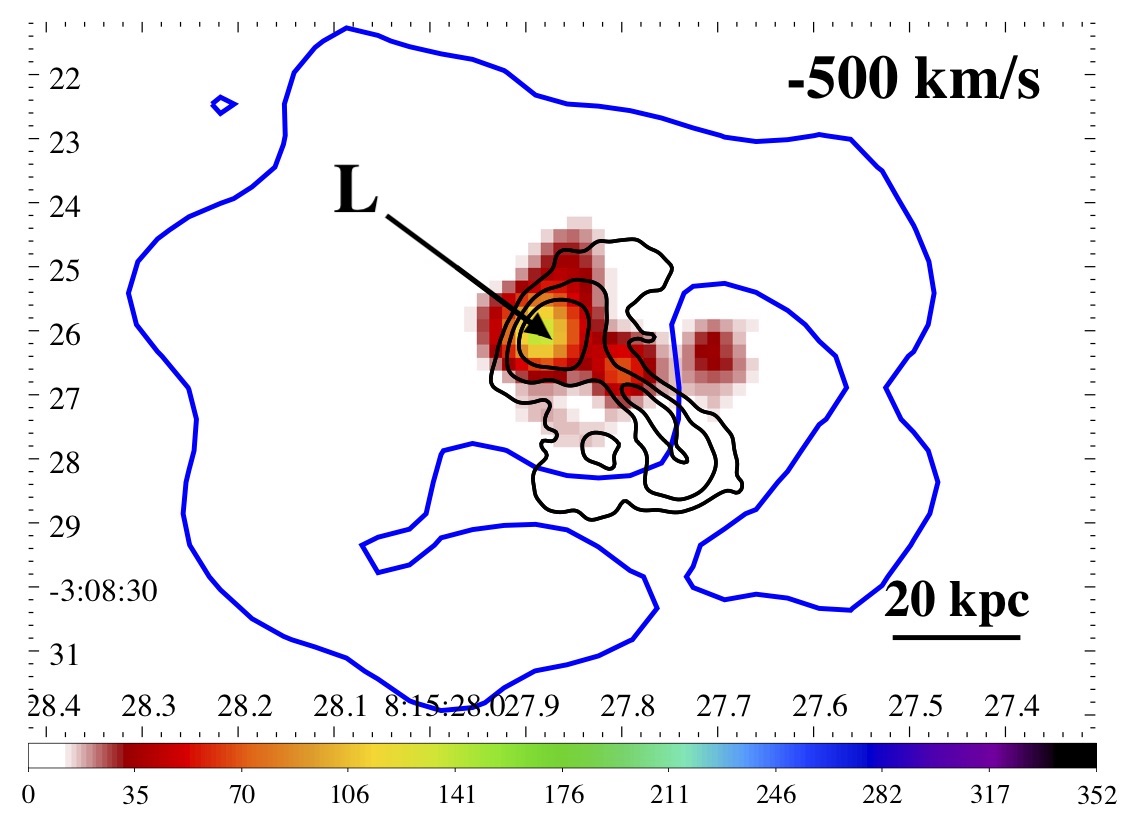}
\includegraphics[height=3cm,angle=0]{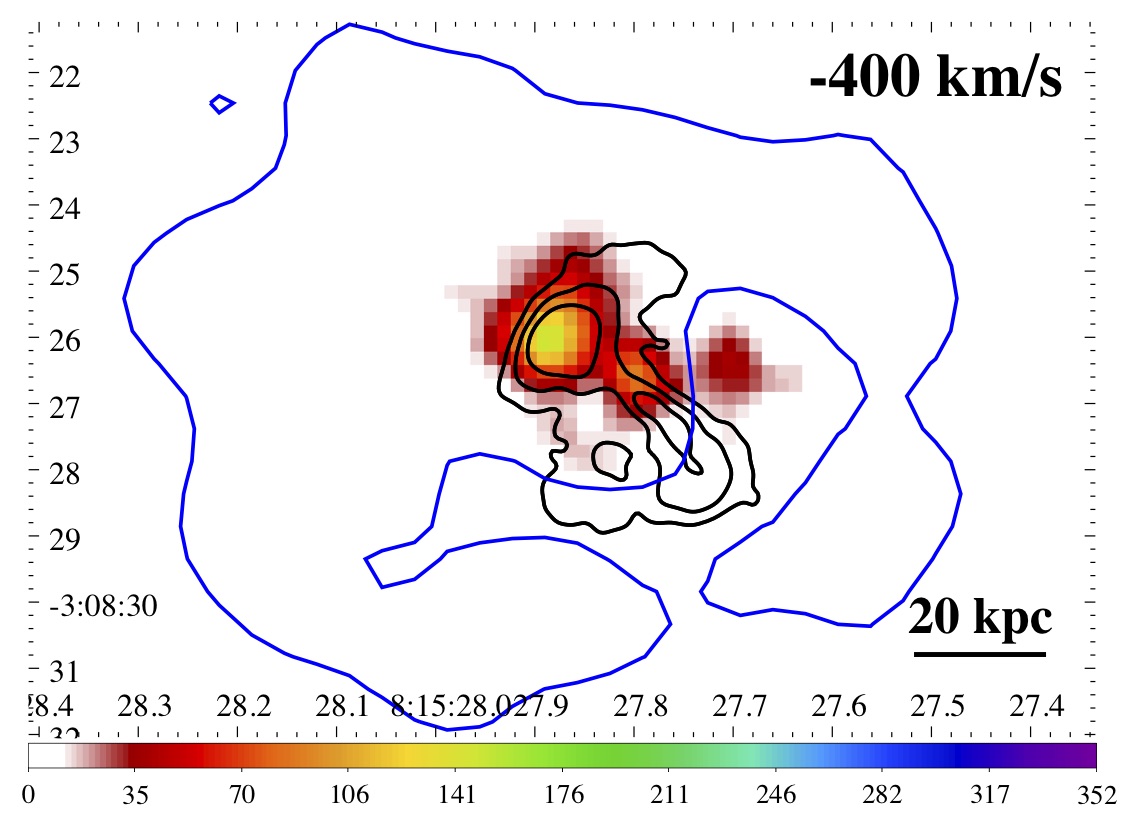}
\includegraphics[height=3cm,angle=0]{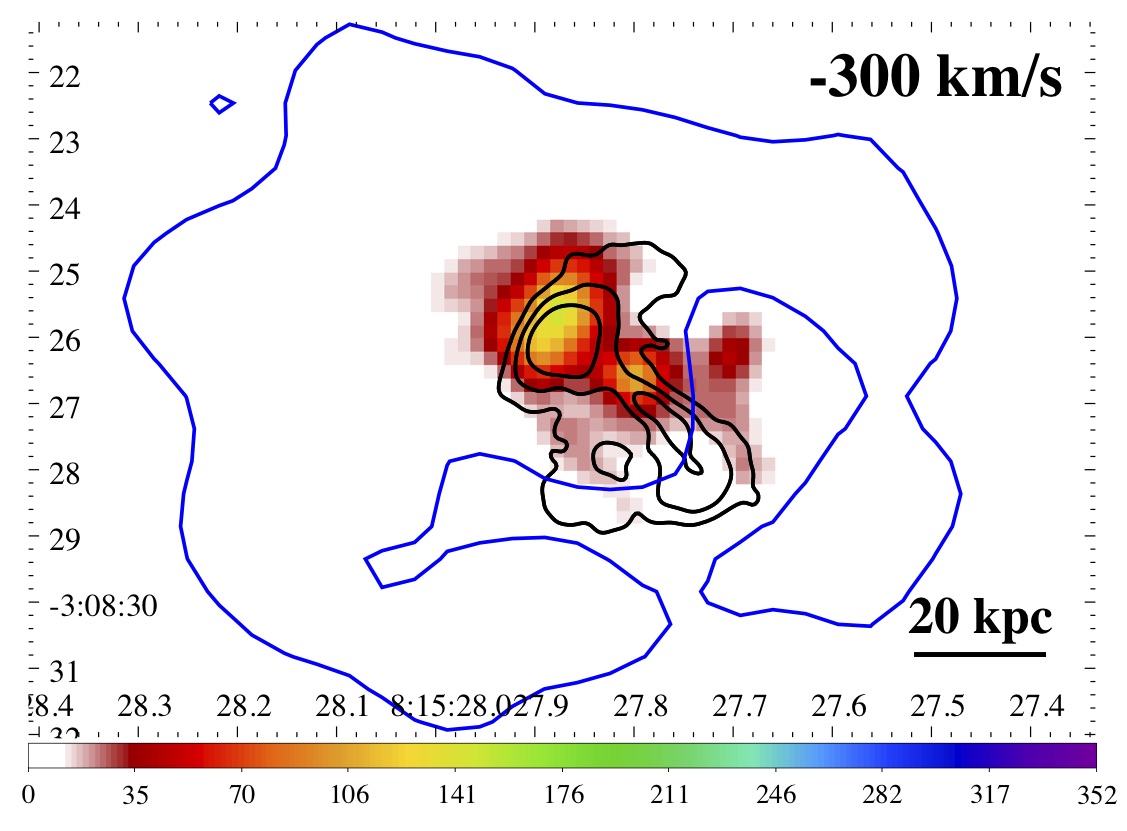}
\includegraphics[height=3cm,angle=0]{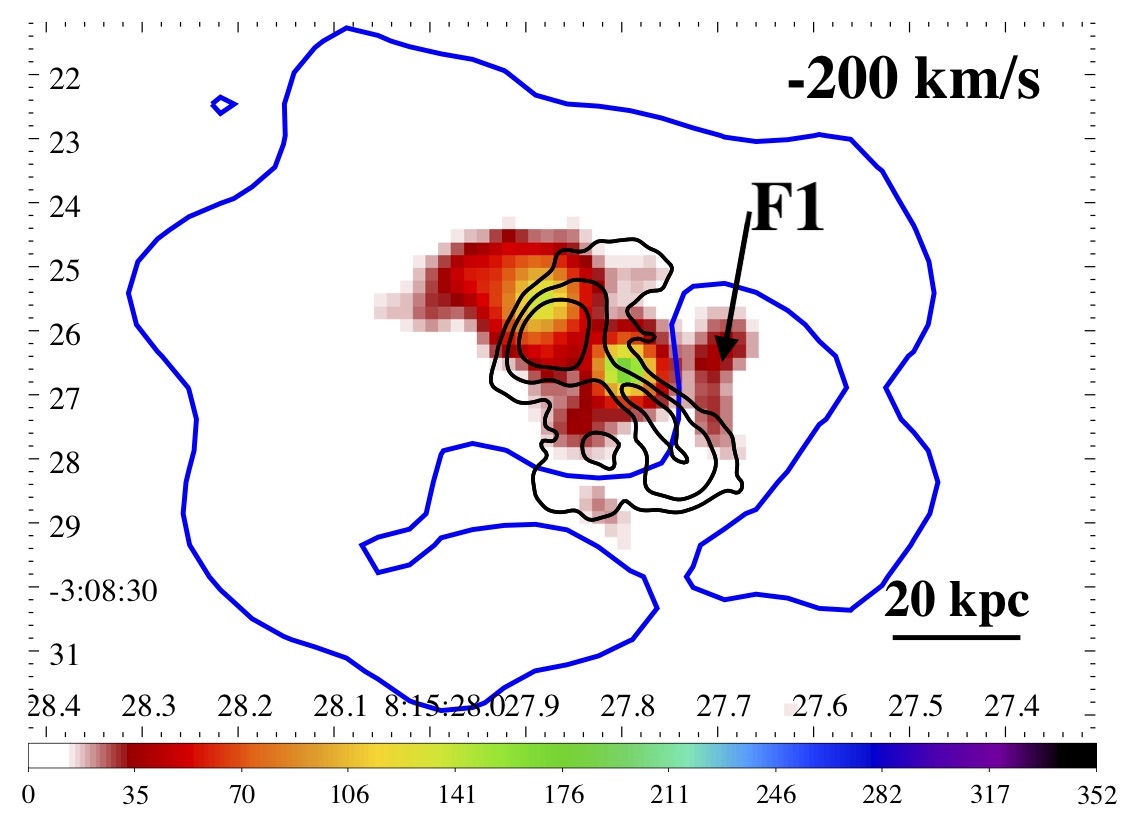}
\includegraphics[height=3cm,angle=0]{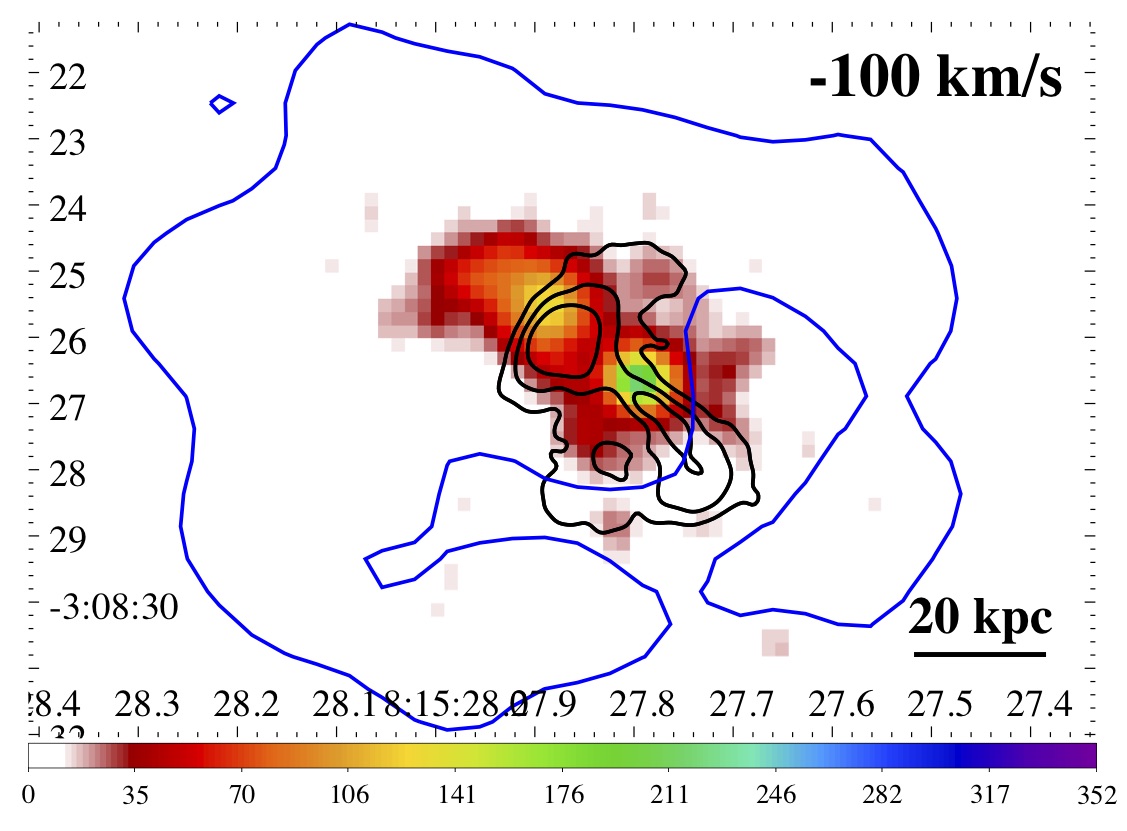}
\includegraphics[height=3cm,angle=0]{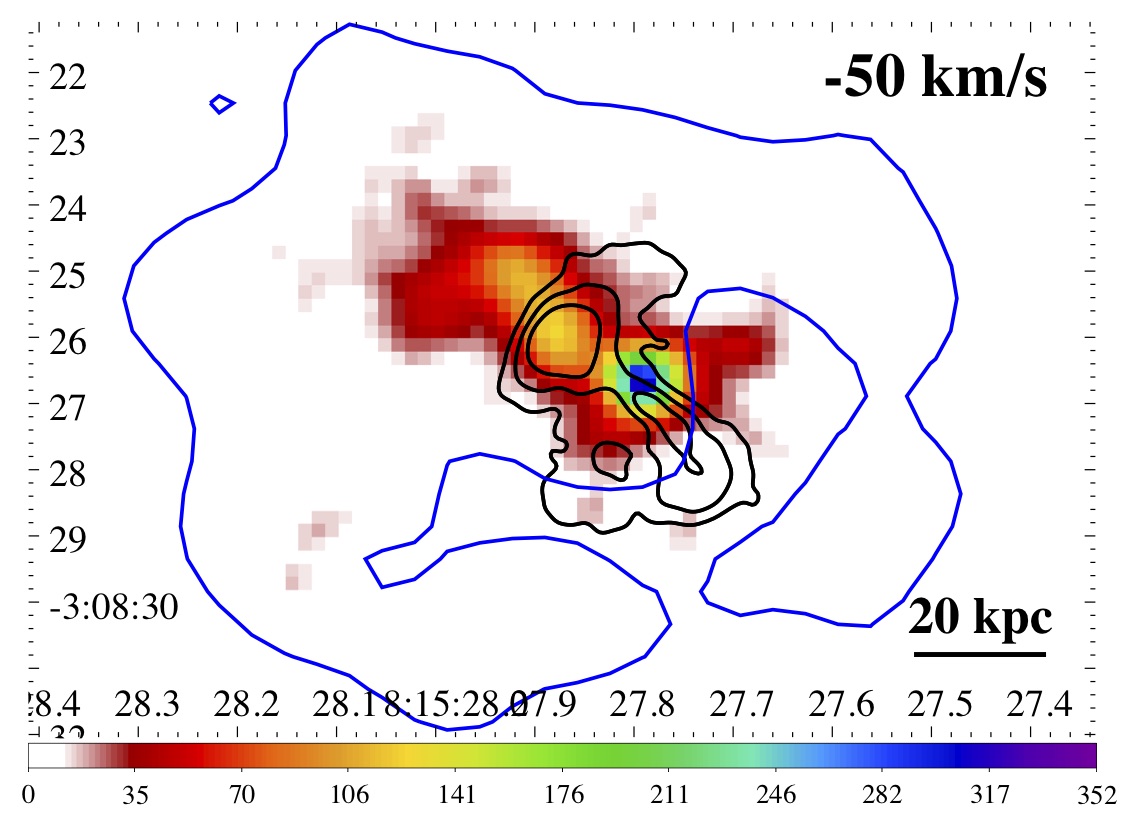}
\includegraphics[height=3cm,angle=0]{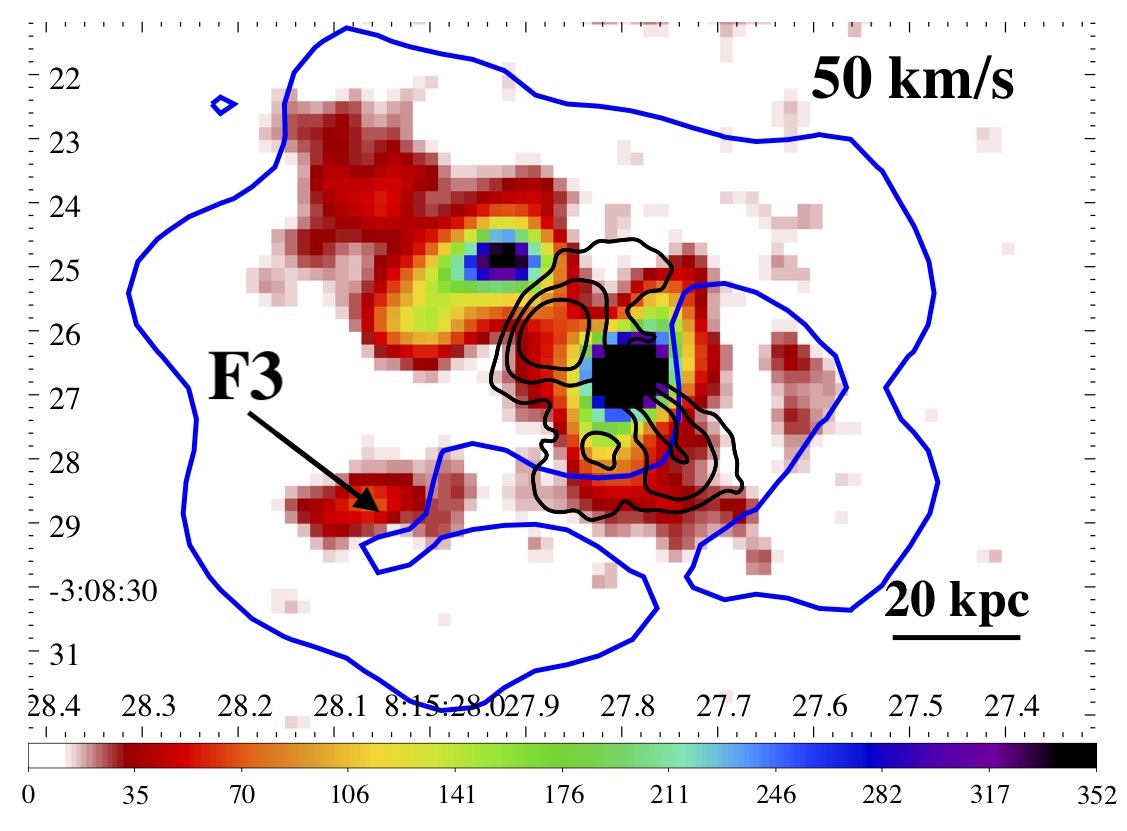}
\includegraphics[height=3cm,angle=0]{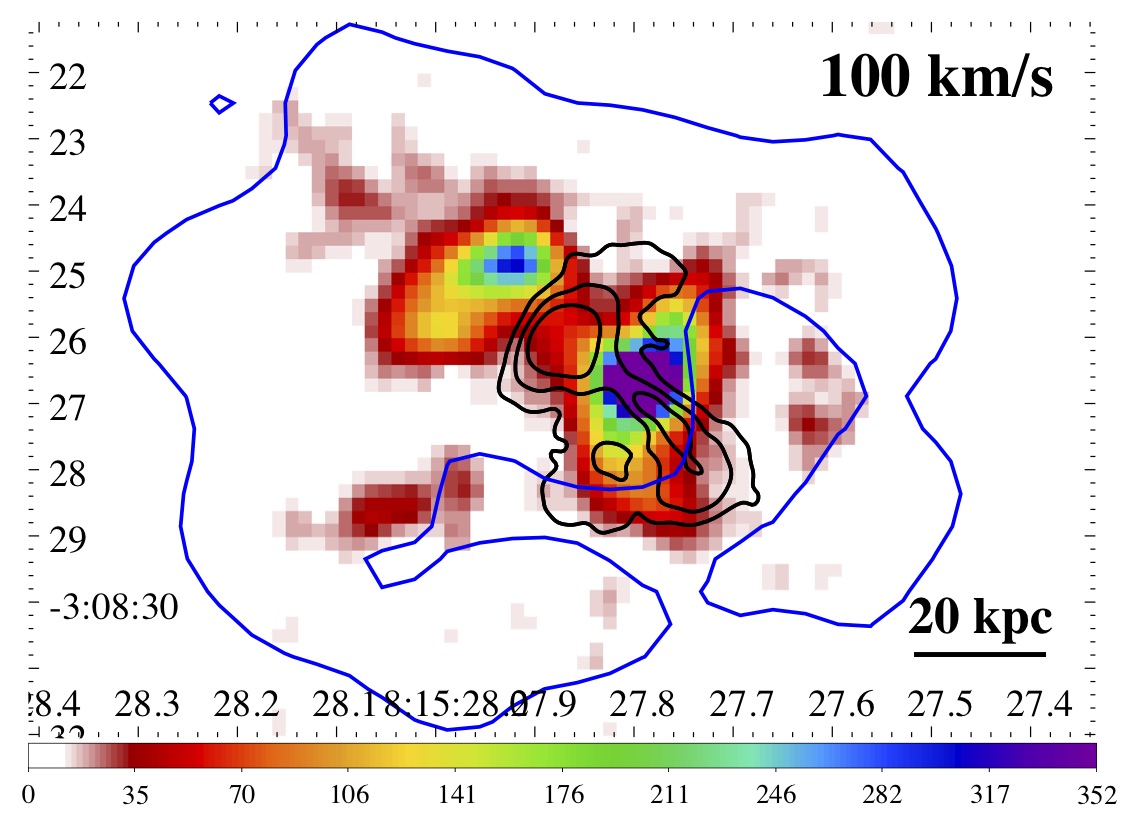}
\includegraphics[height=3cm,angle=0]{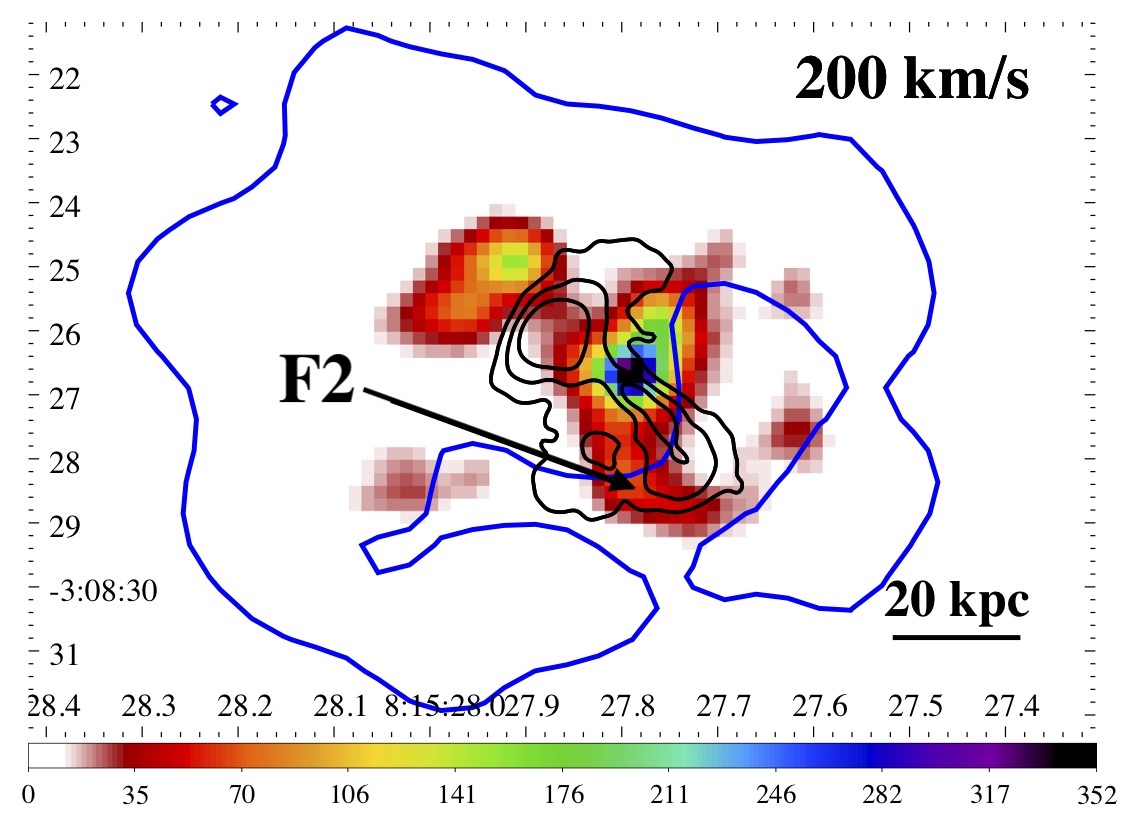}
\includegraphics[height=3cm,angle=0]{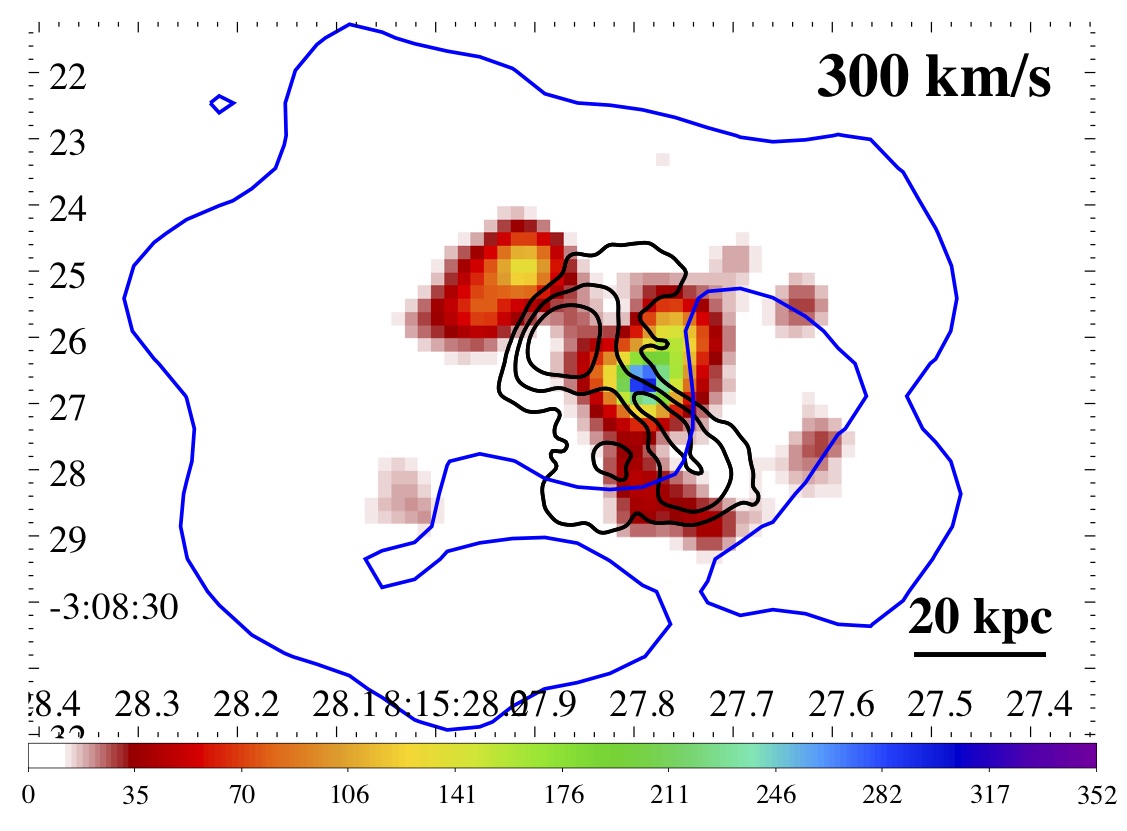}
\includegraphics[height=3cm,angle=0]{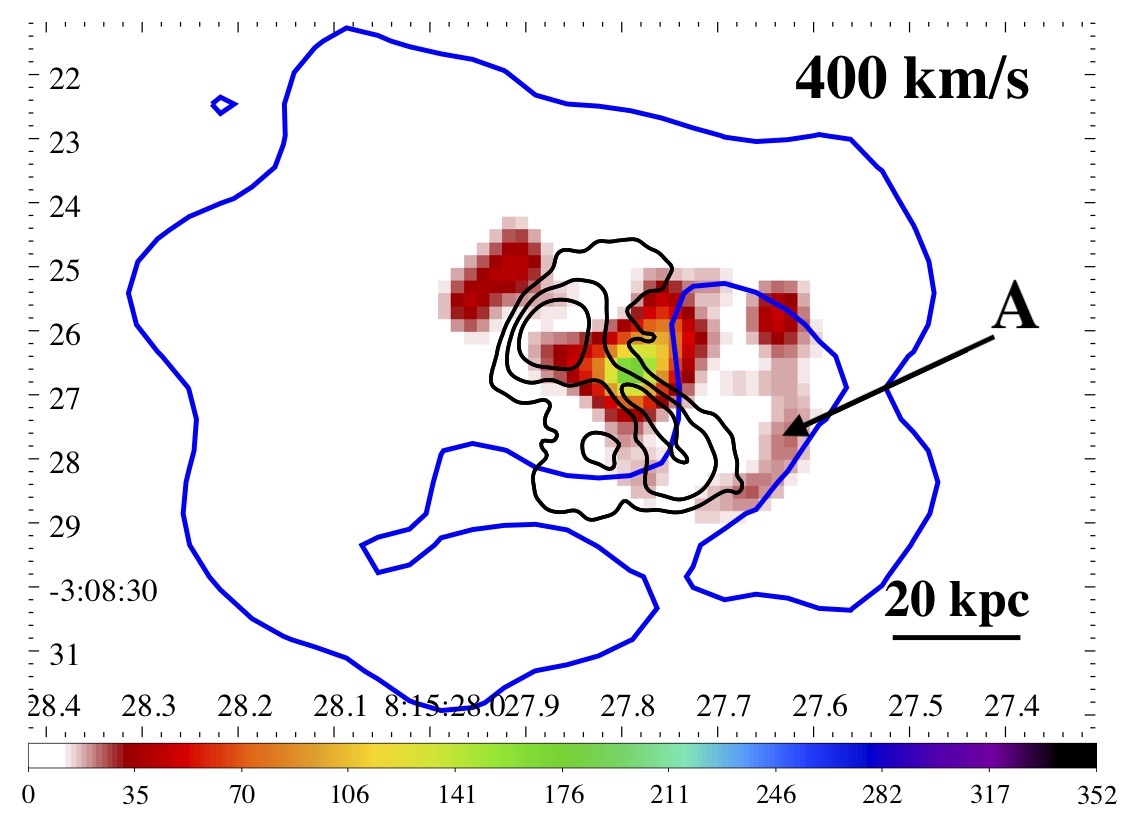}
\includegraphics[height=3cm,angle=0]{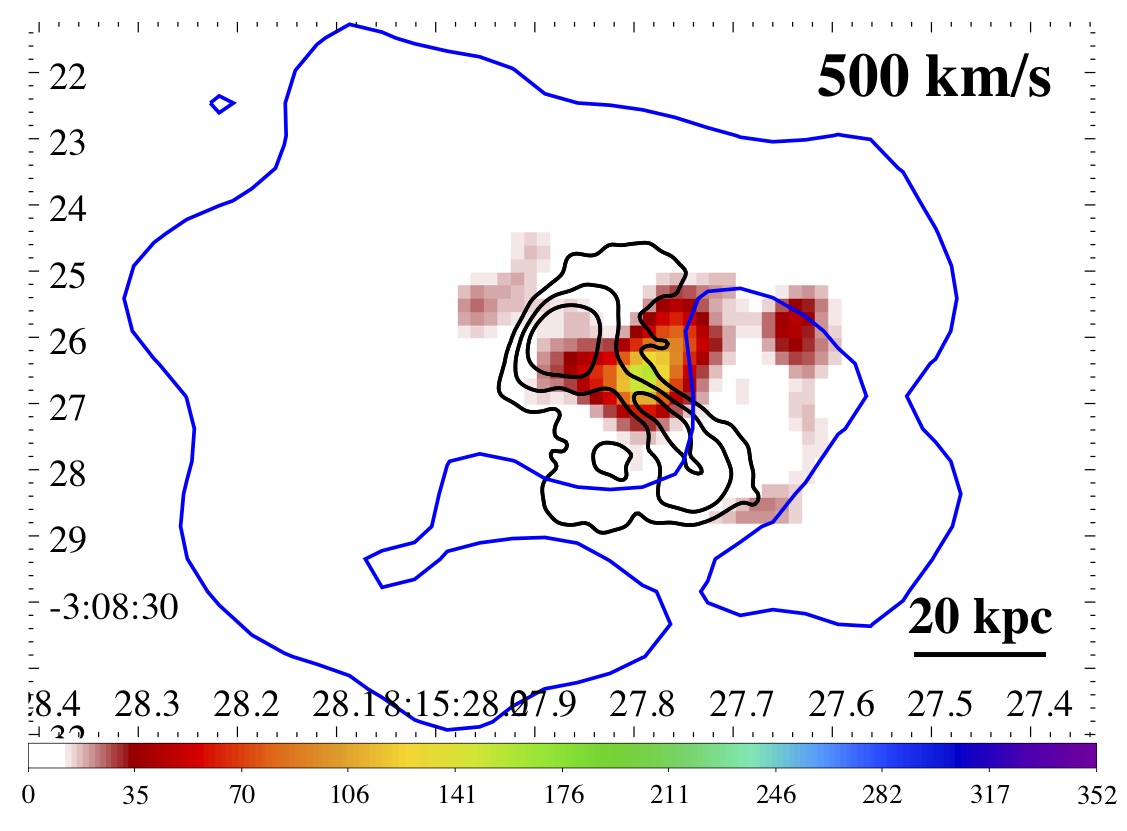}
\includegraphics[height=3cm,angle=0]{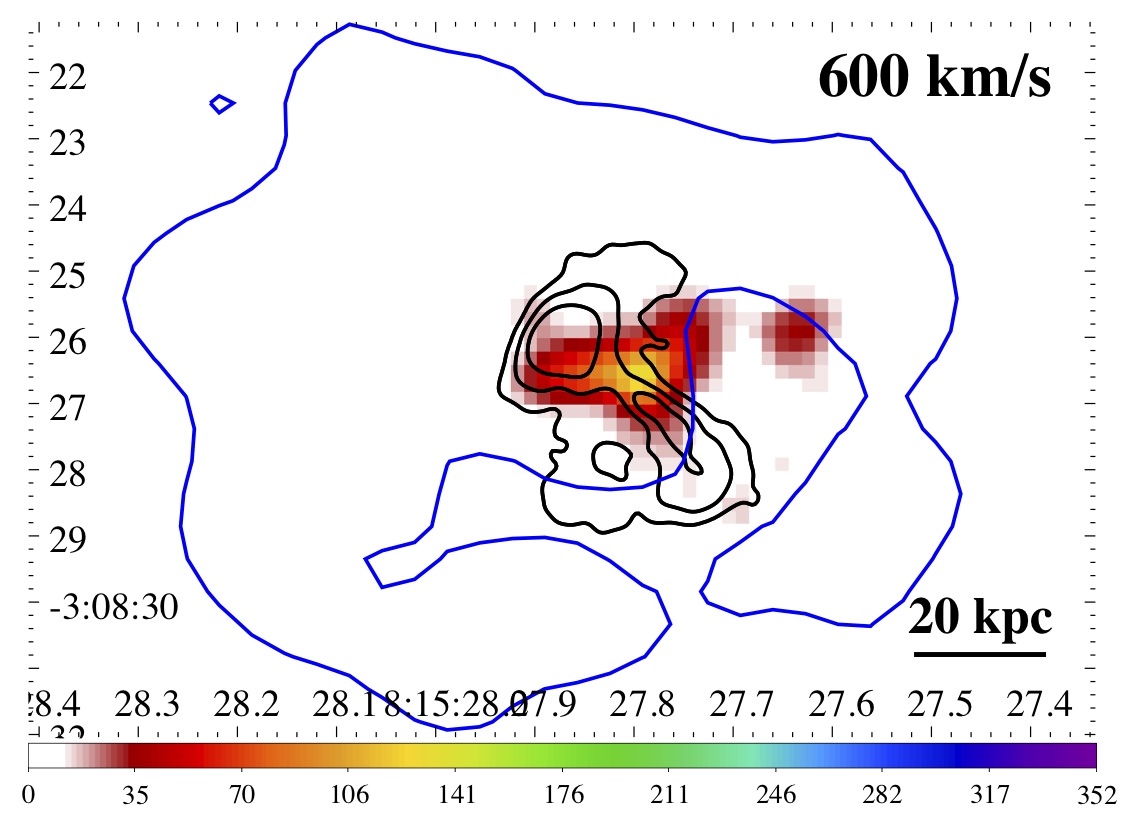}
\includegraphics[height=3cm,angle=0]{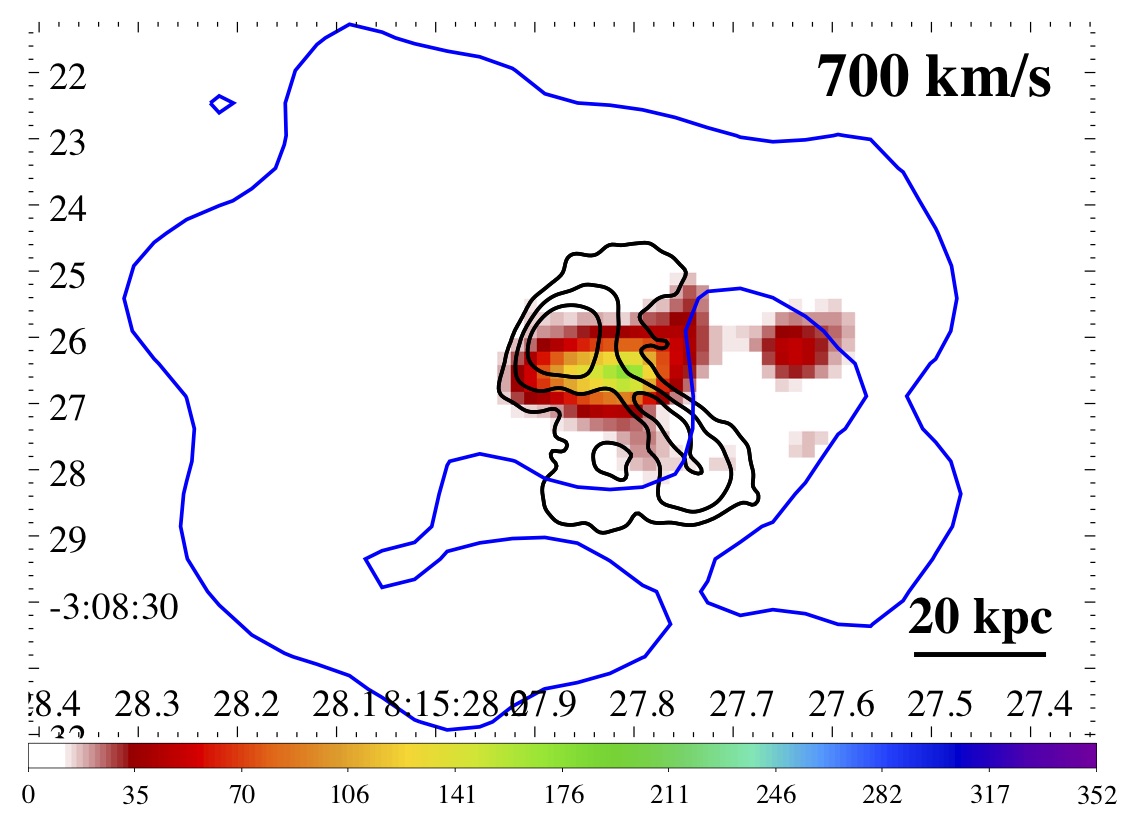}
\includegraphics[height=3cm,angle=0]{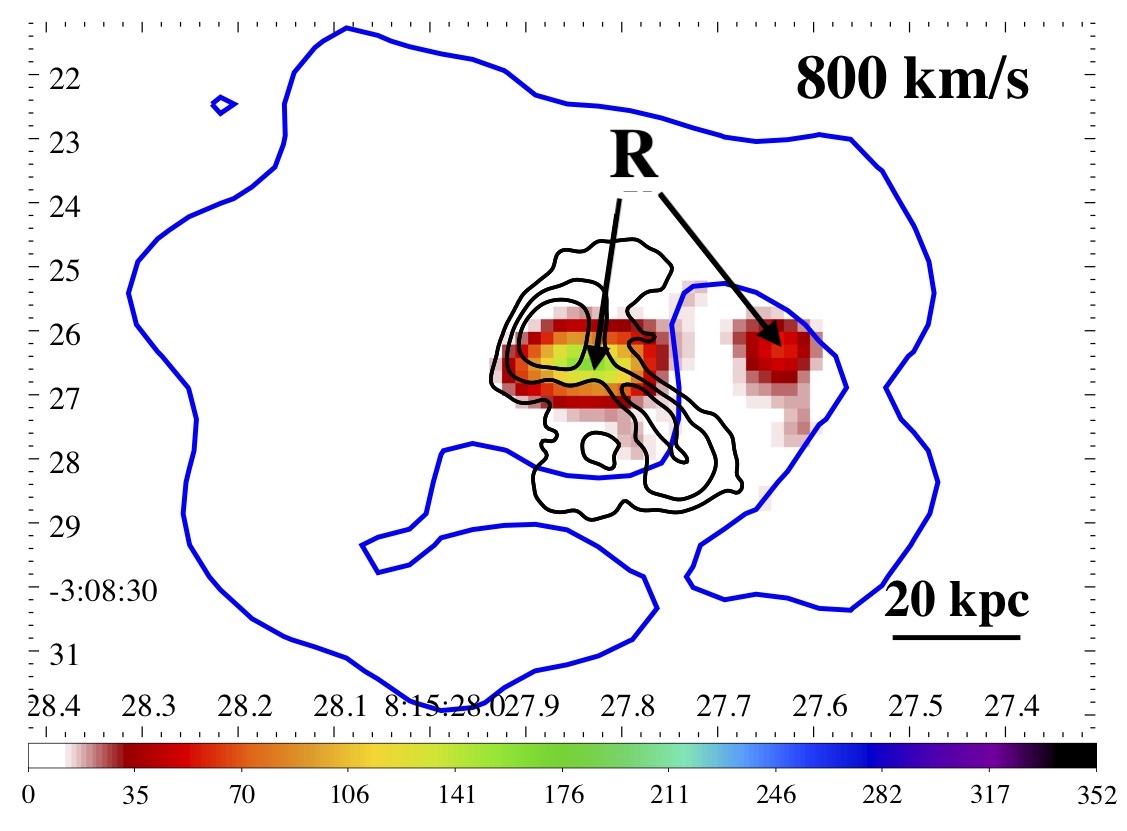}
\includegraphics[height=3cm,angle=0]{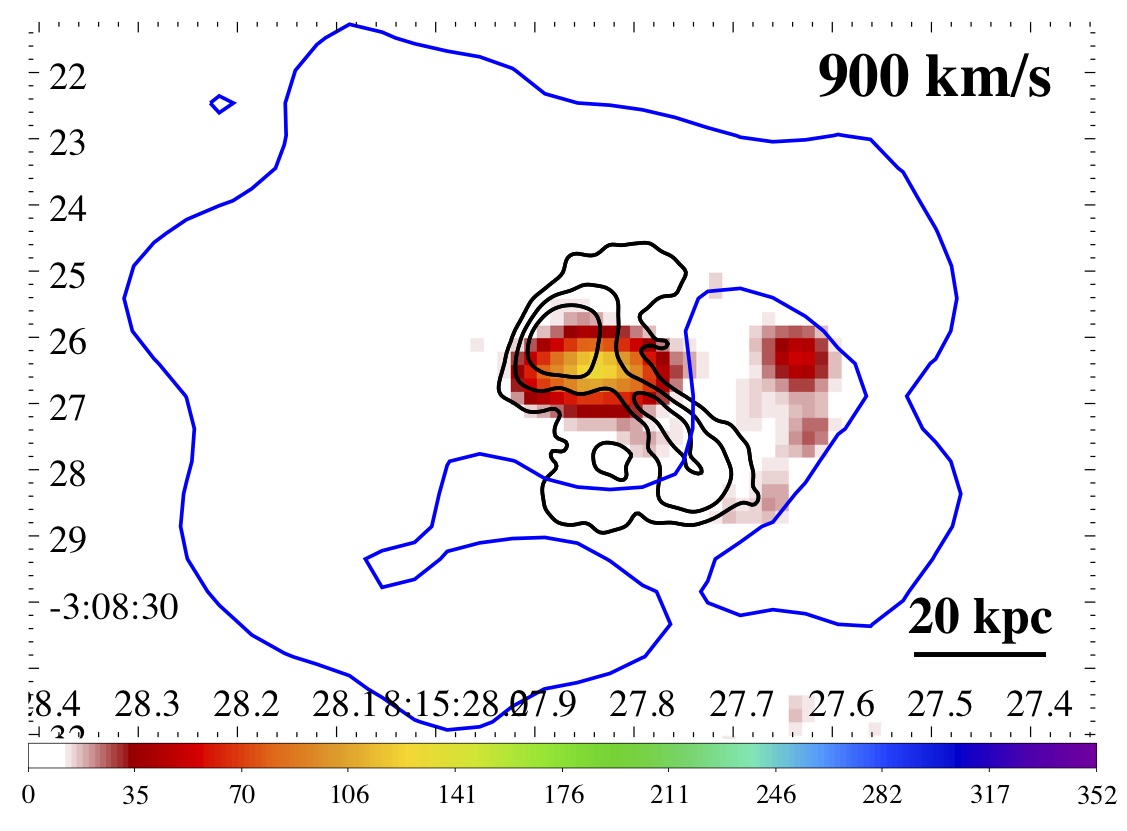}
\caption{H$\alpha$ + [N II]$\lambda6584$ flux in velocity bins of 100 km$\,$s$^{-1}$, with 8.4 GHz VLA (black) and exposure-corrected 0.7 - 2 keV {\it Chandra} (blue) contours overlaid. MUSE images have a pixel size of 0.2 arcsec/pixel. 8.4 GHz VLA contours were drawn at 5, 20, and 50 times the rms level of the background. $\textit{Chandra}$ contours were smoothed with a 1.5\arcsec\ Gaussian kernel radius and drawn in red at 0.25$\,\cdot\,$10$^{-15}$ erg$\,$cm$^{-2}\,$s$^{-1}$. Centers of the velocity bins are shown on the top right corners. Main features are labeled in figures, with L, the region co-spatial with the northeastern radio lobe, F1 and F2, the blue and red filaments that appear to surround the southwestern radio lobe, A, the arched emission that appears to trace the inner edge of the X-ray cavity, F3, an additional filament not co-spatial with the X-ray cavity, and R, the region polluted by the redward component.}
\label{fig:velocity}
\end{center}
\end{figure*}

We decided to give an overview of the gas kinematics via velocity channel maps, as shown in Fig. \ref{fig:velocity} since the blending of the spectral components (see Fig. \ref{fig:cloudspec}) and the low signal-to-noise in the cavity region prevent us from drawing strong quantitative conclusions from the study of the kinematics of each component. This approach is commonly used for isolated lines. Thus, to avoid contamination due to emission lines being part of a triplet, we considered different reference lines for blue and red-shifted emission. We considered as blue-shifted, emission blueward of the [N II]$\lambda6548$ rest frame line and, as red-shifted, emission redward of the [N II]$\lambda6584$ rest frame. Therefore, the central wavelength range including the emission of the H$\alpha$+[N II]$\lambda6584$, dominated by spectral blending, is not shown in the channel maps. Channel maps are shown in increments of 100 km$\,$s$^{-1}$, which correspond to two resolution elements, with the exception of the central channels that are shown in increments of 50 km$\,$s$^{-1}$.

Velocity channel maps show that the blue-shifted emission is co-spatial with the north-eastern radio lobe (region L). Additionally, there seems to be a filament of blue-shifted emission on the outer edge of the southern radio lobe (F1). The red-shifted emission, up to $\sim$300 km$\,$s$^{-1}$, shows some extended emission that appears to follow the north-eastern radio jet and a filament that seems to wrap around the southern lobe (F2). There also appears to be some emission connecting filaments F1 and F2, which is most visible at $\sim$400 km$\,$s$^{-1}$ (A). This arched feature is apparently tracing the inner edge of the X-ray cavity. Lastly, at the reddest velocities ($>$500 km$\,$s$^{-1}$), the only emission that is detected corresponds to the redward [N II]$\lambda6584$ shown in the right panel of Fig. \ref{fig:chandra} and in the spectrum of region {\it b} in Fig. \ref{fig:spectra} (region R).

All extended ionized gas features, i.e., lobe and red and blue-shifted filaments (regions L, F1, and F2, marked in Fig. \ref{fig:other}) present LINER-like ionization states, as shown in the BPT diagram in Fig. \ref{fig:bpt} (see, e.g., \citealt{Baldwin1981} and \citealt{Kewley2006}). The only region shown in Fig. \ref{fig:bpt} with a Seyfert-like ionization state is the region corresponding to the optical host position. The position in the BPT diagram of the extra redward component shown in Fig. \ref{fig:chandra}, right panel, could not be established since the spectral and spatial blending does not allow us to obtain estimates of the line ratios in this region with the current signal-to-noise ratio of our MUSE data. All extraction regions used for the BPT diagram were selected on the basis of specific morphology/kinematics of the source and are shown in Fig. \ref{fig:other}. Additionally, all regions have a size of $\sim$1 arcsec$^2$, to match the seeing of the observation.

\section{Discussion}
\label{sec:discussion}

Multiphase gas halos, surrounding AGN hosted in BCGs harbored in cool core galaxy clusters, represent one of the keys to understanding AGN feedback mechanisms. Currently, the main hypothesis for the origin of ionized gas in these halos is the cooling and condensation of the hot, X-ray emitting plasma, often in the shape of i) filaments (see e.g., \citealt{Fabian2008, Fabian2011, Fabian2012}, \citealt{Gaspari2017,Gaspari2018}, \citealt{Qiu2020, Qiu2021}, \citealt{Jimenez2021}), ii) rotating disks (see e.g., \citealt{Wilman2005} \citealt{Hamer2014}), or iii) plumes (see e.g., \citealt{Hamer2012, Hamer2016}, \citealt{Pasini2019}). One of the most remarkable examples of ionized gas filaments in a cool core cluster can be found in NGC 1275 in the center of the Perseus galaxy cluster (\citealt{Lynds1970}, \citealt{Conselice2001} and \citealt{Fabian2008}), where ionized gas filaments are spatially associated with an X-ray excess (\citealt{Fabian2011}). The same situation occurs for other BCGs of cool core clusters, where a tight connection between X-ray and optical filamentary emission was found, such as the cases of A1795 (\citealt{Crawford2005}), A1644 (\citealt{Mcdonald2010}), A2597 (\citealt{Tremblay2016}), and MKW\,3s (\citealt{Jimenez2021}).

Ionized gas filaments have also been found surrounding X-ray cavities, i.e., co-spatial with the rims of X-ray cavities (see e.g., works on A2052 by \citealt{Blanton2011} and \citealt{Balmaverde2018}, as well as works by \citealt{Lim2008}, \citealt{Tremblay2015} and \citealt{Olivares2019}). However, ionized gas spatially associated, and thus, potentially filling an X-ray cavity, is a rare occurrence with no clear explanation. In this work, we present the potential first detection of ionized gas spatially associated with an X-ray cavity, using H$\alpha$ + [N II]$\lambda6584$ emission as a tracer. Although the scenario where this ionized gas is actually overlaid to the X-ray cavity cannot be conclusively excluded, it appears unlikely. If that was the case, we would expect to see ionized gas spatially associated with the outer borders of the X-ray cavity, where its column density would be the highest, instead of with the X-ray cavity.

%In contrast with the H$\alpha$ + [N II]$\lambda6584$ emission, which extends from the northeast to the southwest of the nucleus, the detected H$\beta$ and [O III]$\lambda\lambda4960,5008$ emission is restricted to the central and northeastern regions (see left panel of Fig. \ref{fig:other}), which suggests a low ionization state.

\subsection{The north-eastern plume}

Similar to other cool core galaxy clusters such as A1991, A3444, Ophiuchus (\citealt{Hamer2012}), A111, A133, A2415 (\citealt{Hamer2016}), and A2495 (\citealt{Pasini2019}), the peak of the X-ray emission surrounding 3CR\,196.1 does not correspond to the position of the BCG, presenting an offset of $\sim$2\arcsec\ (i.e., $\sim$6.5 kpc; see left panel of Fig. \ref{fig:spectra}). Such offsets have been interpreted as caused by sloshing of the ICM due to events such as minor mergers. Typically, in those cases, the ionized gas forms plumes either following the X-ray peak or connecting the X-ray peak and the BCG. However, in the case of 3CR\,196.1, we argue that the X-ray peak seen in the ICM is most likely caused by the interaction of the north-eastern radio jet with the ICM, as seen in 3CR\,171 and 3CR\,305 (\citealt{Hardcastle2010, Hardcastle2012}), due to the alignment between the north-eastern radio lobe and the X-ray peak, as well as to the ``hammer-head" shape of the north-eastern radio lobe and the ionized gas plume extending in the direction opposite to the position of the BCG (see Fig. \ref{fig:other}). Thus, the ionized gas corresponding to region L could originate from the cooling of the jet-heated ICM. This origin is compatible with the ionized gas presenting a LINER-like ionization state, as shown in Fig. \ref{fig:bpt} since LINER-like emission in extranuclear regions is often associated with large-scale outflows and shocks or with radio jet-shocked regions (see \citealt{Kewley2006} and references therein).

\subsection{Ionized gas filaments around 3CR\,196.1}

One of the most likely mechanisms responsible for the formation of ionized gas filaments in cool core galaxy clusters is chaotic cold accretion (CCA, see \citealt{Gaspari2012, Gaspari2013, Gaspari2015,Gaspari2017,Gaspari2018}), i.e., the condensation of the hot plasma in the ICM and its precipitation onto the black hole. ICM cooling due to non-linear thermal instabilities was first introduced by \citet{Pizzolato2005} as ``cold feedback". More recently, \citet{Gaspari2018} adopted the ratio of the cooling time and the eddy turnover time, $C=t_{cool}/t_{eddy}$, as a tool to assess the multiphase state of a system, with $0.6<C<1.8$ marking the extent of the condensation region. We applied this criterion to 3CR\,196.1, where we assumed $t_{cool}\sim$500 Myr, as obtained by \citet{Ricci2018} for the inner region (i.e., 10\arcsec\ or $\sim$30 kpc radius), since currently available X-ray observations did not allow us to carry out a more detailed spectral analysis. We obtained the eddy turnover time, as a function of the distance to the galaxy cluster core, as $t_{eddy}=2\pi \frac{r^{2/3}L^{1/3}}{\sigma_{v,L}}$ (see \citealt{Gaspari2018}), where $L\sim$10 kpc is the injection scale, traced by the diameter of the X-ray cavity, and $\sigma_{v,L}\sim$212 km s$^{-1}$ is the velocity dispersion at the injection scale. Thus, significant precipitation is expected to occur within a region of $r\sim$9 - 48 kpc radius, where $C\approx1$. Since the blue and red filaments at the edges of the southern radio lobe (F1 and F2 in the right panel of Fig. \ref{fig:other}) extend up to 10 kpc (where $C\sim$1.7) and the small ionized gas filament, F3, extends up to 15 kpc (where $C\sim$1.3) to the south-east (see Fig. \ref{fig:velocity}), the origin of ionized gas filaments surrounding 3CR\,196.1 is compatible with a condensation rain. Moreover, the velocity and velocity dispersion along filament F2, $\sim$200 km s$^{-1}$ and $\sim$100 km s$^{-1}$, respectively, are consistent with those literature observations compared with CCA simulations by \citet{Gaspari2018}. Therefore, 3CR\,196.1 presents possible signs of black hole feeding.

Alternatively, as proposed e.g. in simulations by \citet{Qiu2020, Qiu2021}, ionized gas filaments in 3CR\,196.1 could be due to AGN outflows uplifting warm ($10^4<T\leq10^7$ K) gas from the central 2 kpc of the galaxy cluster, which then experiences radiative cooling. The energy needed to inflate adiabatically an ionized gas bubble with a radius $r\sim$5 kpc, i.e., the size of the arched structure {\it A} in Fig. \ref{fig:velocity}, and velocity $v\sim$400 km s$^{-1}$, in an ambient medium with density $n_0=0.098$ cm$^{-3}$, as obtained by \citet{Ricci2018} for the inner $\sim$30 kpc of 3CR\,196.1, can be obtained following the approach described by \citet{Nesvadba2006}:
\begin{equation}
    \dot E \approx 1.5\cdot10^{46}\,r^2_{10}\,v^3_{1000}\,n_0\,\text{erg s}^{-1}
\end{equation} 
where $r_{10}$ is the radius of the bubble in units of 10 kpc, and $v_{1000}$ is the velocity in units of 1000 km s$^{-1}$. Thus, the minimum energy needed to inflate such a bubble, $\dot E\sim$2.4$\cdot 10^{43}$ erg s$^{-1}$, is an order of magnitude below the jet power obtained by \citet{Ricci2018}, $P_{jet}\sim$1.9$\cdot 10^{44}$ erg s$^{-1}$, and, therefore, the ionized gas spatially associated with the X-ray cavity is consistent with a bubble being inflated by the radio jet.

The electron density in filaments F1 and F2 can be estimated using the [S II]$\lambda\lambda6718,6733$ duplet ratio as described in \citet{Sanders2016}. We selected two regions of the size of the PSF ($\sim$1.0\arcsec) over the filaments to compute the electron density. Thus, the electron density within the cavity is $<350$ e$^{-}\,$cm$^{-3}$. This density is comparable with that of the filaments found surrounding an X-ray cavity in A2052 by \citet{Balmaverde2018}, which were interpreted as being caused by an expanding bubble, as well as with other filaments in cool core galaxy clusters (see e.g., \citealt{Mcdonald2012}, \citealt{Jimenez2021}). Additionally, extended ionized gas features, corresponding to filaments F1 and F2, could be interpreted as being caused by the interaction of the southern radio jet with its surrounding medium, since the filaments present LINER-like ionization states, as shown in Fig. \ref{fig:bpt}.

Lastly, similarly to the expanding bubble in A2052, filaments F1 and F2 in 3CR\,196.1 are blue and redshifted, respectively, and a potential region with split lines that could be due to the overlap of the emission of both filaments, a typical sign of an expanding bubble (\citealt{Balmaverde2018}), is also found to be co-spatial with the X-ray cavity (region R in Fig. \ref{fig:velocity}). If that is the case, the expansion velocity of the bubble, which can only be estimated indirectly through X-ray observations, could be measured as the maximum velocity observed at the arched structure, A, i.e., $\sim$400 - 500 km s$^{-1}$. Nevertheless, due to the low signal-to-noise ratio in the cavity region of current MUSE observations, the kinematics of the ionized gas filaments cannot be firmly derived and, thus, this scenario cannot be confirmed.

\subsection{The central region of 3CR\,196.1}
During this analysis, we discovered that, in the central and X-ray cavity regions, all spectral features, except for the H$\beta$ and [O III]$\lambda\lambda4960,5008$ and [O I]$\lambda6300$ lines, have an additional component at $\sim$1000 km$\,$s$^{-1}$ to the red of rest-frame (see right panel of Fig. \ref{fig:chandra} and Figs. \ref{fig:spectra} and \ref{fig:cloudspec}). This additional redward component could be due to a background gas cloud since there does not appear to be a continuous range of velocities from rest-frame to the redward component that would indicate the presence of an outflow. This emission could also be due to filaments of ionized gas being accreted onto the black hole through CCA, similarly to filament F3. Despite that, the spatial resolution ($\sim$1\arcsec) of the current observations together with the blending of lines in the H$\alpha$ + [N II]$\lambda6584$ complex due to the limited spectral resolution ($\sim$50 km$\,$s$^{-1}$) prevent an in-depth analysis on the kinematics and ionization state of the gas in the central and X-ray cavity regions with current signal-to-noise.
%We estimated the total mass of ionized gas as
%\begin{equation}
% M=7.5\cdot10^{-3}\,\left(\frac{10^4}{n_e}\,\frac{L_{H\beta}}{L_\odot}\right) \,M_\odot
%\end{equation}

%with $L_{H\beta}$, the luminosity of the H$\beta$ line (see \citealt{Osterbrock1989}) and adopting $n_e \sim$20 e$^{-}\,$cm$^{-3}$. The ionized gas mass obtained is $M \sim$10$^7$ M$_\odot$, which is consistent with the masses found by \citet{Baldi2019} for a sample of 19 3C radio galaxies. 

\section{Summary and conclusions}
\label{sec:conclus}
We carried out a comparison of optical VLT/MUSE and X-ray {\it Chandra} observations of 3CR\,196.1, the central radio galaxy associated with the cool core galaxy cluster CIZA J0815.4-0303, which exhibits a ``butterfly-shaped'' X-ray cavity at $\sim$10 kpc toward the southwest of the nucleus. Through this analysis, we detected the presence of ionized gas spatially associated with an X-ray cavity.

The main possible explanations for detecting H$\alpha$ + [N II]$\lambda6584$ emission spatially associated with an X-ray cavity instead of with its rim include:
\begin{enumerate}
    \item The ionized gas is actually concentrated in filaments that wrap around the X-ray cavity, such as the ones shown in Fig. \ref{fig:velocity}. However, due to projection effects, as well as to the presence of an additional redward component in the cavity region (see right panel of Fig. \ref{fig:chandra} and Fig. \ref{fig:spectra}), the ionized gas appears to be co-spatial with the X-ray cavity. We tend to disfavor this scenario since it implies that we would be seeing the gas where it has the lowest column density along the line of sight, instead of at the edges, where its column density would be the highest.
\item The ionized gas is actually filling the X-ray cavity. In that case, there could be two possible explanations: (i) the gas has undergone multiple ionization events due to different AGN outbursts, or (ii) the ionized gas is forming filaments originating from the cooling of warm ($10^4<T\leq10^7$ K) AGN outflows. Scenario (i) is compatible with an AGN outburst occurring after the one responsible for the X-ray cavity, since typical cycling times between quiescence and activity in radio galaxies are $\sim$10$^7$-10$^8$ yr (see \citealt{Shabala2008} and references therein) and the estimated age of the X-ray cavity is 12 Myr (\citealt{Ricci2018}). On the other hand, scenario (ii) is compatible with simulations by \citet{Qiu2020}, which show that warm ($10^4<T\leq10^7$ K) gas present in the central 2 kpc of galaxy clusters (due to radiative feedback) could be lifted by AGN outflows and radiatively cooled, originating ionized gas filaments. Additional simulations in \citet{Qiu2021} showed that these filaments could be shaped as rings perpendicular to the direction of the outflow, which would explain the arched feature (A in Fig. \ref{fig:velocity}), that appears to connect filaments F1 and F2 (surrounding the southwestern radio lobe in Fig. \ref{fig:other}, right panel) and that seems to trace the inner edge of the X-ray cavity.
\end{enumerate}

To distinguish between these scenarios, additional observations would allow us to characterize, in more detail, the temperature and density profiles of the central ICM emission. To characterize the ionized gas, we would also need deeper and higher spatial resolution MUSE observations, with, e.g., the adaptive optic-assisted MUSE narrow-field-mode. Alternatively, the new ERIS instrument at the VLT, or the NIRSPEC IFU on the James Webb Space Telescope, could help us understand the origin of the additional redward component found in the spectra, as well as to derive the kinematics of the ionized gas spatially associated with the X-ray cavity. Such observations would allow us to measure the expansion velocity in the case in which the ionized gas is tracing an expanding bubble.

\acknowledgments % @ack
We thank the anonymous referee for their useful comments that led to the substantial improvement of the paper.
A. J. thanks M. Gaspari for his feedback on Chaotic Cold Accretion.
F. M. is in debt with S. Bianchi for his valuable input on X-ray photoionization scenarios.
This work is supported by the ``Departments of Excellence 2018 - 2022’’ Grant awarded by the Italian Ministry of Education, University and Research (MIUR) (L. 232/2016).
This research has made use of resources provided by the Ministry of Education, Universities and Research for the grant MASF\_FFABR\_17\_01.
This investigation is supported by the National Aeronautics and Space Administration (NASA) grants GO9-20083X and GO0-21110X.
A. J. acknowledges the financial support (MASF\_CONTR\_FIN\_18\_01) from the Italian National Institute of Astrophysics under the agreement with the Instituto de Astrofisica de Canarias for the ``Becas Internacionales para Licenciados y/o Graduados Convocatoria de 2017’’.
F.R. acknowledges support from PRIN MIUR 2017 project ``Black Hole winds and the Baryon Life Cycle of Galaxies: the stone-guest at the galaxy evolution supper”, contract $\#$2017PH3WAT.
G.V. acknowledges support from ANID program FONDECYT Postdoctorado 3200802.
A.P. acknowledges financial support from the Consorzio Interuniversitario per la fisica Spaziale (CIFS) under the agreement related to the grant MASF\_CONTR\_FIN\_18\_02.
W.F. and R.K. acknowledge support from the Smithsonian Institution and the Chandra High Resolution Camera Project through NASA contract NAS8-03060.
S.B. and C.O. acknowledge support from the Natural Sciences and Engineering Research Council (NSERC) of Canada.

\appendix

\section{Atrometric registration}
\label{app:regis}

\begin{figure*}
\begin{center}
\includegraphics[height=6.3cm,angle=0]{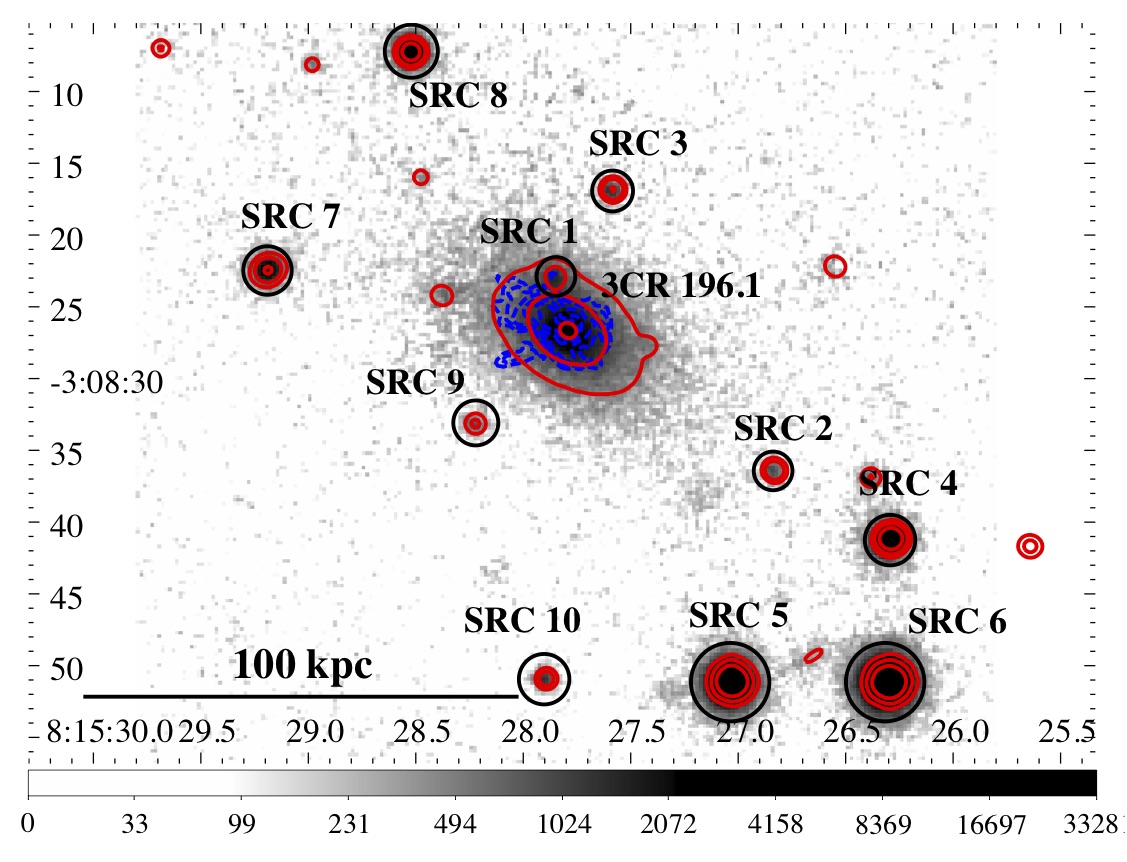}

\caption{\textit{r}-band PanSTARRS image with MUSE 4800 - 6800 $\AA$ (red) and MUSE H$\alpha$ + [N II]$\lambda6584$ (blue) contours overlaid. Sources SRC1 to SRC4 were used to compute the shift applied, while sources SRC5 to SRC8 allowed us to verify the registration. MUSE images have a pixel size of 0.2\arcsec. 4800 - 6800 $\AA$ contours were drawn at 20, 50, and 200 times the root mean square (rms) level of the background. H$\alpha$+[N II]$\lambda6584$ contours were drawn at 5, 10, 20, 50, and 100 times the rms level of the background. We found a good alignment between the Pan-STARRS and the MUSE emission applying a shift o 3.1\arcsec\ (i.e., $\sim$10 kpc).
}
\label{fig:registration}
\end{center}
\end{figure*}

All details related to the strategy adopted to carry out the astrometric registration of MUSE images are reported in the following. We measured the centroids of the optical counterpart of 3CR\,196.1 and sources SRC1 to SRC4, in Fig. \ref{fig:registration}, in the Pan-STARRS and MUSE white light collapsed images and applied the average of the shifts to the MUSE data cube. The final shift was 3.1\arcsec\ (corresponding to $\sim$10 kpc, which is expected since it is a known issue that MUSE astrometry can be off by several arcseconds, as reported by \citealt{Prieto2016}, \citealt{Balmaverde2019,Balmaverde2021}, and \citealt{Tucker2021}). We then verified that the registration adopted was appropriate for sources SRC5 to SRC8, in Fig. \ref{fig:registration}, by measuring the difference between their centroids in the Pan-STARRS and MUSE white light collapsed images after the shift. The adopted shift yielded an rms of 0.18\arcsec. The final alignment between Pan-STARRS and MUSE data for 3CR\,196.1 and the rest of the field sources is shown in Fig. \ref{fig:registration}, for the MUSE white light image (red contours) and the H$\alpha$+[N II]$\lambda6584$ emission (blue contours).

\newpage

\section{Ionized gas kinematics}
\label{sec:maps}

\begin{figure*}
\begin{center}
\includegraphics[height=3.5cm,angle=0]{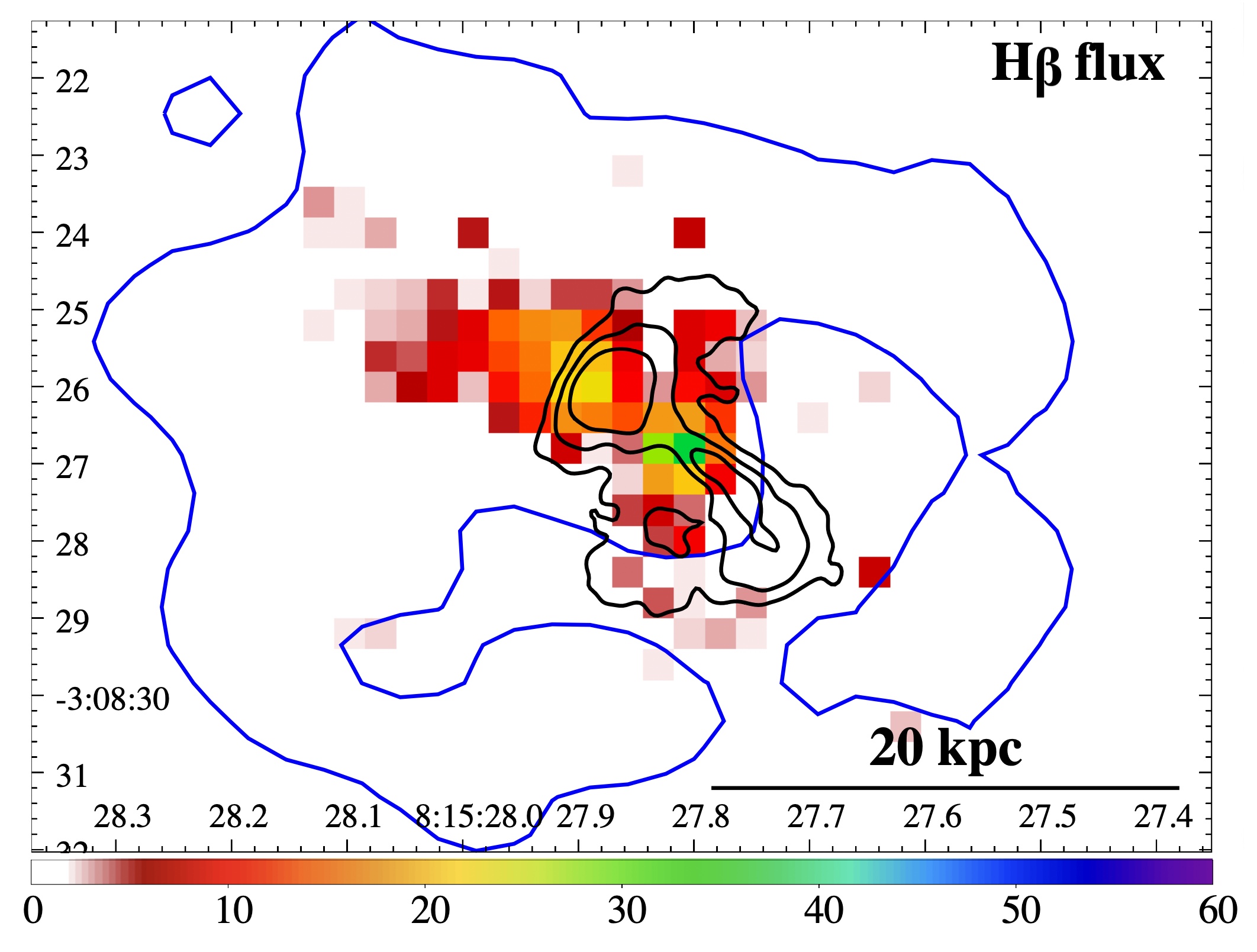}
\includegraphics[height=3.5cm,angle=0]{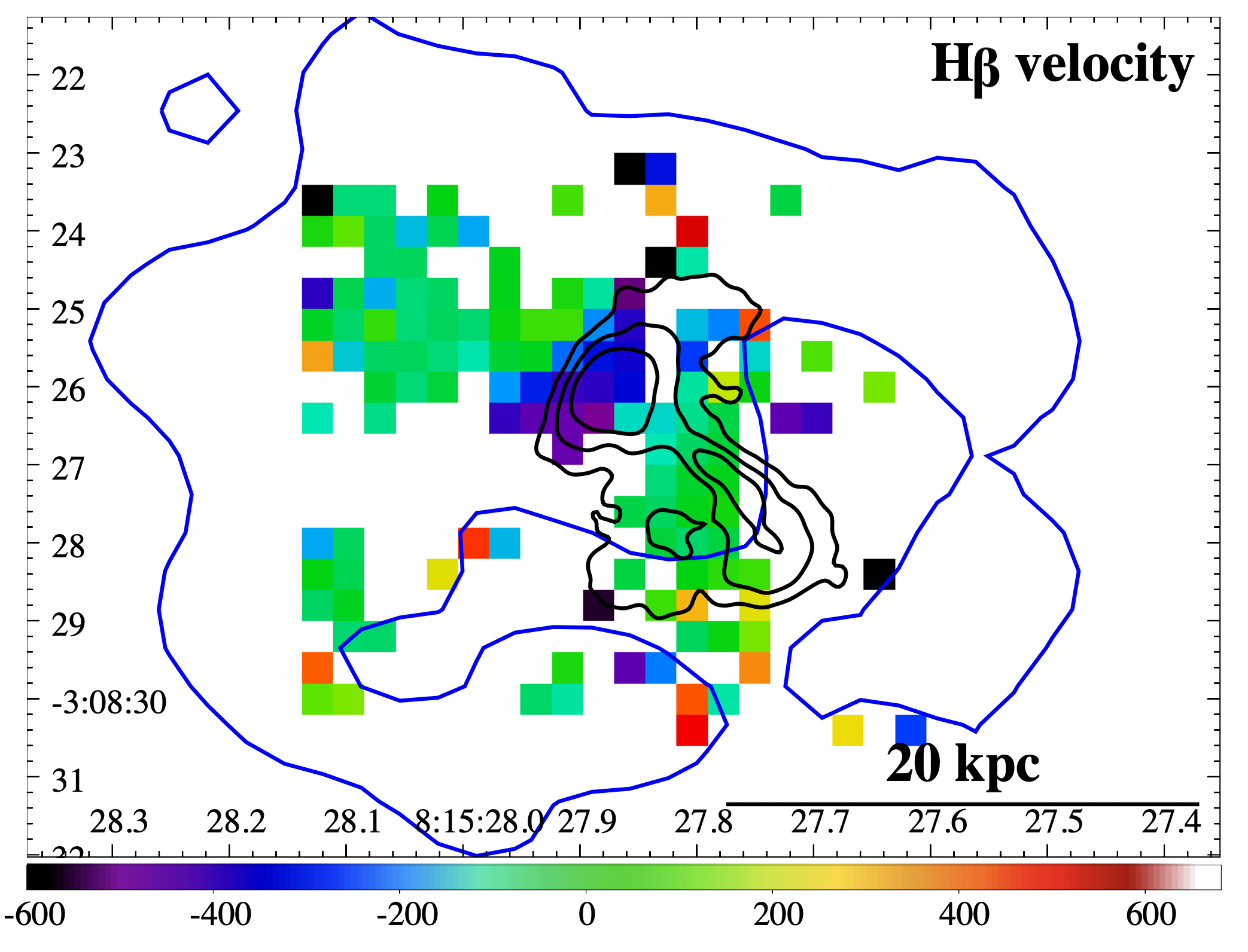}
\includegraphics[height=3.5cm,angle=0]{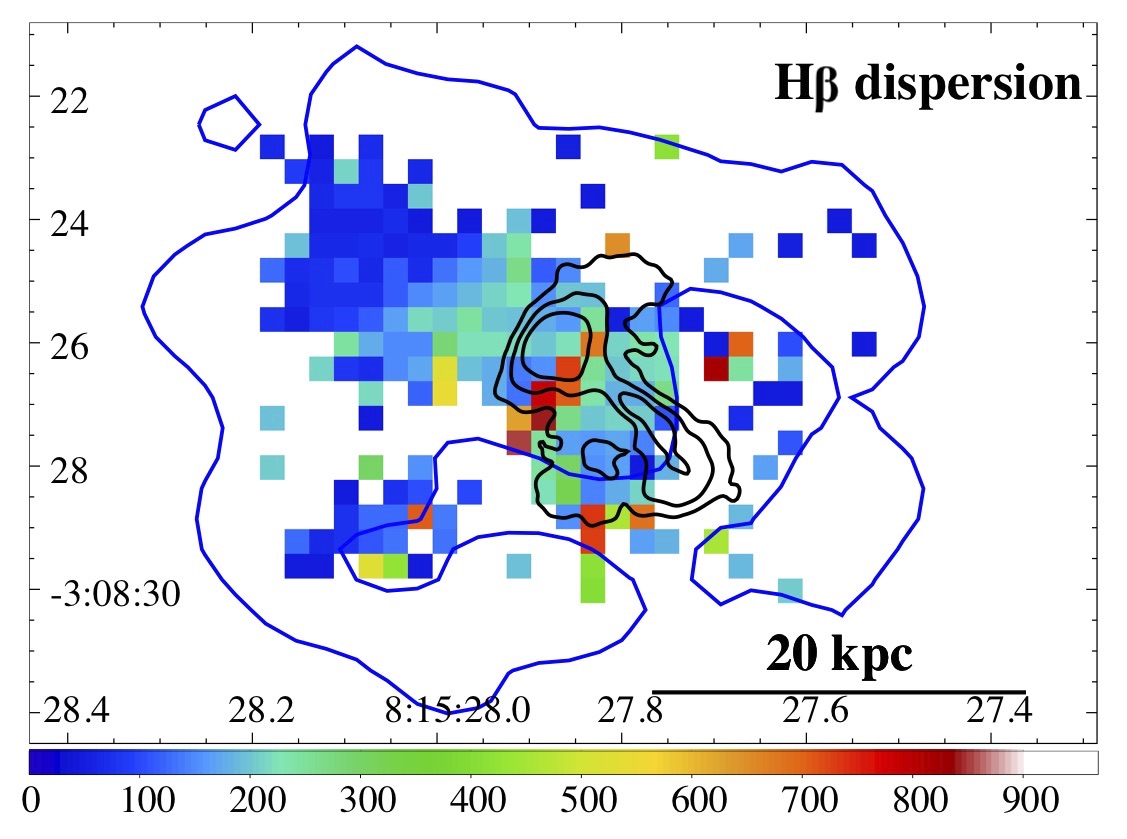}\\
\includegraphics[height=3.5cm,angle=0]{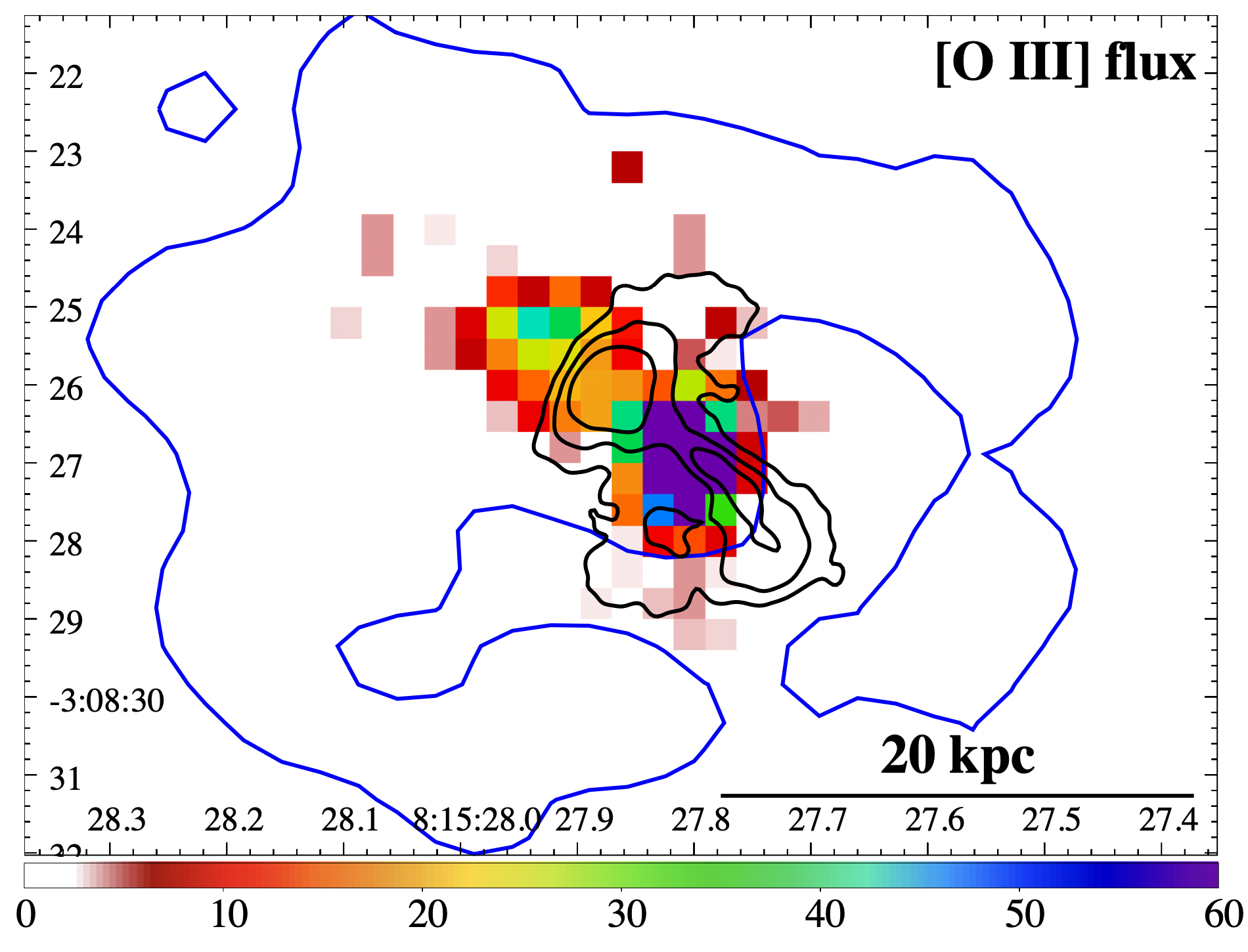}
\includegraphics[height=3.5cm,angle=0]{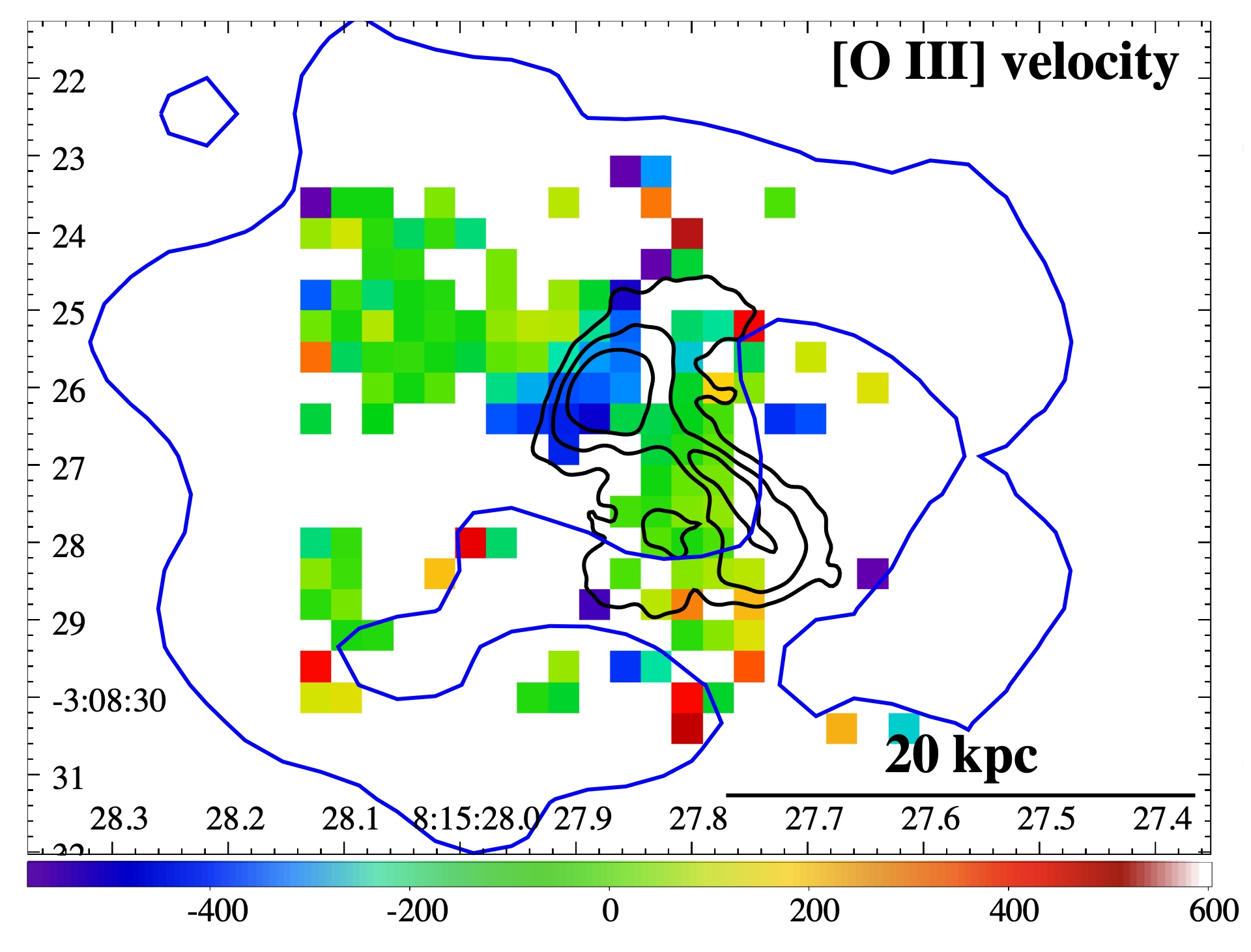}
\includegraphics[height=3.5cm,angle=0]{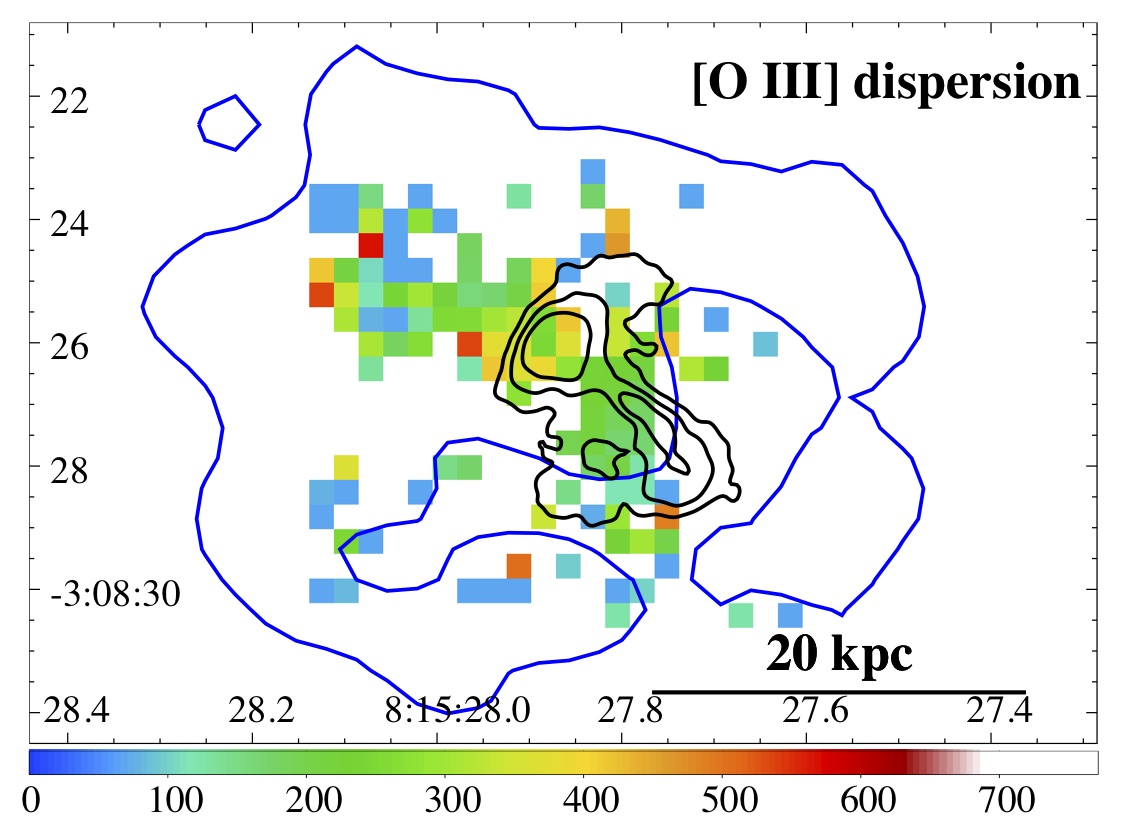}\\
\includegraphics[height=3.5cm,angle=0]{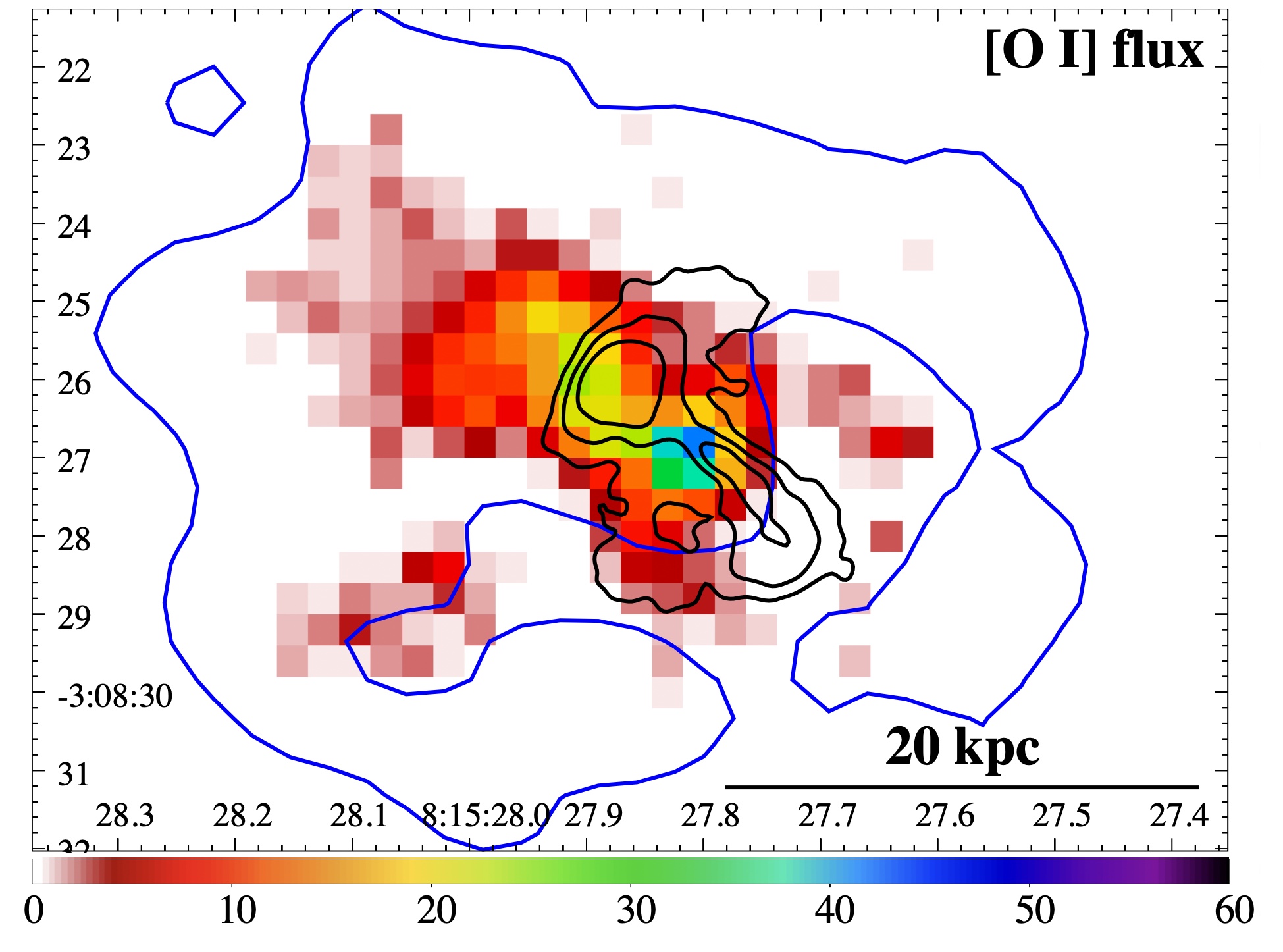}
\includegraphics[height=3.5cm,angle=0]{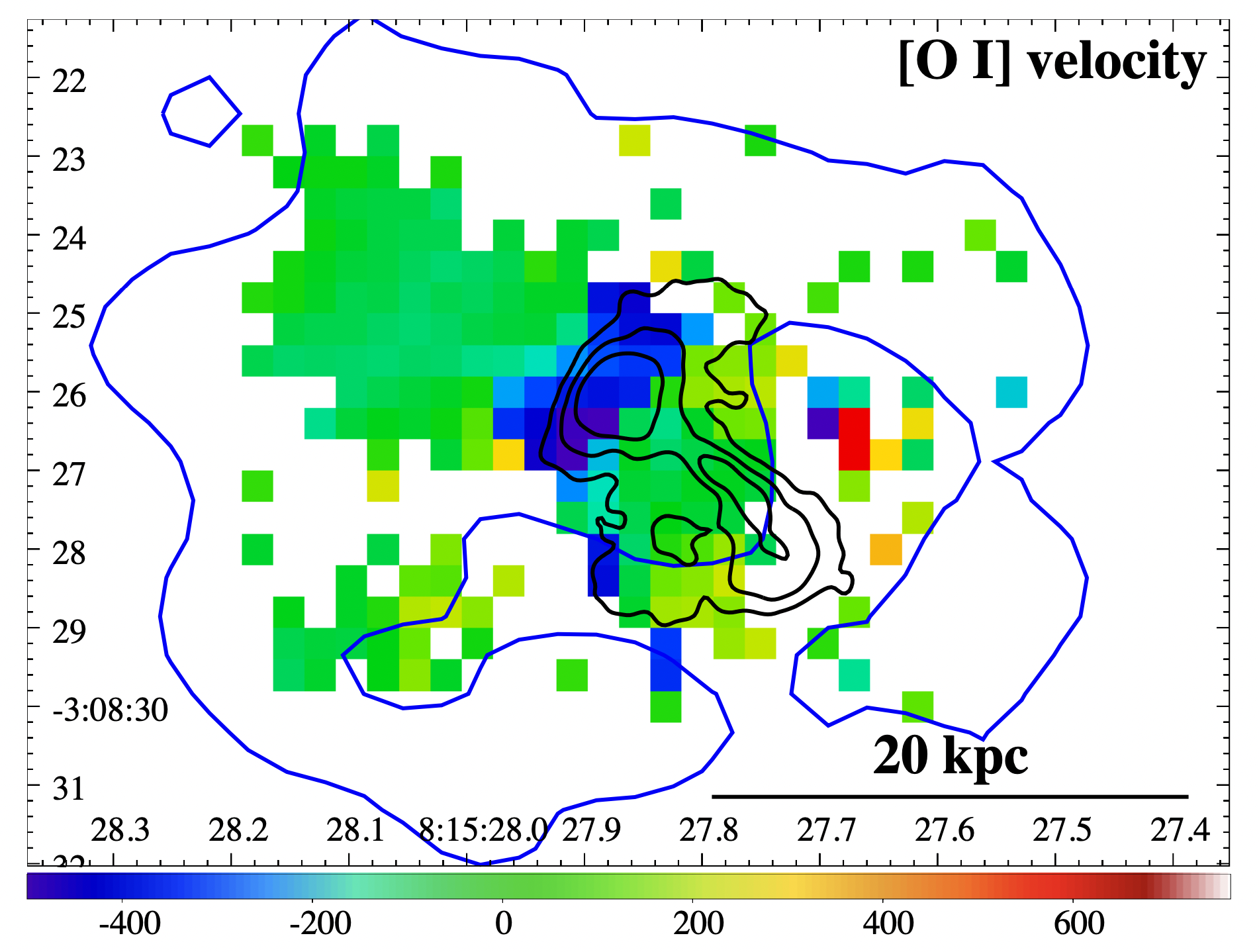}
\includegraphics[height=3.5cm,angle=0]{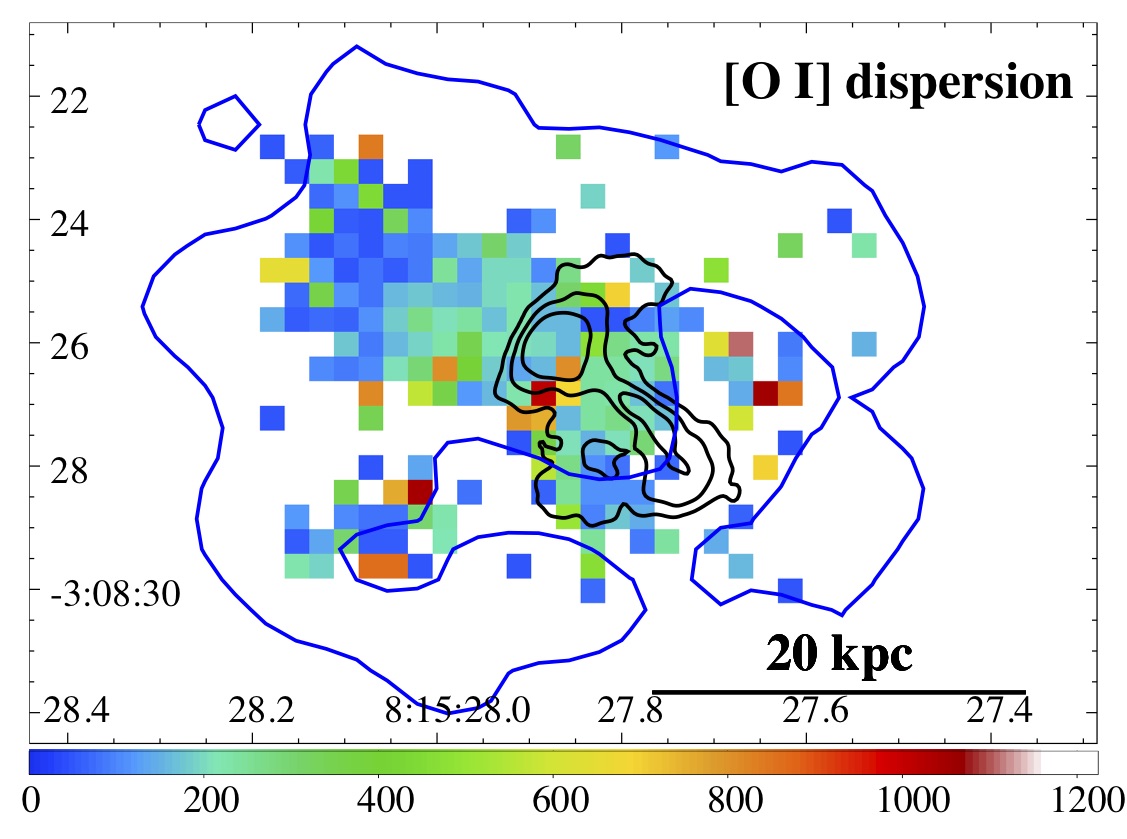}\\
\includegraphics[height=3.5cm,angle=0]{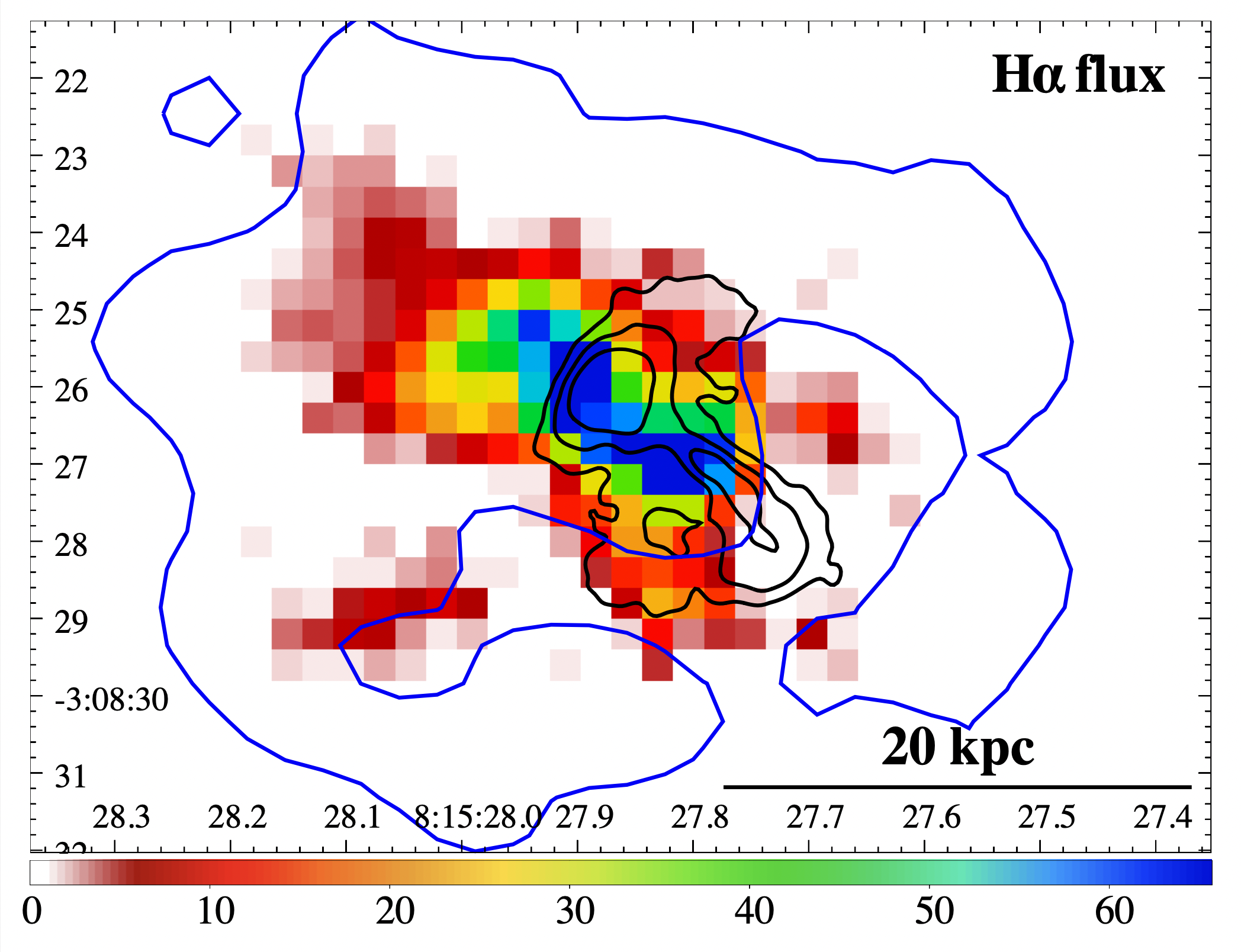}
\includegraphics[height=3.5cm,angle=0]{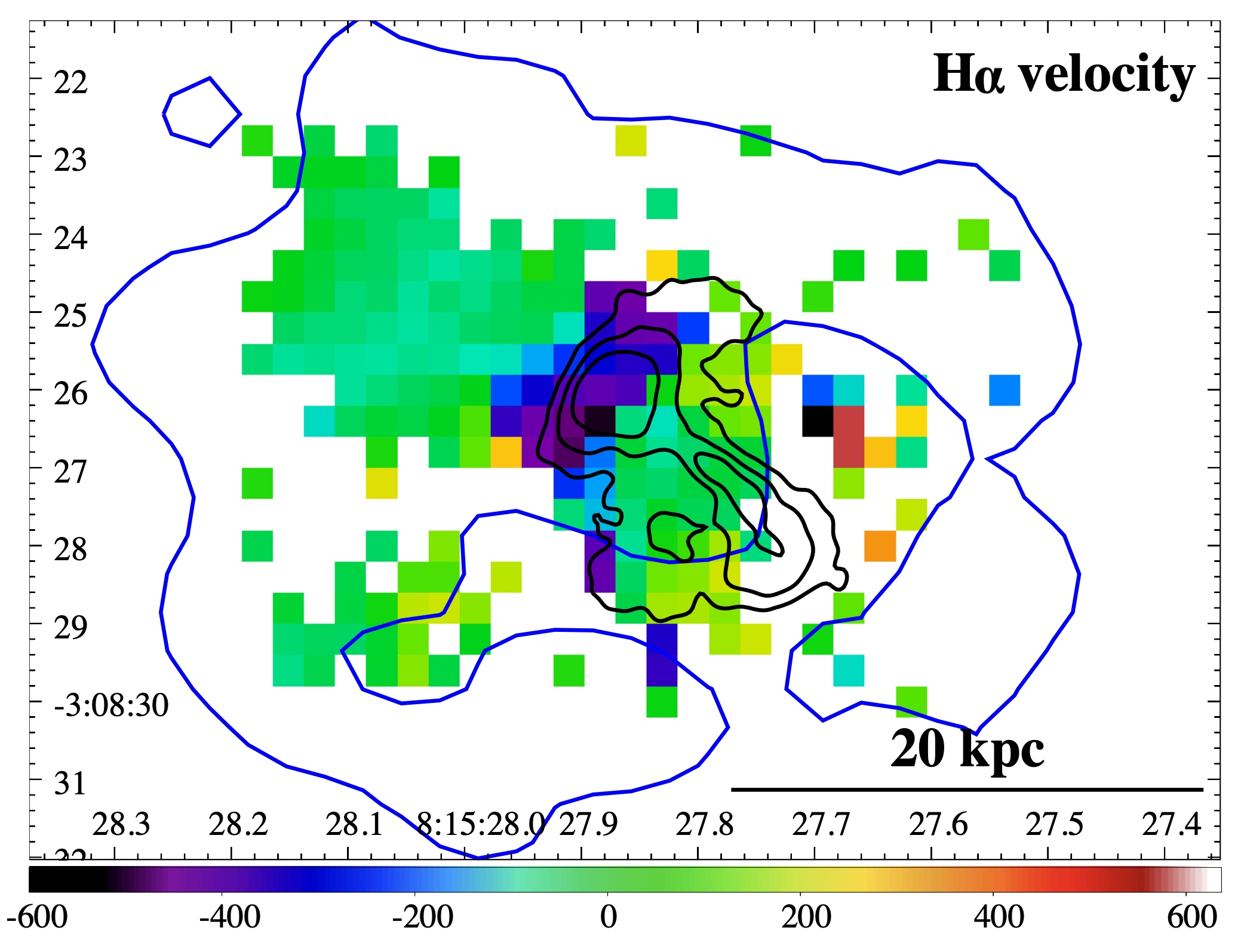}
\includegraphics[height=3.5cm,angle=0]{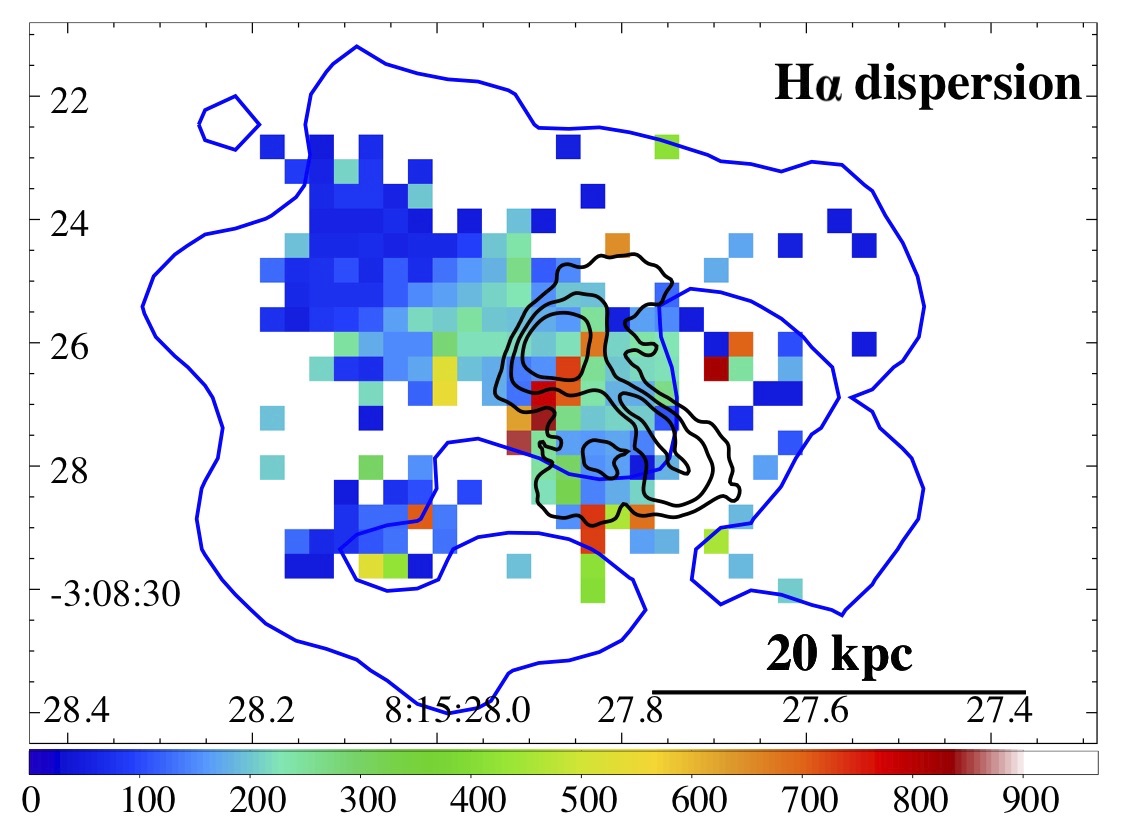}\\
\includegraphics[height=3.5cm,angle=0]{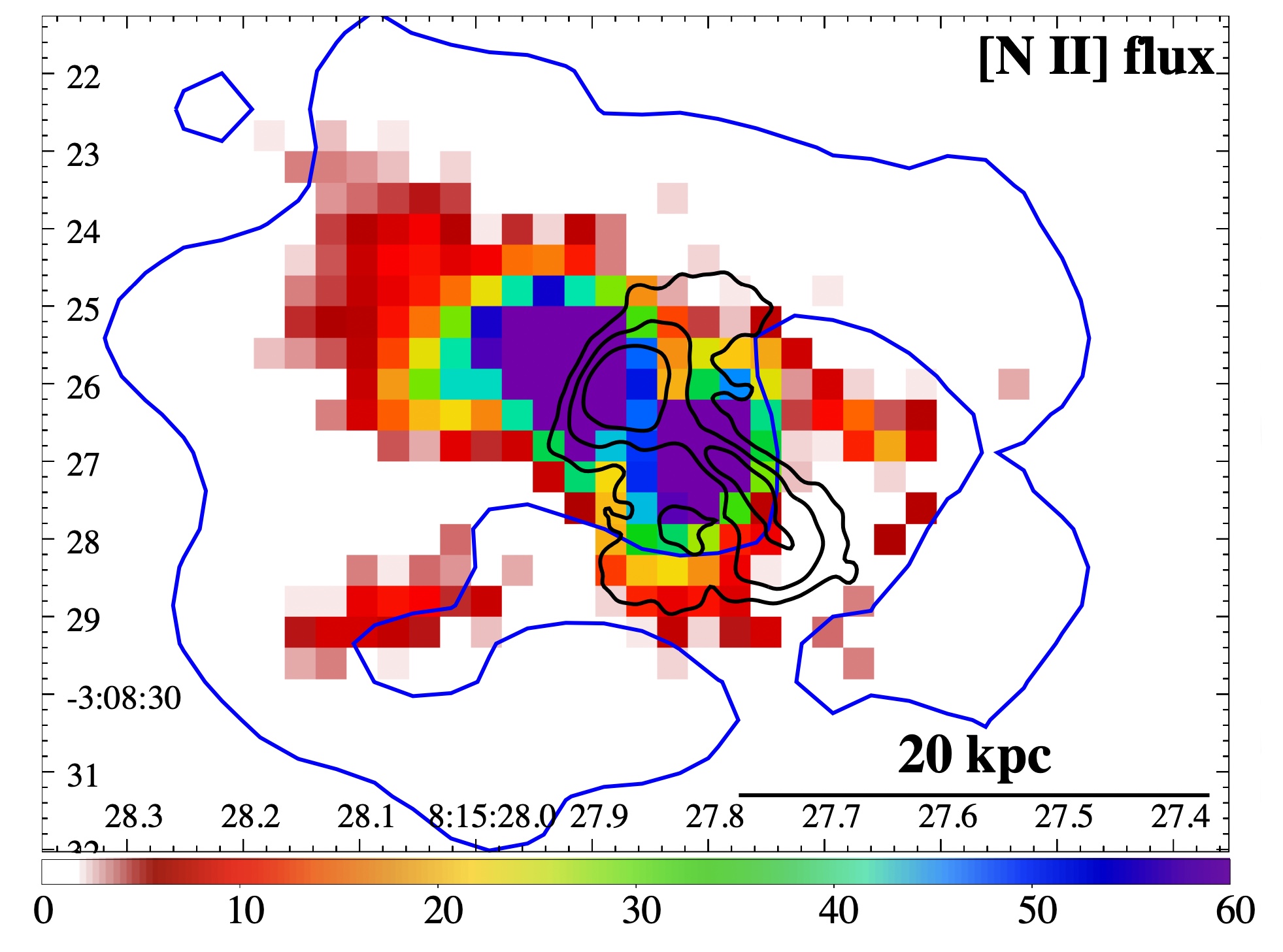}
\includegraphics[height=3.5cm,angle=0]{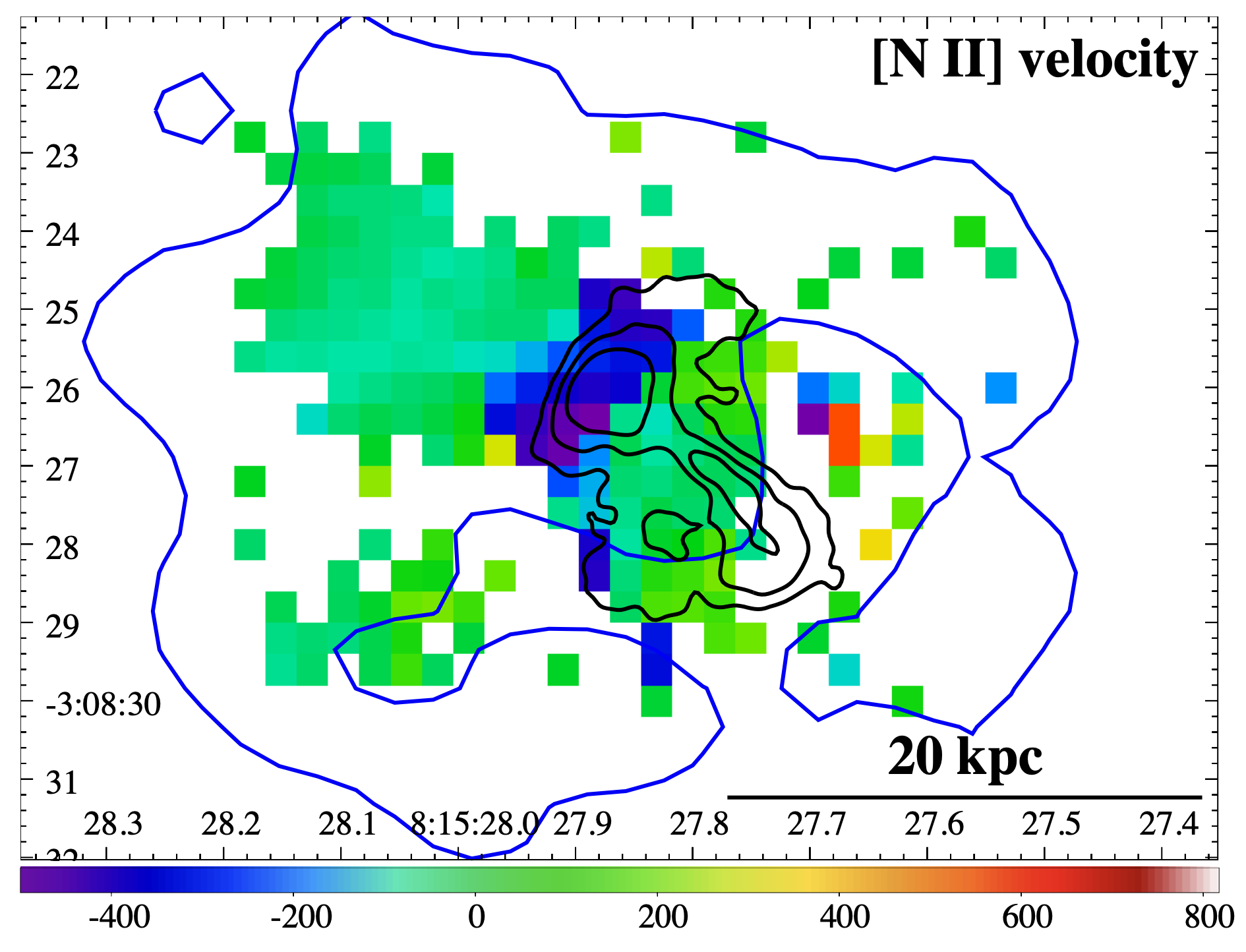}
\includegraphics[height=3.5cm,angle=0]{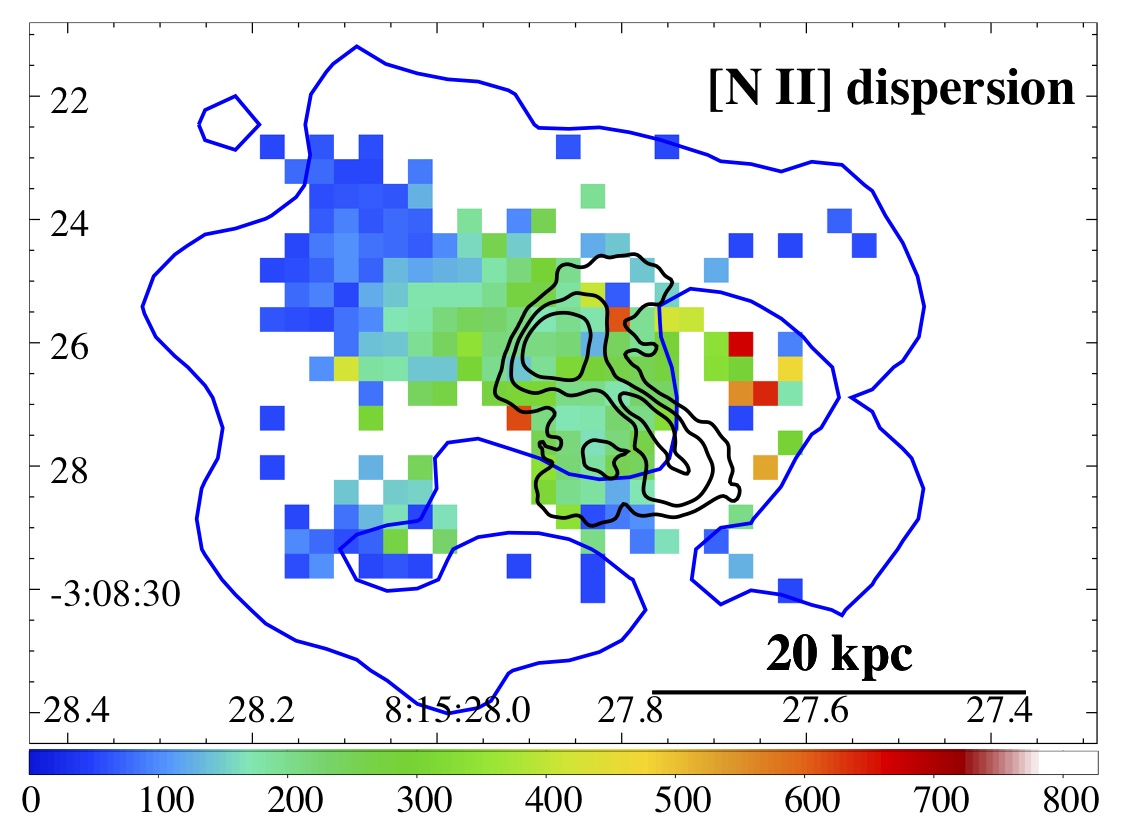}\\
\includegraphics[height=3.5cm,angle=0]{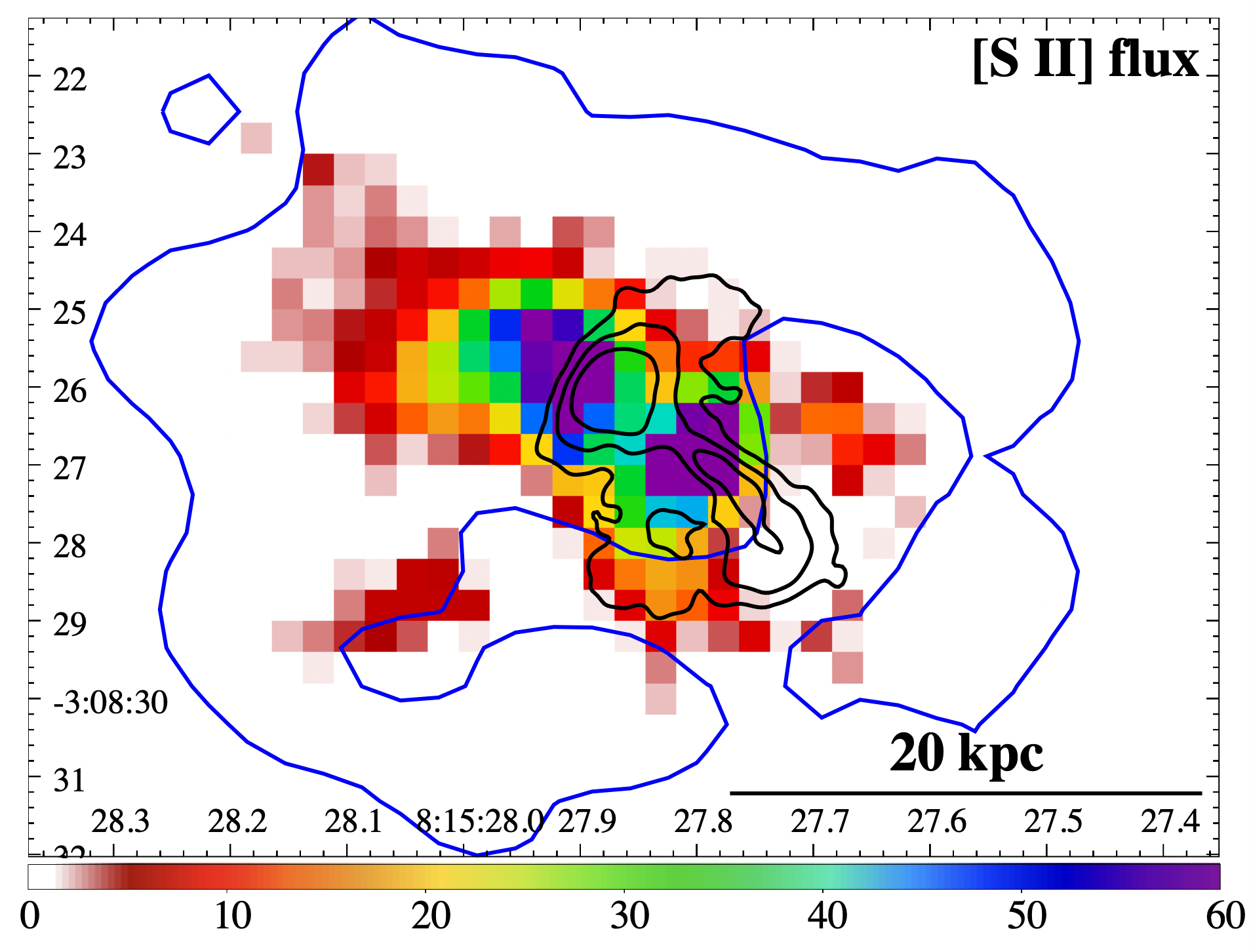}
\includegraphics[height=3.5cm,angle=0]{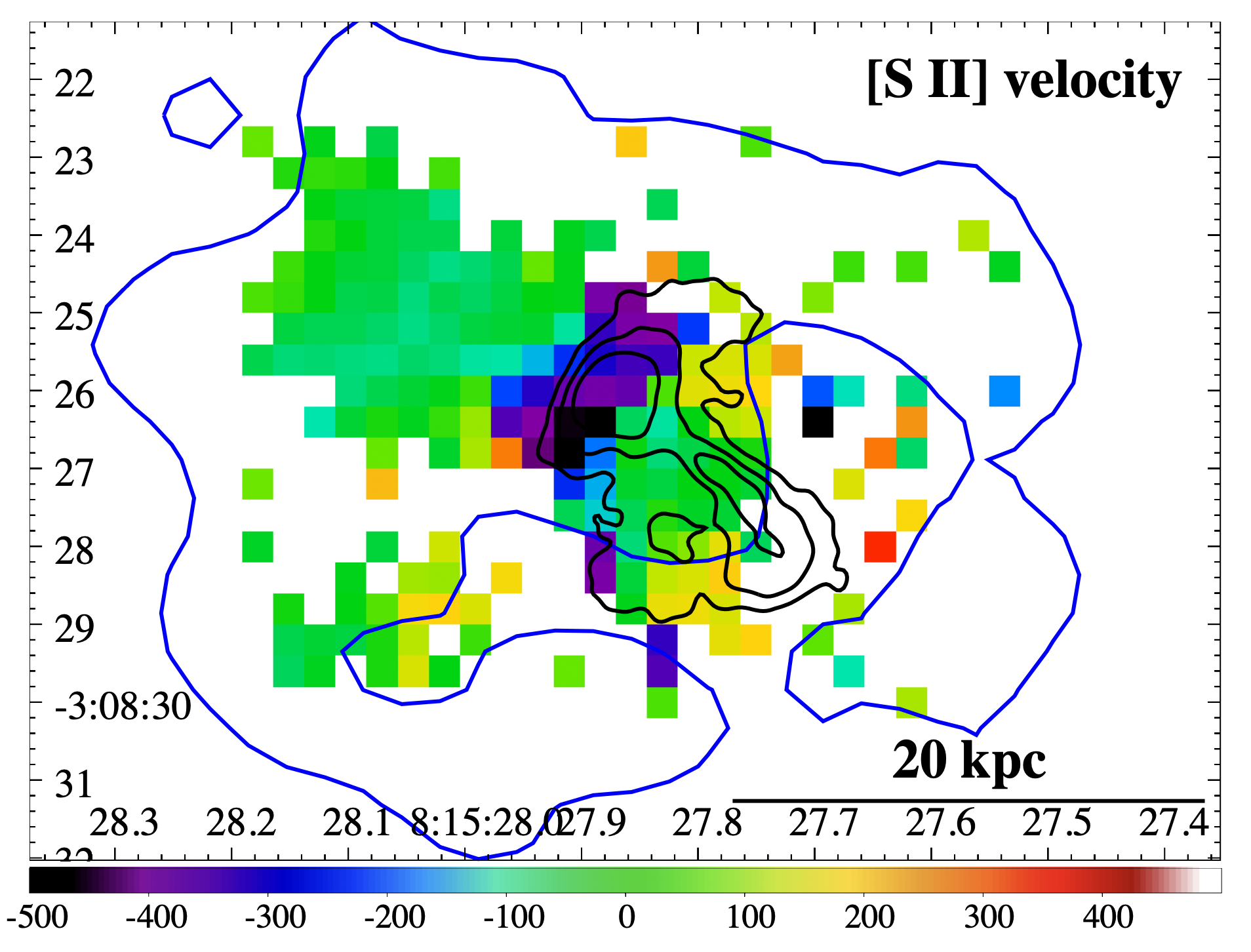}
\includegraphics[height=3.5cm,angle=0]{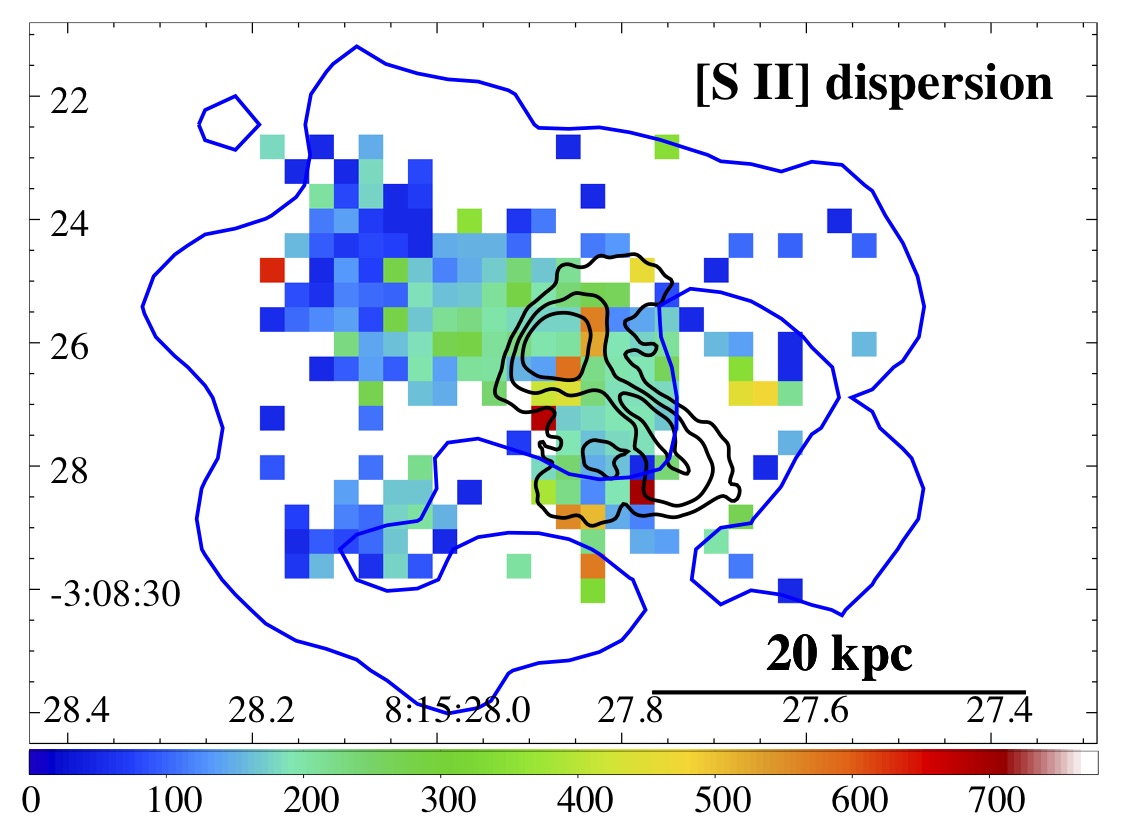}
\caption{Flux (left), velocity (middle), and velocity dispersion (right) maps with 8.4 GHz VLA (black) and exposure-corrected 0.7 - 2 keV {\it Chandra} (blue) contours overlaid for the H$\beta$, [O III]$\lambda\lambda4960,5008$, [O I]$\lambda6300$, H$\alpha$, [N II]$\lambda\lambda6548,6584$, and [S II]$\lambda\lambda6718,6733$ emission lines. MUSE images have a pixel size of 0.4 arcsec/pixel. 8.4 GHz VLA contours were drawn at 5, 20, and 50 times the rms level of the background. $\textit{Chandra}$ contours were smoothed with a 1.5\arcsec\ Gaussian kernel radius and drawn in blue at 0.25$\,\cdot\,$10$^{-15}$ erg$\,$cm$^{-2}\,$s$^{-1}$. Flux maps have units of 10$^{-17}$ erg$\,$cm$^{-2}\,$s$^{-1}$, while velocity and velocity dispersion maps have units of km s$^{-1}$.}
\label{fig:maps}
\end{center}
\end{figure*}

Here we report flux, velocity, and velocity dispersion maps for the H$\beta$, [O III]$\lambda\lambda4960,5008$, [O I]$\lambda6300$, H$\alpha$, [N II]$\lambda\lambda6548,6584$, and [S II]$\lambda\lambda6718,6733$ emission lines (see Fig. \ref{fig:maps}), obtained following the fitting procedure described in \S~\ref{sec:musedata}, with \textit{Chandra} and 8.4 GHz contours overlaid in blue and black, respectively. All maps were spatially binned by a factor of 2, resulting in a pixel size of 0.4 arcsec/pixel, to spatially Nyquist-sample the observation, as the image quality of MUSE data is of the order of 1\arcsec. Additionally this binning allows us to better match the resolution of {\it Chandra} and to increase the signal-to-noise ratio.

All emission features show similar morphologies and kinematics. The largest blueshift ($\sim$-400 km s$^{-1}$) can be seen at the termination of the northeastern radio lobe, pointing to the cooling of jet-heated ICM as the origin of ionized gas in the north-east direction. Lastly, the velocity dispersion of all lines is higher ($\sim$200 - 300 km s$^{-1}$) in the central $\sim$10 kpc of 3CR\,196.1.

\begin{thebibliography}{40}
\expandafter\ifx\csname natexlab\endcsname\relax\def\natexlab#1{#1}\fi
\bibitem[{{Bacon} {et~al.}(2010)}]{Bacon2010} {Bacon}, R., {Accardo}, M., {Adjali}, L., {et~al.} 2010, in Society of Photo-Optical Instrumentation Engineers (SPIE) Conference Series, Vol. 7735, Proc. SPIE, 773508
\bibitem[{{Baldi} {et~al.}(2019)}]{Baldi2019} {Baldi}, R.~D., {Rodr{\'\i}guez Zaur{\'\i}n}, J., {Chiaberge}, M., {et~al.} 2019, \apj, 870, 53
\bibitem[{{Baldwin} {et~al.}(1981)}]{Baldwin1981} {Baldwin}, J.~A., {Phillips}, M.~M. \& {Terlevich}, R. 1981, \pasp, 93, 5
\bibitem[{{Balmaverde} {et~al.}(2018)}]{Balmaverde2018} {Balmaverde}, B., {Capetti}, A., {Marconi}, A., {et~al.} 2018, \aap, 612, A19
\bibitem[{{Balmaverde} {et~al.}(2019)}]{Balmaverde2019} {Balmaverde}, B., {Capetti}, A., {Marconi}, A., {et~al.} 2019, \aap, 632, A124
\bibitem[{{Balmaverde} {et~al.}(2021)}]{Balmaverde2021} {Balmaverde}, B., {Capetti}, A., {Marconi}, A., {et~al.} 2021, \aap, 645, A12
\bibitem[{{Baum} {et~al.}(1988)}]{Baum1988} {Baum}, S.~A., {Heckman}, T.~M., {Bridle}, A., {et~al.} 1988, \apjs, 68, 643
\bibitem[{{Baum} {et~al.}(1990)}]{Baum1990} {Baum}, S.~A., {Heckman}, T. \& {van Breugel}, W. 1990, \apjs, 74, 389
\bibitem[{{Bennett}(1962)}]{Bennett1962} {Bennett}, A.~S. 1962, \mnras, 125, 75
\bibitem[{{Bennett} {et~al.}(2014)}]{bennett14} {Bennett}, C.~L., {Larson}, D., {Weiland}, J.~L. \& {Hinshaw}, G. 2014, \apj, 794, 135
\bibitem[{{B{\^\i}rzan} {et~al.}(2004)}]{Birzan2004} {B{\^\i}rzan}, L., {Rafferty}, D.~A., {McNamara}, B.~R., {et~al.} 2004, \apj, 607, 800
\bibitem[{{Blanton} {et~al.}(2011)}]{Blanton2011} {Blanton}, E.~L., {Randall}, S.~W., {Clarke}, T.~E., {et~al.} 2011, \apj, 737, 99
\bibitem[{{B\"ohringer} {et~al.}(1996)}]{Boringer1996} {B\"ohringer}, H., {Neumann}, D.~M., {Schindler}, S., {et~al.} 1996, \apj, 467, 168
\bibitem[{{Buttiglione} {et~al.}(2010)}]{Buttiglione2010} {Buttiglione}, S., {Capetti}, A., {Celotti}, A., {et~al.} 2010, \aap, 509, A6
\bibitem[{{Capetti} \& {Baldi}(2011)}]{Capetti2011} {Capetti}, A. \& {Baldi}, R.~D. 2011, \aap, 529, A126
\bibitem[{{Cappellari}(2017)}]{Cappellari2017} {Cappellari}, M. 2017, \mnras, 466, 798
\bibitem[{{Cavagnolo} {et~al.}(2010)}]{Cavagnolo2010} {Cavagnolo}, K.~W., {McNamara}, B.~R., {Nulsen}, P.~E.~J., {et~al.} 2010, \apj, 720, 1066
\bibitem[{{Churazov} {et~al.}(2000)}]{Churazov2000} {Churazov}, E., {Forman}, W., {Jones}, C., {et~al.} 2000, \aap, 356, 788
\bibitem[{{Conselice} {et~al.}(2001)}]{Conselice2001} {Conselice}, C.~J., {Gallagher}, J.~S., III \& {Wyse}, R.~F.~G. 2001, \aj, 122, 2281
\bibitem[{{Crawford} {et~al.}(2005)}]{Crawford2005} {Crawford}, C.~S., {Sanders}, J.~S. \& {Fabian}, A.~C. 2005, \mnras, 361, 17
\bibitem[{{de Koff} {et~al.}(1996)}]{deKoff1996} {de Koff}, S., {Baum}, S.~A., {Sparks}, W.~B., {et~al.} 1996, \apjs, 107, 621
\bibitem[{{Edge} {et~al.}(1959)}]{Edge1959} {Edge}, D.~O., {Shakeshaft}, J.~R., {McAdam}, W.~B., {et~al.} 1959, \memras, 69, 37
\bibitem[{{Fabian} {et~al.}(2003)}]{Fabian2003} {Fabian}, A.~C., {Sanders}, J.~S., {Allen}, S.~W., {et~al.} 2003, \mnras, 344, L43
\bibitem[{{Fabian} {et~al.}(2006)}]{Fabian2006} {Fabian}, A.~C., {Sanders}, J.~S., {Taylor}, G.~B., {et~al.} 2006, \mnras, 366, 417
\bibitem[{Fabian} {et al.}(2008)]{Fabian2008} {Fabian}, A.~C., {Johnstone}, R.~M., {Sanders}, J.~S., {et~al.} 2008, \nat, 454, 968
\bibitem[{Fabian} {et al.}(2011)]{Fabian2011} {Fabian}, A.~C., {Sanders}, J.~S., {Williams}, R.~J.~R., {et~al.} 2011, \mnras, 417, 172
\bibitem[{{Fabian}(2012)}]{Fabian2012}{Fabian}, A.~C. 2012, \araa, 50, 455
\bibitem[{{Fanaroff} \& {Riley}(1974)}]{Fanaroff1974}
{Fanaroff}, B.~L. \& {Riley}, J.~M. 1974, MNRAS, 167, 31P
\bibitem[{{Forman} {et~al.}(2005)}]{Forman2005} {Forman}, W., {Jones}, C., {Churazov}, E., {et~al.} 2005, \apj, 665, 1057
\bibitem[{{Forman} {et~al.}(2017)}]{Forman2017} {Forman}, W., {Churazov}, E., {Jones}, C., {et~al.} 2017, \apj, 844, 122
\bibitem[{{Fruscione} {et~al.}(2006)}]{Fruscione2006} {Fruscione}, A., {McDowell}, J.~C., {Allen}, G.~E., {et~al.} 2006, Society of Photo-Optical Instrumentation Engineers (SPIE) Conference Series, 6270, 62701V
\bibitem[{{Fosbury}(1986)}]{Fosbury1986} {Fosbury}, R.~A.~E. 1986, in Astrophysics and Space Science Library Vol. 121, Structure and Evolution of Active Galactic Nuclei, ed. G. Giuricin, M. Mezzetti, M. Ramella, \& F. Mardirossian (Dordrecht: Reidel), 297
\bibitem[{{Gaspari} {et~al.}(2012)}]{Gaspari2012} {Gaspari}, M., {Ruszkowski}, M. \&{Sharma}, P. 2012, \apj, 746, 94
\bibitem[{{Gaspari} {et~al.}(2013)}]{Gaspari2013} {Gaspari}, M., {Ruszkowski}, M. \& {Oh}, S.~Peng 2013, \mnras, 432, 3401
\bibitem[{{Gaspari} {et~al.}(2015)}]{Gaspari2015} {Gaspari}, M., {Brighenti}, F. \& {Temi}, P. 2015, \aap, 579, A62
\bibitem[{{Gaspari} {et~al.}(2017)}]{Gaspari2017} {Gaspari}, M., {Temi}, P. \& {Brighenti}, F. 2017, \mnras, 466, 677
\bibitem[{{Gaspari} {et~al.}(2018)}]{Gaspari2018} {Gaspari}, M., {McDonald}, M., {Hamer}, S.~L., {et~al.} 2018, \apj, 854, 167
\bibitem[{{Gitti} {et~al.}(2012)}]{Gitti2012} {Gitti}, M., {Brighenti}, F. \& {McNamara}, B.~R. 2012, AdAst, 2012, 950641
\bibitem[{{Gopal-Krishna} \& {Wiita}(2000)}]{Gopal2000} {Gopal-Krishna} \& {Wiita}, P.~J. 2000, \aap, 363, 507
\bibitem[{{Graham} {et~al.}(2008)}]{Graham2008} {Graham}, J., {Fabian}, A.~C. \& {Sanders}, J.~S. 2008, \mnras, 386, 278
\bibitem[{{Hamer} {et~al.}(2012)}]{Hamer2012} {Hamer}, S.~L., {Edge}, A.~C., {Swinbank}, A.~M., {et~al.} 2012, \mnras, 421, 3409
\bibitem[{{Hamer} {et~al.}(2014)}]{Hamer2014} {Hamer}, S.~L., {Edge}, A.~C., {Swinbank}, A.~M., {et~al.} 2014, \mnras, 437, 862
\bibitem[{{Hamer} {et~al.}(2016)}]{Hamer2016} {Hamer}, S.~L., {Edge}, A.~C., {Swinbank}, A.~M., {et~al.} 2016, \mnras, 460, 1758
\bibitem[{{Hansen} {et~al.}(1987)}]{Hansen1987} {Hansen}, L., {Norgaard-Nielsen}, H.~U. \& {Jorgensen}, H.~E. 1987, \aaps, 71, 465
\bibitem[{{Hardcastle} {et~al.}(2010)}]{Hardcastle2010}
{Hardcastle}, M.~J., {Massaro}, F. \& {Harris}, D.~E., 2010, \mnras, 401, 2697
\bibitem[{{Hardcastle} {et~al.}(2012)}]{Hardcastle2012}
{Hardcastle}, M.~J., {Massaro}, F. \& {Harris}, D.~E., 2012, \mnras, 424, 1774
%\bibitem[{{Hudson} {et~al.}(2010)}]{Hudson2010} {Hudson}, D.~S., {Mittal}, R., {Reiprich}, T.~H., {et~al.} 2010, \aap, 513, A37
\bibitem[{{Jimenez-Gallardo} {et~al.}(2021)}]{Jimenez2021} {Jimenez-Gallardo}, A., {Massaro}, F., {Balmaverde}, B., {et~al.} 2021, \apjl, 912, L25
\bibitem[{{Jones} {et~al.}(2002)}]{Jones2002} {Jones}, C., {Forman}, W.,{Vikhlinin}, A., {et~al.} 2002, \apjl, 567, L115
\bibitem[{{Kauffmann} {et~al.}(2003)}]{Kauffmann2003} {Kauffmann}, G., {Heckman}, T.~M., {Tremonti}, C., {et~al.} 2003, \mnras, 346, 1055
\bibitem[{{Kewley} {et~al.}(2001)}]{Kewley2001} {Kewley}, L.~J., {Heisler}, C.~A., {Dopita}, M.~A, {et~al.} 2001, \apjs, 132, 37
\bibitem[{{Kewley} {et~al.}(2006)}]{Kewley2006} {Kewley}, L.~J., {Groves}, B., {Kauffmann}, G., {et~al.} 2006, \mnras, 372, 961
\bibitem[{{Kocevski} {et~al.}(2007)}]{Kocevski2007} {Kocevski}, D.~D., {Ebeling}, H., {Mullis}, C.~R., {et~al.} 2007, \apj, 662, 224
\bibitem[{{Kraft} {et~al.}(2012)}]{Kraft2012} {Kraft}, R.~P., {Birkinshaw}, M., {Nulsen}, P.~E.~J., {et~al.} 2012, \apj, 749, 19
\bibitem[{{Laing} {et~al.}(1983)}]{Laing1983} {Laing}, R.~A., {Riley}, J.~M. \& {Longair}, M.~S. 1983, \mnras, 204, 151
\bibitem[{{Lim} {et~al.}(2008)}]{Lim2008} {Lim}, J., {Ao}, YiPing \& {Dinh-V-Trung} 2008, \apj, 672, 252
\bibitem[{{Lynds}(1970)}]{Lynds1970} {Lynds}, R. 1970, \apjl, 159, L151
\bibitem[{{Madrid} {et~al.}(2006)}]{Madrid2006} {Madrid}, J.~P., {Chiaberge}, M., {Floyd}, D., {et~al.} 2006, \apjs, 164, 307
\bibitem[{{Markevitch} \& {Vikhlinin}(2007)}]{Markevitch2007} {Markevitch}, M. \& {Vikhlinin}, A. 2007, \physrep, 443, 1
\bibitem[{{Massaro} {et~al.}(2011)}]{Massaro2011} {Massaro}, F., {Harris}, D.~E. \& {Cheung}, C.~C. 2011, \apjs, 197, 24
\bibitem[{{Massaro} {et~al.}(2012)}]{Massaro2012} {Massaro}, F., {Tremblay}, G.~R., {Harris}, D.~E., {et~al.} 2012, \apjs, 203, 31
\bibitem[{{Massaro} {et~al.}(2015)}]{Massaro2015}
{Massaro}, F., {Harris}, D.~E., {Liuzzo}, E., {et~al.} 2015, \apjs, 220, 5
\bibitem[{{McCarthy}(1988)}]{McCarthy1988}{McCarthy}, P.~J. 1988, PhD thesis, AA (California Univ., Berkeley)
\bibitem[{{McDonald} {et~al.}(2010)}]{Mcdonald2010} {McDonald} M., {Veilleux}, S., {Rupke}, D.~S.~N., {et~al.} 2010, \apj, 721, 1262
\bibitem[{{McDonald} {et~al.}(2012)}]{Mcdonald2012} {McDonald} M., {Veilleux}, S., {Rupke}, D.~S.~N., {et~al.} 2012, \apj, 746, 153
\bibitem[{{McNamara} {et~al.}(2005)}]{McNamara2005} {McNamara}, B.~R., {Nulsen}, P.~E.~J., {Wise}, M.~W., {et~al.} 2005, \nat, 433, 7021
\bibitem[{{McNamara} \& {Nulsen}(2007)}]{McNamara2007} {McNamara}, B.~R. \& {Nulsen}, P.~E.~J. 2007, \araa, 45, 117
\bibitem[{{McNamara} {et~al.}(2009)}]{McNamara2009} {McNamara}, B.~R., {Kazemzadeh}, F., {Rafferty}, D.~A., {et~al.} 2009, \apj, 698, 594
\bibitem[{{McNamara} \& {Nulsen}(2012)}]{McNamara2012} 
{McNamara} B.~R. \& {Nulsen}, P.~E.~J. 2012, New Journal of Physics, 14, 5
\bibitem[{{Nesvadba} {et~al.}(2006)}]{Nesvadba2006} {Nesvadba}, N.~P.~H., {Lehnert}, M.~D., {Eisenhauer}, F., {et~al.} 2006, \apj, 650, 693
\bibitem[{{Nulsen} {et~al.}(2005a)}]{Nulsen2005a} {Nulsen}, P.~E.~J., {McNamara}, B.~R., {Wise}, M.~W., {et~al.} 2005a, \apj, 628, 629
\bibitem[{{Nulsen} {et~al.}(2005b)}]{Nulsen2005b} {Nulsen}, P.~E.~J., {Hambrick}, D.~C., {McNamara}, B.~R., {et~al.} 2005b, \apjl, 625, L9
\bibitem[{{Olivares} {et~al.}(2019)}]{Olivares2019} {Olivares}, V., {Salome}, P., {Combes}, F., {et~al.} 2019, \aap, 631, A22
\bibitem[{{Osterbrock} \& {Ferland}(2006)}]{Osterbrock2006} {Osterbrock}, D.~E. \& {Ferland}, G.~J. 2006, Astrophysics of gaseous nebulae and active galactic nuclei (2nd ed.; Sausalito, CA: Science Books)
\bibitem[{{Pasini} {et~al.}(2019)}]{Pasini2019} {Pasini}, T., {Gitti}, M., {Brighenti}, F., {et~al.} 2019, \apj, 885, 111
\bibitem[{{Pasini} {et~al.}(2021)}]{Pasini2021} {Pasini}, T., {Gitti}, M., {Brighenti}, F., {et~al.} 2021, \apj, 911, 66
\bibitem[{{Pizzolato} \& {Soker}(2005)}]{Pizzolato2005} {Pizzolato}, F. \& {Soker}, N. 2005, \apj, 632, 821
\bibitem[{{Prieto} {et~al.}(2016)}]{Prieto2016} {Prieto}, J.~L., {Kr{\"u}hler}, T., {Anderson}, J.~P., {et~al.} 2016, \apjl, 830, L32
\bibitem[{{Qiu} {et~al.}(2020)}]{Qiu2020} {Qiu}, Y., {Bogdanovi{\'c}}, T., {Li}, Y., {et~al.} 2020, Nature Astronomy, 4, 900
\bibitem[{{Qiu} {et~al.}(2021)}]{Qiu2021} {Qiu}, Y., {Hu}, H., {Inayoshi}, K., {et~al.} 2021, \apjl, 917, L7
\bibitem[{{Randall} {et~al.}(2011)}]{Randall2011} {Randall}, S.~W., {Forman}, W.~R., {Giacintucci}, S., {et~al.} 2011, \apj, 726, 86
\bibitem[{{Randall} {et~al.}(2015)}]{Randall2015} {Randall}, S.~W., {Nulsen}, P.~E.~J., {Jones}, C., {et~al.} 2015, \apj, 805, 112
\bibitem[{{Ricci} {et~al.}(2018)}]{Ricci2018} {Ricci}, F., {Lovisari}, L., {Kraft}, R.~P., {et~al.} 2018, \apj, 867, 35
\bibitem[{{Sanders} {et~al.}(2016)}]{Sanders2016} {Sanders}, Ryan L., {Shapley}, A.~E., {Kriek}, M., {et~al.} 2016, \apjl, 825, L23
\bibitem[{{Shabala} {et~al.}(2008)}]{Shabala2008} {Shabala}, S.~S., {Ash}, S., {Alexander}, P., {et~al.} 2008, \mnras, 388, 625
\bibitem[{{Speranza} {et~al.}(2021)}]{Speranza2021} {Speranza}, G., {Balmaverde}, B., {Capetti}, A., {et~al.} 2021, \aap, 653, A150
\bibitem[{{Spinrad} {et~al.}(1985)}]{Spinrad1985} {Spinrad}, H., {Djorgovski}, S., {Marr}, J. \& {Aguilar}, L. 1985, \pasp, 97, 932
\bibitem[{{Tremblay} {et~al.}(2009)}]{Tremblay2009} {Tremblay}, G.~R., {Chiaberge}, M., {Sparks}, W.~B., {et~al.} 2009, \apjs, 183, 278
\bibitem[{{Tremblay} {et~al.}(2012)}]{Tremblay2012} {Tremblay}, G.~R., {O'Dea}, C.~P., {Baum}, S.~A., {et~al.} 2012, \mnras, 424, 1026
\bibitem[{{Tremblay} {et~al.}(2015)}]{Tremblay2015} {Tremblay}, G.~R., {O'Dea}, C.~P., {Baum}, S.~A., {et~al.} 2015, \mnras, 451, 3768
\bibitem[{{Tremblay} {et~al.}(201)}]{Tremblay2016} {Tremblay}, G.~R., {Oonk}, J.~B.~R., {Combes}, F., {et~al.} 2016, \nat, 534, 218
\bibitem[{{Tucker} {et~al.}(2021)}]{Tucker2021} {Tucker}, M.~A., {Shappee}, B.~J., {Hinkle}, J.~T., {et~al.} 2021, \mnras, 506, 6014
\bibitem[{{Wilman} {et~al.}(2005)}]{Wilman2005} {Wilman}, R.~J., {Edge}, A.~C. \& {Johnstone}, R.~M. 2005, \mnras, 359, 755
\bibitem[{{Zirbel}(1996)}]{Zirbel1996} {Zirbel}, E.~L. 1996, \apj, 473, 713




\end{thebibliography}
\end{document}